\documentclass[11pt,a4paper,fontset=fandol]{ctexart}

\usepackage{textcomp}

\usepackage[margin=1in]{geometry}
\usepackage{parskip}
\usepackage{amsmath,amssymb,amsfonts}
\usepackage{graphicx}
\usepackage{booktabs}
\usepackage{tabularx}
\usepackage{subcaption}
\usepackage{microtype}
\usepackage{xcolor}
\usepackage[section]{placeins}
\usepackage{float}
\usepackage[sort,compress]{cite}
\usepackage{authblk}

\renewcommand\Affilfont{\small\itshape}

\renewcommand\Authands{, and }
\makeatletter
\def\@maketitle{%
  \newpage
  \null
  \vskip 2em%
  \noindent
  \begin{minipage}{\textwidth}%
    \let\footnote\thanks
    {\LARGE\@title\par}%
    \vskip 1.5em%
    {\normalsize\@author\par}%
    \vskip 1em%
    {\normalsize\@date}%
  \end{minipage}%
  \par
  \vskip 1.5em}
\let\bilingual@maketitle\maketitle
\let\bilingual@internalmaketitle\@maketitle
\let\bilingual@title\title
\let\bilingual@author\author
\let\bilingual@affil\affil
\let\bilingual@date\date
\let\bilingual@thanks\thanks
\let\bilingual@and\and
\makeatother

\usepackage{hyperref}
\hypersetup{
  colorlinks=true,
  breaklinks=true,
  linkcolor=blue!50!black,
  citecolor=blue!50!black,
  urlcolor=blue!60!black,
  pdfauthor={Xi Wang, Jun Lu, Cheng Zhang, Xiaoying Zhuang, Timon Rabczuk, Hehua Zhu, Wei Wu, Peinan Li, Jidong Zhao, Zhenyu Yin, Hao Sun},
  pdftitle={OpenPINN/PINO (English + Chinese)},
}

\makeatletter
\newcommand{\bibcitezh}[2]{%
  \@newl@bel{b}{#1.zh}{\hyper@@link[cite]{}{cite.#1.zh}{#2}}%
}
\makeatother

\begin{document}

\title{OpenPINN/PINO: An Extensible Open-Source Framework for Agent-Assisted Physics-Informed Neural Networks and Neural Operators}
\author[1,2]{Xi Wang}
\author[1]{Jun Lu}
\author[3]{Cheng Zhang}
\author[1,4]{Xiaoying Zhuang\thanks{Corresponding author. Email: \texttt{xiaoying.zhuang@gmail.com}}}
\author[4,5]{Timon Rabczuk}
\author[1]{Hehua Zhu}
\author[1]{Wei Wu}
\author[3]{Peinan Li}
\author[6]{Jidong Zhao}
\author[2]{Zhenyu Yin\thanks{Corresponding author. Email: \texttt{zhenyu.yin@polyu.edu.hk}}}
\author[7]{Hao Sun\thanks{Corresponding author. Email: \texttt{haosun@ruc.edu.cn}}}
\affil[1]{College of Civil Engineering, Tongji University, Shanghai 200092, China}
\affil[2]{Department of Civil and Environmental Engineering, The Hong Kong Polytechnic University, Hong Kong, 999077, China}
\affil[3]{College of Environmental Science and Engineering, Donghua University, Shanghai 201620, China}
\affil[4]{Institute of Computational Mechanics x Artificial Intelligence, Fudan University, Shanghai, 200433, China}
\affil[5]{Institute of Structural Mechanics, Bauhaus-Universit\"at Weimar, 99423 Weimar, Germany}
\affil[6]{Department of Civil and Environmental Engineering, The Hong Kong University of Science and Technology, Hong Kong, 999077, China}
\affil[7]{Gaoling School of Artificial Intelligence, Renmin University of China, Beijing 100872, China}

\date{}

\maketitle

\renewcommand{\abstractname}{Abstract}
\renewcommand{\figurename}{Figure}
\renewcommand{\tablename}{Table}
\begin{abstract}
Substantial challenges in physics-informed learning have motivated many improvements, but fragmented implementations hinder systematic organization, integration, and comparison.
We propose OpenPINN/PINO, an extensible open-source PyTorch framework for physics-informed neural networks and neural operators.
Two shared specifications connect problem setup, local computation, and evaluation.
Agents support problem formulation, solver planning, and diagnosis.
ProblemSpec defines geometry, governing equations, constitutive relations, parameters, initial and boundary conditions, and optional data.
MethodSpec configures the model, sampling, loss, optimization, and differentiation.
Researchers can register new problem templates and method components, then combine or compare strategies such as Fourier features, residual-adaptive sampling, and self-adaptive loss weighting for a fixed problem.
Agents translate natural-language descriptions into specifications, flag missing information, and propose solver plans.
The framework validates these specifications and deterministically generates local training and testing code.
A diagnostic agent reviews training logs and evaluation results to identify problems and suggest revisions.
Three examples show how method choices and agent diagnostics affect accuracy.
For viscous Burgers, Fourier features and residual-adaptive sampling each reduced relative $L^2$ error by about $75\%$ from a shared baseline.
For periodic Poisson, spectral differentiation reduced mean-free relative $L^2$ error by about $68\%$ across 16 held-out forcings.
For a plate with a hole, successive agent diagnoses guided nondimensionalization and optimization.
After initial training failure, all three runs met the $10\%$ target, with relative $L^2$ displacement errors of $3.51\%$--$6.31\%$.

\end{abstract}


\section{Introduction}
\label{sec:introduction}

Physics-informed neural networks (PINNs) embed governing equations into neural-network training through automatic differentiation and have become a standard mesh-free approach for forward and inverse partial differential equations (PDEs)~\cite{raissi2019pinn,lu2021deepxde}.
Neural operators extend the same scientific-machine-learning agenda from a single PDE instance to maps over families of instances~\cite{lu2021deeponet,li2021fno}.
Despite rapid progress, assembling a complete workflow remains difficult for many groups.
Researchers must define the physical problem and numerical method, and choose among growing collections of models, samplers, losses, and optimizers.

Open-source libraries such as SciANN showed that compact, mathematics-aligned APIs make PINNs teachable and extensible~\cite{haghighat2021sciann}.
Operator ecosystems such as DeepXDE and NeuralOperator matured in parallel~\cite{lu2021deepxde,kossaifi2026neuraloperator}.
A newer line of work uses language models and multi-agent systems to draft solver code from natural-language prompts.
These threads are still only loosely coupled in practice: problem definitions, method variants, PINN and operator backends, and agent-generated scripts often remain in separate libraries and workflows.

We present OpenPINN/PINO, an open-source PyTorch framework for agent-assisted physics-informed learning.
Shared \texttt{ProblemSpec} and \texttt{MethodSpec} contracts capture the physical problem and the solution method, feed PINN, neural-operator, and custom backends that train on the user's machine, and support postprocessing into metrics, figures, and experiment records.
A method ecosystem organizes registered components along five axes: models, sampling, losses, optimization, and differentiation.
Researchers can swap these axes across backends and register new components through the same interfaces.
An agent-assisted workflow spans problem understanding, solver planning, specification validation, deterministic code building, and diagnostic evaluation.
Users may enter through natural language when the upstream agents are available, or through a documentation-embedded configurator and YAML projects that begin at validation and code building.

\section{PINNs and Neural Operators}
\label{sec:algorithms}


\subsection{Deep neural networks}
A feed-forward network realizes a compositional map from coordinates or sensors to field values through stacked affine transforms and elementwise nonlinearities.
Weights and biases are the trainable parameters.
Physics-informed solvers and operator learners reuse this network algebra as a building block while targeting different maps.
A PINN approximates a single-instance field \(u(\mathbf{x})\).
A neural operator approximates a map \(a\mapsto u\) over a family of instances.

\subsection{Physics-informed neural networks}
Consider a generic steady or time-dependent PDE on a domain $\Omega$,
\begin{equation}
  \mathcal{N}[u](\mathbf{x}) = f(\mathbf{x}),
  \qquad \mathbf{x} \in \Omega,
\end{equation}
together with boundary or initial conditions $\mathcal{B}[u]=g$ on $\partial\Omega$ or at $t=0$.
A PINN replaces the unknown field $u$ by a neural network $\hat{u}(\cdot;\boldsymbol{\theta})$.
Automatic differentiation evaluates exact derivatives of network outputs with respect to inputs by applying the chain rule to the computational graph.
The method then penalizes equation and condition residuals on collocation sets $\mathcal{T}_f\subset\Omega$ and $\mathcal{T}_b\subset\partial\Omega$:
\begin{equation}
  \mathcal{L}(\boldsymbol{\theta})
  =
  w_f\,\mathcal{L}_f(\boldsymbol{\theta};\mathcal{T}_f)
  +
  w_b\,\mathcal{L}_b(\boldsymbol{\theta};\mathcal{T}_b)
  +
  w_d\,\mathcal{L}_d(\boldsymbol{\theta};\mathcal{T}_d),
\end{equation}
where $\mathcal{L}_f$ measures the PDE residual, $\mathcal{L}_b$ the boundary or initial mismatch, and $\mathcal{L}_d$ an optional data term on observations $\mathcal{T}_d$.
Training minimizes $\mathcal{L}$ with first-order or quasi-Newton optimizers.
Forward solves use known coefficients and conditions.
Inverse or hybrid settings leave selected parameters free and identify them jointly with $\boldsymbol{\theta}$ when data terms are present~\cite{raissi2019pinn}.

Relative to mesh-based discretizations, the PINN template is mesh-free at the residual level, shares one code path for many PDE types, and absorbs inverse tasks by extending the loss.
Accuracy still depends on sampling, loss weights, network capacity, and optimization, so method choices along those axes matter in practice.

\subsection{Neural operators}
A neural operator approximates a map between function spaces,
\begin{equation}
  \mathcal{G}:\; a \mapsto u,
\end{equation}
where $a$ may encode coefficients, forcing, or boundary data and $u$ is the corresponding solution field.
After training on a family of instances, evaluation on a new $a$ yields a surrogate solve without restarting a full PINN optimization from scratch.

DeepONet factors the operator through a branch network that encodes $a$ and a trunk network that encodes query coordinates, with their latent features combined to predict $u$~\cite{lu2021deeponet}.
The Fourier neural operator parameterizes global integral kernels in Fourier space and is naturally paired with structured grids~\cite{li2021fno}.
On such grids, derivatives for physics-informed training may be formed with spectral or finite-difference operators on structured arrays, as an alternative to pointwise automatic differentiation.
Physics-informed operator learning adds residual penalties on collocation grids or points, so training may use data terms, physics terms, or both.

\section{Related Work}
\label{sec:related}


\subsection{Physics-informed and operator-learning libraries}
Open-source libraries have made neural PDE solvers practical to write, teach, and extend.
Table~\ref{tab:library-survey} summarizes representative libraries in chronological order, together with publication year, framework, and documented emphasis.

\begin{table}[htbp]
  \centering
  \small
  \caption{Representative physics-informed and operator-learning libraries.
  Years follow the first public appearance used in this survey (journal online date, proceedings online date, or initial arXiv posting).
  NVIDIA Modulus succeeds SimNet in the same product line.}
  \label{tab:library-survey}
  \setlength{\tabcolsep}{3pt}
  \begin{tabularx}{\linewidth}{@{}l c l >{\raggedright\arraybackslash}X@{}}
    \toprule
    Library & Year & Framework & Emphasis \\
    \midrule
    NeuroDiffEq~\cite{chen2020neurodiffeq} & 2020 & Python / PyTorch & Neural differential-equation solvers \\
    SciANN~\cite{haghighat2021sciann} & 2020 & Keras / TensorFlow & Declarative scientific and PINN API \\
    DeepXDE~\cite{lu2021deepxde,lu2021deeponet} & 2021 & TensorFlow / PyTorch / JAX & Strong-form PINNs, DeepONet, geometry API \\
    TensorDiffEq~\cite{mcclenny2021tensordiffeq} & 2021 & TensorFlow & Multi-GPU forward and inverse PINNs \\
    SimNet / Modulus~\cite{hennigh2021simnet} & 2021 & PyTorch & Production backends, multiphysics templates \\
    IDRLnet~\cite{peng2021idrlnet} & 2021 & PyTorch & Modular PINN problem and model components \\
    NeuralPDE~\cite{zubov2021neuralpde} & 2021 & Julia / SciML & Automated PINN construction \\
    JAXPI~\cite{wang2023jaxpi} & 2023 & JAX & Training practices for stable PINNs \\
    NeuralOperator~\cite{kossaifi2026neuraloperator,li2021fno} & 2024 & PyTorch & Function-space operators, FNO family \\
    \bottomrule
  \end{tabularx}
\end{table}

NeuroDiffEq packages neural differential-equation solvers in Python~\cite{chen2020neurodiffeq}.
SciANN provides a Keras and TensorFlow wrapper for scientific and physics-informed deep learning with a declarative style~\cite{haghighat2021sciann}.
DeepXDE exposes geometry, residuals, boundary and initial conditions, and training loops behind a compact API aligned with the mathematical statement of a problem, and it covers strong-form PINNs together with operator-learning paths such as DeepONet~\cite{lu2021deepxde,lu2021deeponet}.
TensorDiffEq targets scalable multi-GPU forward and inverse PINN training~\cite{mcclenny2021tensordiffeq}.
NVIDIA SimNet, later evolved into Modulus, emphasizes production backends and multiphysics templates for physics-ML workflows~\cite{hennigh2021simnet}.
IDRLnet offers a PyTorch-oriented PINN library with modular problem and model components~\cite{peng2021idrlnet}.
Within the Julia SciML ecosystem, NeuralPDE automates PINN construction with associated error-approximation tooling~\cite{zubov2021neuralpde}.
Training-oriented toolkits such as JAXPI accompany documented best practices for stable PINN optimization~\cite{wang2023jaxpi}.
NeuralOperator provides a PyTorch library for learning maps between function spaces, including Fourier neural operators and related architectures for parametric PDE families~\cite{kossaifi2026neuraloperator,li2021fno}.

Collectively, these libraries lowered the barrier to single-backend PINN and operator experiments.

\subsection{Agent and language-model assisted solvers}
A newer line of systems uses language models and multi-agent workflows to draft, formalize, or tune physics-informed and broader SciML solvers from high-level problem descriptions.
Table~\ref{tab:agent-survey} summarizes representative efforts in chronological order.

\begin{table}[htbp]
  \centering
  \small
  \caption{Representative agent and language-model assisted systems for PDE and physics-informed learning.
  Years follow the first public appearance used in this survey.}
  \label{tab:agent-survey}
  \setlength{\tabcolsep}{3pt}
  \begin{tabularx}{\linewidth}{@{}l c >{\raggedright\arraybackslash}X@{}}
    \toprule
    System & Year & Emphasis \\
    \midrule
    PINNsAgent~\cite{wuwu2025pinnsagent} & 2025 & LLM-driven automated PDE surrogation with PINNs \\
    PDE-Controller~\cite{soroco2025pdecontroller} & 2025 & LLM autoformalization and reasoning for PDEs \\
    Lang-PINN~\cite{he2025langpinn} & 2025 & Multi-agent path from language to PINN pipelines \\
    ATHENA~\cite{toscano2025athena} & 2025 & Agentic teams for hierarchical numerical algorithms \\
    PDE-Agent~\cite{liu2025pdeagent} & 2025 & Toolchain-augmented multi-agent PDE solving \\
    OpInf-LLM~\cite{wang2026opinfllm} & 2026 & LLM-assisted parametric PDE solving via operator inference \\
    AgenticSciML~\cite{jiang2026agenticsciml} & 2026 & Collaborative agents for SciML architecture discovery \\
    EvoPINN~\cite{yin2026evopinn} & 2026 & Agentic discovery of executable PINN algorithms \\
    \bottomrule
  \end{tabularx}
\end{table}

PINNsAgent automates PDE surrogation with LLMs built around PINN training as the numerical substrate~\cite{wuwu2025pinnsagent}.
PDE-Controller uses LLMs for autoformalization and reasoning over PDE-governed systems~\cite{soroco2025pdecontroller}.
Lang-PINN coordinates agents that interpret natural-language problem statements into PDE formulations, architectures, losses, and training pipelines~\cite{he2025langpinn}.
ATHENA organizes agentic teams for hierarchical evolutionary design of numerical algorithms~\cite{toscano2025athena}.
PDE-Agent frames PDE solving as toolchain invocation through a multi-agent loop~\cite{liu2025pdeagent}.
OpInf-LLM couples language models with operator inference for parametric PDE families~\cite{wang2026opinfllm}.
AgenticSciML studies collaborative multi-agent discovery of SciML architectures, losses, and training strategies~\cite{jiang2026agenticsciml}.
EvoPINN targets agentic search for executable PINN algorithm designs~\cite{yin2026evopinn}.

\section{The OpenPINN/PINO Framework}
\label{sec:openpinn}


\subsection{Overview}
Relative to the libraries and agent systems surveyed in Section~\ref{sec:related}, OpenPINN/PINO emphasizes shared \texttt{ProblemSpec} and \texttt{MethodSpec} contracts across PINN and operator backends, registered method components for controlled comparison, and agents that validate specifications and build auditable local training projects.
The design follows an overall architecture of preprocessing, computation, and postprocessing, together with a five-module agent-assisted workflow on the same contracts.
The method ecosystem and qualitative trade-offs follow in Section~\ref{sec:methods}.

\subsection{Overall architecture}
OpenPINN/PINO is organized in the same broad stages used by commercial numerical workflows: preprocessing, computation, and postprocessing.
Preprocessing again carries most of the scientific decisions.
Computation is local training and testing on a chosen backend.
Postprocessing turns run artifacts into metrics, figures, and reusable records.
The same machine-readable contracts serve all three stages so that agent outputs, the documentation-embedded configurator, and YAML projects remain interchangeable.
Figure~\ref{fig:arch} summarizes the three stages, the shared \texttt{ProblemSpec} and \texttt{MethodSpec} contracts, the three computation backends, and the postprocessing products.

\begin{figure}[htbp]
  \centering
  \includegraphics[width=\linewidth]{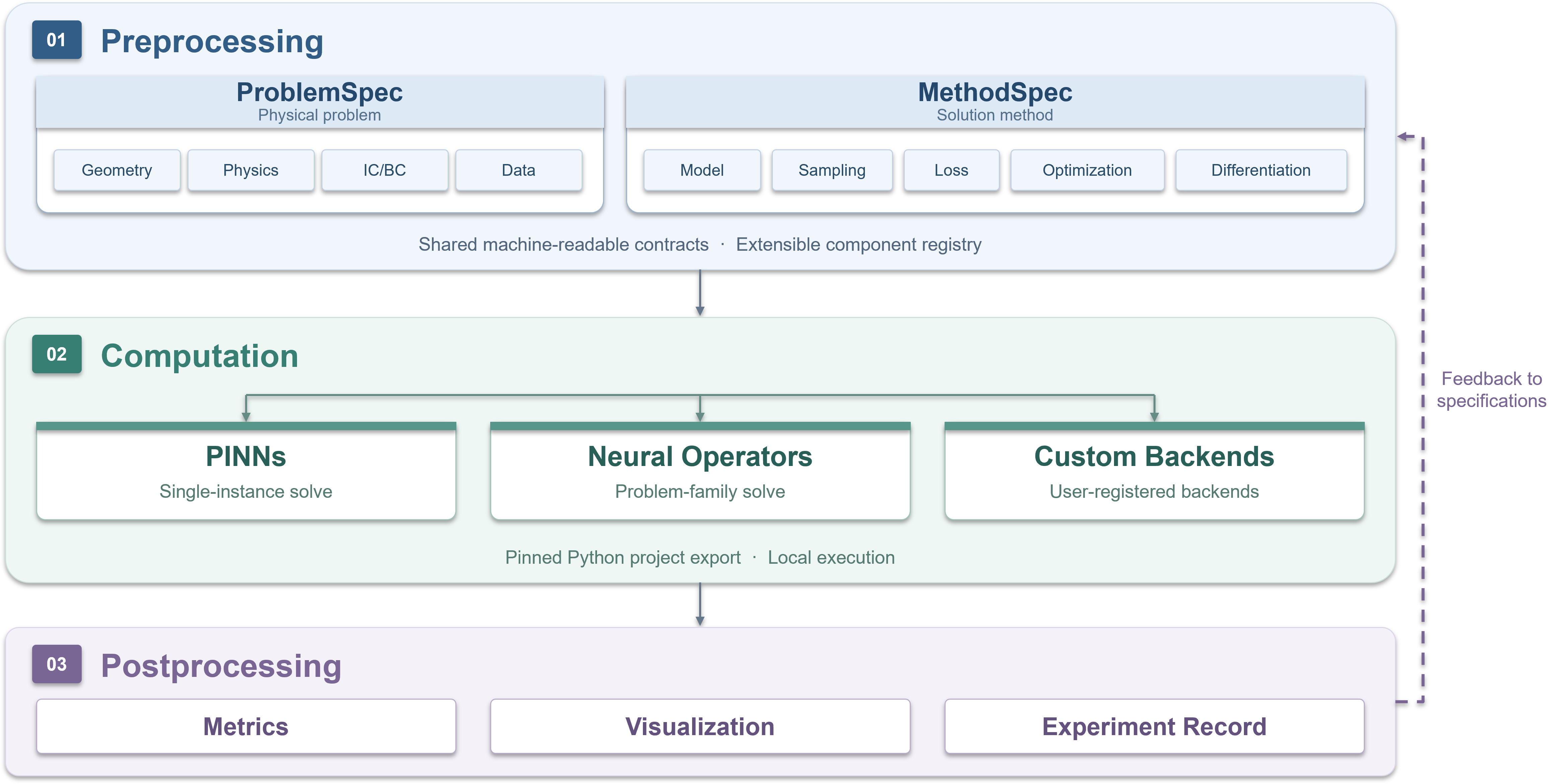}
  \caption{OpenPINN/PINO overall architecture.
  Preprocessing records the physical problem and solution method in shared specifications.
  Computation runs locally on a PINN, neural-operator, or custom backend.
  Postprocessing produces metrics, visualizations, and experiment records, with optional feedback to the specifications.}
  \label{fig:arch}
\end{figure}

\paragraph{Physical problem setting.}
The physical problem is recorded in \texttt{ProblemSpec} as mutually independent field groups.
\begin{itemize}
  \item \textbf{Geometry} defines the computational domain, coordinate conventions, and mesh or point-set hints.
  \item \textbf{Physics} defines governing equations, constitutive models, and material or equation parameters, including parameters left free for identification.
  \item \textbf{IC/BC} define initial conditions, boundary conditions, loads, and other constraints that close the problem.
  \item \textbf{Data} define optional measured or simulated observations used for inverse, data-driven, or hybrid settings, and as held-out references for testing and comparison.
\end{itemize}
These groups travel unchanged across PINN, operator, and custom backends.
New physics enters by extending the corresponding group or by registering a reusable template, without rewriting the rest of the architecture.

\paragraph{Solution method setting.}
The numerical method is recorded in \texttt{MethodSpec} as a second set of independent groups.
\begin{itemize}
  \item \textbf{Model} selects the network family, width and depth, activations, and hard-constraint wrappers.
  \item \textbf{Sampling} selects collocation, boundary, and data samplers, including fixed or adaptive strategies.
  \item \textbf{Loss} selects residual, boundary, initial, and data terms together with their weights or balancing rules.
  \item \textbf{Optimization} selects optimizers, learning-rate schedules, iteration budgets, and multi-stage training protocols.
  \item \textbf{Differentiation} selects how derivatives are formed, including coordinate automatic differentiation for pointwise models and structured-grid derivative operators for grid-native models.
\end{itemize}
Each group is backed by a component registry.
Users can swap a sampler, loss, or derivative path through the registries, which keeps the method surface general and extensible.

\paragraph{Computation backends.}
Training and testing run locally through one of three backends that consume the same validated specifications.
\begin{itemize}
  \item \textbf{PINN} trains a field network on one physical problem instance against equation and condition residuals, then evaluates that network on the same instance.
  \item \textbf{Neural operators} train a map over a family of instances and evaluate the learned operator on new inputs without restarting a full single-instance optimization.
  \item \textbf{Custom} accepts user-registered solvers, models, or training loops that implement the required interfaces and still pass through specification validation and deterministic export when possible.
\end{itemize}
Training, testing, checkpointing, and artifact write-out all run on the user's local machine.
Exported projects are ordinary Python packages with pinned specifications and seeds, so computation stays portable across machines and batch studies.

\paragraph{Postprocessing.}
Postprocessing reads the artifacts produced by local training and testing and turns them into auditable scientific products.
The designed surface has three groups: metrics and reference comparisons, visualization and export of plots or tables, and an experiment record that stores configuration pins, key metrics, and decision notes.
Application-specific derived quantities are evaluated when the corresponding postprocessors are available.

\subsection{Agent-assisted workflow}
The agent-assisted workflow is organized into five modules that take a problem from intake to locally runnable training code, with an optional post-run evaluation stage.
Figure~\ref{fig:workflow} summarizes the five modules, side inputs, local runtime, and the feedback path for interpretable adjustments.

\begin{figure}[htbp]
  \centering
  \includegraphics[width=\linewidth]{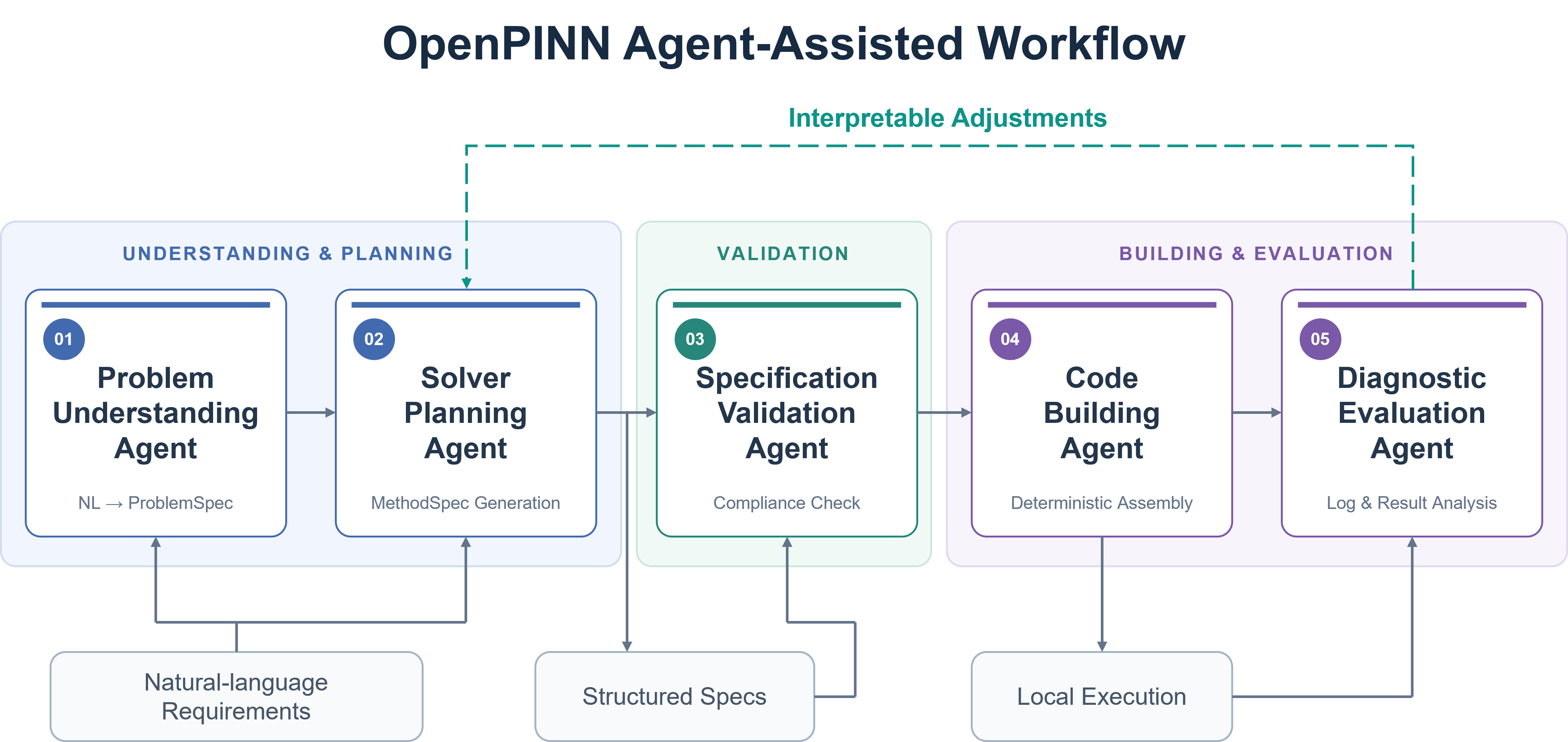}
  \caption{OpenPINN/PINO agent-assisted workflow.
  Natural-language and method requirements enter understanding and planning.
  Configurator or YAML specifications may join at specification validation.
  Code building exports local training and testing code.
  Diagnostic evaluation parses run artifacts, returns adjustments, or writes an experiment report.}
  \label{fig:workflow}
\end{figure}

\begin{enumerate}
  \item \textbf{Problem Understanding Agent.}
  This module accepts a natural-language description of the problem to be solved and produces a structured problem definition aligned with \texttt{ProblemSpec} fields such as geometry, physics, and IC/BC.

  \item \textbf{Solver Planning Agent.}
  Given that structured understanding, this module proposes a solution plan and recommends suitable models, sampling, losses, and optimizers.
  It can also ingest explicit user requirements for the solving scheme and encode them as a structured method configuration aligned with \texttt{MethodSpec}.

  \item \textbf{Specification Validation Agent.}
  This module checks whether the outputs of problem understanding and solver planning are complete and consistent enough for subsequent training.
  The checks act as a pre-submission compliance gate on the structured specifications before any numerical job is launched.
  The shipped validator covers this role for configurator and YAML inputs that already provide structured specifications.

  \item \textbf{Code Building Agent.}
  After successful validation, this module assembles registered templates and components into training and testing code that the user runs locally.
  Assembly is deterministic once the validated specifications are fixed, so that equivalent inputs yield auditable, repeatable exports.

  \item \textbf{Diagnostic Evaluation Agent.}
  After local training and testing, this module parses training-process logs and test-result files.
  When the parsed outcomes miss stated targets, it proposes interpretable adjustments to the problem or method specifications for another pass through planning or validation.
  When the targets are met, it compiles an experiment report that records the configuration, key metrics, and reproducibility pins for the run.
\end{enumerate}

Users may enter the agent-assisted workflow through natural language once understanding and planning are available, or they may supply structured specifications directly through the configurator or YAML and begin at specification validation and code building.

\label{sec:methods}


\subsection{Method ecosystem}
OpenPINN/PINO records the numerical method in \texttt{MethodSpec} as five independent groups: model, sampling, loss, optimization, and differentiation.
Each group is backed by a component registry so that axes can be swapped independently.
New models, samplers, losses, optimizers, or differentiation paths can be registered through the same interfaces, enlarging the method ecosystem over time.
This subsection describes these method groups and the practical trade-offs that guide method choice.
Quantitative comparisons appear with the demonstration cases in Section~\ref{sec:examples}.

\subsubsection{Model}
The model group admits pointwise field networks for single-instance PINNs and operator backbones for maps over problem families.
Table~\ref{tab:model-families} summarizes classical families that the registry is intended to host, together with the settings where each family is most often used.

\begin{table}[htbp]
  \centering
  \small
  \caption{Classical model families for the OpenPINN/PINO model axis.
  Entries are literature archetypes for an extensible registry, not a release inventory.}
  \label{tab:model-families}
  \setlength{\tabcolsep}{3pt}
  \renewcommand{\tabularxcolumn}[1]{m{#1}}
  \begin{tabularx}{\linewidth}{@{}l l >{\raggedright\arraybackslash}X@{}}
    \toprule
    Family & Typical forms & Common setting \\
    \midrule
    Pointwise MLP & MLP & Single-instance strong-form PINNs~\cite{raissi2019pinn} \\
    Spectral embedding & Fourier features & Multi-scale or high-frequency fields~\cite{tancik2020fourierfeatures,wang2021fourierbias} \\
    Hard constraints & Trial-solution / distance wrappers & Exact selected boundary conditions~\cite{lagaris1998annpde,lu2021hardconstraints} \\
    Domain decomposition & cPINN, XPINN-style multi-network & Large or multi-region domains~\cite{jagtap2020cpinn,jagtap2020xpinn} \\
    Branch--trunk operators & DeepONet & Irregular sensors, query coordinates~\cite{lu2021deeponet} \\
    Graph operators & GNO, multipole GNO & Irregular meshes, multi-scale graphs~\cite{li2020gno,li2020mgno} \\
    Spectral / kernel operators & FNO, low-rank kernels & Structured grids, parametric families~\cite{li2021fno,kovachki2023neuraloperator} \\
    Emerging bases & Physics-informed KANs & Alternative univariate bases in PINNs~\cite{wang2025pikan} \\
    \bottomrule
  \end{tabularx}
\end{table}

For PINNs, a plain multilayer perceptron remains the baseline map from coordinates to field values~\cite{raissi2019pinn}.
Fourier feature embeddings change the input representation of the same network family and enlarge the representable frequency content on multi-scale fields~\cite{tancik2020fourierfeatures,wang2021fourierbias}.
Hard-constraint wrappers change the trial representation so that selected boundary conditions are built into the network output, following classical ANN-PDE constructions and later PINN hard-constraint designs, at the cost of geometry-specific construction~\cite{lagaris1998annpde,lu2021hardconstraints}.
Domain-decomposition layouts such as conservative and extended PINNs replace a single global network by coupled subnetworks on space or space--time subdomains~\cite{jagtap2020cpinn,jagtap2020xpinn}.
Physics-informed Kolmogorov--Arnold networks illustrate how newer univariate bases can enter the same pointwise slot~\cite{wang2025pikan}.

For operator learning, DeepONet factors the map through a branch network on the input function and a trunk network on query coordinates, which suits irregular sensor layouts~\cite{lu2021deeponet}.
Graph neural operators learn integral kernels on graph discretizations and are natural when the spatial mesh is unstructured~\cite{li2020gno}.
Multipole graph neural operators add multi-scale graph coarsening to accelerate long-range interactions on large graphs~\cite{li2020mgno}.
Fourier neural operators parameterize global kernels in Fourier space and pair with structured grids and parametric PDE families~\cite{li2021fno}.
Low-rank kernel parameterizations belong to the same neural-operator family when a compact factorization of the integral kernel is preferred~\cite{kovachki2023neuraloperator}.
When the governing problem admits an energy structure, the deep energy method minimizes the total potential and is often preferable for variational solid-mechanics settings~\cite{samaniego2020dem}.
The same variational idea extends to operator learning in the variational physics-informed neural operator~\cite{eshaghi2025vino} and in physics-informed Transolver pretraining with optional finite-element warm starts~\cite{wang2026pfem}.
Transformer-based operators such as GeoPT, AB-UPT, and Transolver++ target complex geometries that dominate many engineering applications and have reported strong results on fluid and aircraft benchmarks~\cite{wu2026geopt,alkin2025abupt,luo2025transolverpp}.

\subsubsection{Sampling}
Sampling selects the collocation, boundary, and data locations at which residual and constraint terms are evaluated.
Table~\ref{tab:sampling-families} separates fixed non-adaptive designs from adaptive families organized by spatial indicators, and includes temporal curricula that control which time slices are active.

\begin{table}[htbp]
  \centering
  \small
  \caption{Representative sampling families for the OpenPINN/PINO sampling axis.
  Spatial adaptive rows are grouped by the indicator used to place or reweight points.
  Temporal curricula restrict or grow the active time support of the collocation set.}
  \label{tab:sampling-families}
  \setlength{\tabcolsep}{3pt}
  \renewcommand{\tabularxcolumn}[1]{m{#1}}
  \begin{tabularx}{\linewidth}{@{}l l >{\raggedright\arraybackslash}X@{}}
    \toprule
    Family & Typical forms & Common setting \\
    \midrule
    Non-adaptive & Uniform, grid, LHS / QMC & Fixed collocation baselines~\cite{raissi2019pinn} \\
    Residual-based adaptive & RAR, RAD, importance sampling & Concentrate points using PDE residuals~\cite{lu2021deepxde,wu2023adaptivesampling,nabian2021importance} \\
    Gradient-based adaptive & Residual/gradient ASMs & Sharp fronts and large solution gradients~\cite{mao2023gradientadaptive} \\
    Uncertainty-based adaptive & Ensemble-disagreement expansion & Grow the collocation set where models disagree~\cite{haitsiukevich2023ensemble} \\
    Learned adaptive & DAS-style density networks & High-dimensional residual landscapes~\cite{tang2023daspinn} \\
    Temporal curriculum & Time marching, growing time windows & Long-horizon evolution PDEs~\cite{wight2021adaptive,krishnapriyan2021failuremodes} \\
    \bottomrule
  \end{tabularx}
\end{table}

Non-adaptive sampling fixes the collocation set in advance.
Uniform random points, equispaced grids, and low-discrepancy sequences such as Latin hypercube or quasi-Monte Carlo draws are the usual baselines~\cite{raissi2019pinn}.
Residual-based adaptive sampling updates locations using the current PDE residual.
Residual-based adaptive refinement (RAR) keeps existing points and adds new ones where residuals remain large~\cite{lu2021deepxde}.
Residual-based adaptive distribution (RAD) keeps a prescribed point budget and redraws or reweights the set from a residual-informed density~\cite{wu2023adaptivesampling}.
Importance sampling likewise biases mini-batches with residual-informed proposals~\cite{nabian2021importance}.
Gradient-based adaptive sampling uses solution-gradient information, often together with residuals, to place points near sharp features~\cite{mao2023gradientadaptive}.
Uncertainty-based adaptive sampling expands the collocation set where an ensemble of PINNs disagrees away from anchored data or initial conditions~\cite{haitsiukevich2023ensemble}.
Learned adaptive sampling fits a parametric density, as in deep adaptive sampling (DAS), and draws new points from that density in high-dimensional settings~\cite{tang2023daspinn}.
Failure-informed enrichment is a related residual-indicator variant~\cite{gao2023failure}.
Temporal curriculum sampling restricts training to an active time window and then marches or grows that window toward the final time, so that early-time collocation is resolved before later-time points enter the loss~\cite{wight2021adaptive,krishnapriyan2021failuremodes}.
This choice changes which \((x,t)\) samples are present in the active set.
Time-causal residual reweighting that keeps all times visible but down-weights later residuals is recorded on the loss axis~\cite{wang2024causal}.

\subsubsection{Loss}
The loss group assembles residual, boundary, initial, and optional data terms into a scalar objective, and records how those terms are weighted during training.
Strong-form residuals follow the classical PINN template in Section~\ref{sec:algorithms}.
Table~\ref{tab:loss-families} summarizes classical weighting families that an extensible registry is intended to host.

\begin{table}[htbp]
  \centering
  \small
  \caption{Classical loss-weighting families for the OpenPINN/PINO loss axis.
  Entries are literature archetypes for an extensible registry, not a release inventory.}
  \label{tab:loss-families}
  \setlength{\tabcolsep}{3pt}
  \renewcommand{\tabularxcolumn}[1]{m{#1}}
  \begin{tabularx}{\linewidth}{@{}l l >{\raggedright\arraybackslash}X@{}}
    \toprule
    Family & Typical forms & Common setting \\
    \midrule
    Fixed-weight composite & Hand-tuned scalars & Classical PINN baseline~\cite{raissi2019pinn} \\
    Gradient-balanced & Learning-rate annealing & Competing backpropagated gradients~\cite{wang2021gradpathology} \\
    NTK-balanced & Eigenvalue weighting & Spectral imbalance among loss components~\cite{wang2022ntkpinn} \\
    Uncertainty-balanced & Learnable precision, inverse Dirichlet & Automatic term scaling~\cite{xiang2022lossbalanced,maddu2022inversedichlet} \\
    Pointwise self-adaptive & SA-PINN masks, residual attention & Spatially varying soft constraints~\cite{mcclenny2023sapinn,anagnostopoulos2024rba} \\
    Multi-objective schedules & ReLoBRaLo and related rules & Dynamic relative balancing~\cite{bischof2025multiloss} \\
    Causal / curriculum & Time-causal residual weights & Evolution and long-horizon PDEs~\cite{wang2024causal} \\
    \bottomrule
  \end{tabularx}
\end{table}

Fixed weights assign constant coefficients to each term and remain the default baseline on well-scaled problems~\cite{raissi2019pinn}.
Gradient-balanced schemes update coefficients from the magnitudes of backpropagated gradients so that residual and boundary terms compete more evenly~\cite{wang2021gradpathology}.
NTK-based schemes set weights from neural-tangent-kernel eigenvalues that govern the convergence rate of each loss component~\cite{wang2022ntkpinn}.
Uncertainty-style balancers introduce learnable precision parameters or inverse-Dirichlet weights that rescale terms as training progresses~\cite{xiang2022lossbalanced,maddu2022inversedichlet}.
Pointwise self-adaptive methods attach trainable or residual-driven masks to individual collocation points and strengthen soft constraints where local residuals remain large~\cite{mcclenny2023sapinn,anagnostopoulos2024rba}.
Multi-objective schedules such as relative loss balancing with random lookback update weights from recent loss histories without requiring gradient or kernel spectra~\cite{bischof2025multiloss}.
Causal weighting for time-dependent problems down-weights residuals at later times until earlier times are adequately resolved~\cite{wang2024causal}.

\subsubsection{Optimization}
The optimization group records the optimizer, learning-rate schedule, iteration budget, and multi-stage protocol used to minimize the assembled loss.
Table~\ref{tab:optimization-families} organizes classical families primarily by the derivative information used in each update, and lists hybrid schedules as protocols that sequence those families.

\begin{table}[htbp]
  \centering
  \small
  \caption{Classical optimization families for the OpenPINN/PINO optimization axis.
  The first three rows follow derivative order.
  The last row records multi-stage protocols that combine them.
  Entries are literature archetypes for an extensible registry, not a release inventory.}
  \label{tab:optimization-families}
  \setlength{\tabcolsep}{3pt}
  \renewcommand{\tabularxcolumn}[1]{m{#1}}
  \begin{tabularx}{\linewidth}{@{}l l >{\raggedright\arraybackslash}X@{}}
    \toprule
    Family & Typical forms & Common setting \\
    \midrule
    First-order & Adam, AdamW & Stochastic exploration and operator training~\cite{kingma2015adam,loshchilov2019adamw} \\
    Quasi-Newton & L-BFGS & Full-batch refinement of smooth PINN objectives~\cite{liu1989lbfgs,raissi2019pinn} \\
    Second-order / curvature & Gauss--Newton, natural gradient & High-accuracy refinement using Jacobian or Fisher structure~\cite{jnini2026gngd} \\
    Hybrid multi-stage & Adam then L-BFGS & Sequence first-order exploration with quasi-Newton polish~\cite{lu2021deepxde,urban2025pinnopt} \\
    \bottomrule
  \end{tabularx}
\end{table}

First-order methods such as Adam and AdamW update parameters from mini-batch gradients with moment estimates, and they are the usual workhorses for early training and for operator learning on large data sets~\cite{kingma2015adam,loshchilov2019adamw}.
Quasi-Newton methods such as limited-memory BFGS approximate inverse-Hessian information from gradient history without forming the full second-derivative tensor, and they are widely used for full-batch PINN solves once the loss landscape is locally smooth~\cite{liu1989lbfgs,raissi2019pinn}.
L-BFGS favors full-batch or large-batch evaluations and depends on line-search and memory settings.
Second-order and curvature-based methods use residual Jacobians or Fisher-like metrics explicitly, as in Gauss--Newton natural gradient descent, to accelerate high-accuracy refinement on physics-informed objectives~\cite{jnini2026gngd}.
Exact Newton and higher-order tensor methods appear far less often in the PINN literature, mainly because forming and inverting denser derivative tensors is costly at network scale.
Hybrid schedules first run Adam for a fixed number of iterations and then switch to L-BFGS~\cite{lu2021deepxde}.
Studies of the PINN optimization process show that carefully tuned multi-stage quasi-Newton refinement can reach competitive accuracy on classical benchmarks~\cite{urban2025pinnopt}.
Learning-rate schedules and iteration budgets are stored with the chosen optimizer so that the full training recipe remains reproducible.

\subsubsection{Differentiation}
Differentiation selects how spatial or space--time derivatives in the PDE residual are formed from model outputs.
Table~\ref{tab:diff-families} summarizes classical families that an extensible registry is intended to host, spanning pointwise automatic differentiation, discrete and spectral grid operators, weak-form residual constructions, and hybrid numerical--automatic schemes.

\begin{table}[htbp]
  \centering
  \small
  \caption{Classical differentiation families for the OpenPINN/PINO differentiation axis.
  Entries are literature archetypes for an extensible registry, not a release inventory.}
  \label{tab:diff-families}
  \setlength{\tabcolsep}{3pt}
  \renewcommand{\tabularxcolumn}[1]{m{#1}}
  \begin{tabularx}{\linewidth}{@{}l l >{\raggedright\arraybackslash}X@{}}
    \toprule
    Family & Typical forms & Common setting \\
    \midrule
    Coordinate AD & Forward / reverse-mode AD & Pointwise strong-form PINNs and query-based operators~\cite{baydin2018autodiff,raissi2019pinn,wang2021pideeponet} \\
    Numerical differentiation & Finite differences on points or grids & AD substitutes and verification baselines~\cite{chiu2022canpinn} \\
    Spectral grid derivatives & Fourier / spectral differentiation & Periodic structured grids and physics-informed FNO-style training~\cite{li2021fno,li2024pino} \\
    Discrete grid operators & FD / FV-style array stencils & Non-periodic structured grids~\cite{li2024pino,chiu2022canpinn} \\
    Weak / variational residuals & Integration by parts, VPINN & Lower the classical derivative order required of the network~\cite{kharazmi2021hpvpinn} \\
    Hybrid AD--numerical & Coupled automatic--numerical schemes & Mix exact AD with local stencils for speed or stability~\cite{chiu2022canpinn} \\
    \bottomrule
  \end{tabularx}
\end{table}

Coordinate automatic differentiation applies the chain rule on the computational graph to obtain exact derivatives of network outputs with respect to input coordinates, and it is the standard path for strong-form pointwise PINNs~\cite{baydin2018autodiff,raissi2019pinn}.
The same coordinate path extends to operator models that evaluate fields at query locations, including physics-informed DeepONet training~\cite{wang2021pideeponet}.
Over a batch of input functions and many collocation points, treating every coordinate sample as an independent AD leaf creates a many-root, many-leaf derivative layout that is costly for reverse-mode AD and tends to duplicate the computational graph along the function batch~\cite{leng2024zerocoords}.
Operator learning on structured grids often replaces pointwise AD by spectral differentiation in Fourier space or by finite-difference and related array stencils acting on discrete fields, which pairs naturally with FNO-style backbones and physics-informed neural operators~\cite{li2021fno,li2024pino}.
Numerical differentiation on collocation clouds or grids also serves as a stand-alone residual path and as a building block in coupled automatic--numerical schemes such as CAN-PINN~\cite{chiu2022canpinn}.
Weak and variational formulations move derivatives onto test functions through integration by parts, so the network needs fewer classical derivatives in the strong-form sense~\cite{kharazmi2021hpvpinn}.


\section{Demonstration Examples}
\label{sec:examples}


OpenPINN/PINO records a research problem in \texttt{ProblemSpec} and a numerical recipe in \texttt{MethodSpec}, then swaps registered components along the method axes surveyed in Section~\ref{sec:methods}.
Researchers can run controlled comparisons on the problems they study and keep their attention on method choice and scientific outcomes, while specification validation and deterministic code export assemble the locally runnable training path.
This section reports three complementary demonstrations.
The first solves a one-dimensional viscous Burgers equation under five PINN recipes that differ in one registered method component at a time.
The second shows the same style of controlled comparison on a parametric Poisson operator family with an FNO backbone.
The third records an agent-assisted plate-with-hole elasticity workflow from the Problem Understanding Agent through the Diagnostic Evaluation Agent, together with a researcher-led repair path that reaches a finite-element relative $L^2$ below $0.1$.
All quantitative results below come from pinned OpenPINN-upstream reference runs archived with the preprint materials.

\subsection{PINN demonstration}
\label{sec:ex-pinn}
The PINN demonstration uses the viscous Burgers equation
\begin{equation}
  u_t + u\,u_x - \nu u_{xx} = 0,
  \qquad x\in[-1,1],\quad t\in[0,1],
\end{equation}
with viscosity $\nu=0.02$, initial condition $u(x,0)=-\sin(\pi x)$, and homogeneous Dirichlet boundaries.
Five PINN recipes are compared on this fixed Burgers setting.
Four of them start from a shared baseline and change exactly one registered method component.
Table~\ref{tab:ch5-pinn-burgers} summarizes the five recipes.

\begin{center}
\begin{minipage}{\linewidth}
  \footnotesize
  \captionof{table}{PINN recipes on the Burgers problem with $\nu=0.02$.
  Relative $L^2$ is against the Cole--Hopf reference on a held-out $129\times51$ space--time grid.}
  \label{tab:ch5-pinn-burgers}
  \setlength{\tabcolsep}{2.5pt}
  \begin{tabularx}{\linewidth}{@{}l l >{\raggedright\arraybackslash}X c c@{}}
    \toprule
    Recipe & Changed component & Registered change vs \texttt{Pin-B} & Rel.\ $L^2$ & Ratio vs \texttt{Pin-B} \\
    \midrule
    Pin-B & baseline & MLP, uniform collocation, fixed weights, Adam then L-BFGS & $1.16\times10^{-2}$ & $1.00$ \\
    Pin-M & model & random Fourier features, scale $0.25$ & $2.82\times10^{-3}$ & $0.244$ \\
    Pin-S & sampling & residual-adaptive interior sampling & $2.96\times10^{-3}$ & $0.255$ \\
    Pin-L & loss & self-adaptive loss weights & $9.04\times10^{-3}$ & $0.781$ \\
    Pin-O & optimization & Adam only, no L-BFGS & $4.47\times10^{-2}$ & $3.86$ \\
    \bottomrule
  \end{tabularx}
\end{minipage}
\end{center}

\begin{figure}[H]
  \centering
  \captionsetup[subfigure]{skip=1.5pt,font=scriptsize}
  \resizebox{0.78\linewidth}{!}{\begin{minipage}{\linewidth}
  \begin{subfigure}{0.48\linewidth}
    \centering
    \includegraphics[width=\linewidth]{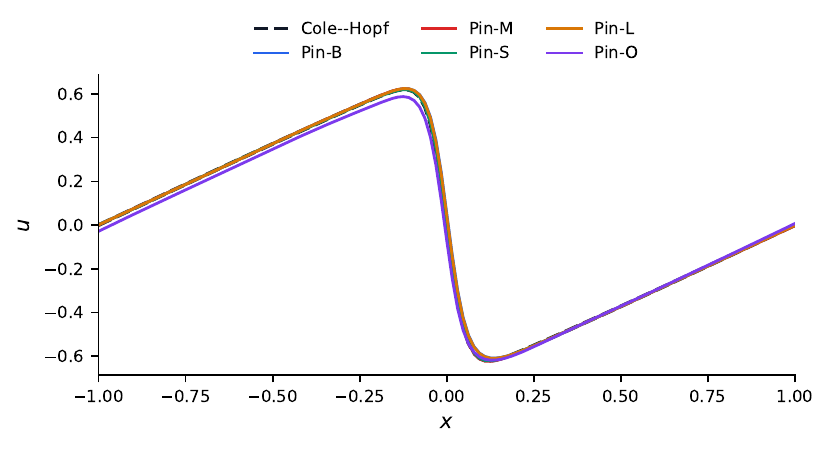}
    \caption{Predicted solutions at $t=1$ against the Cole--Hopf reference.}
    \label{fig:ch5-pinn-a}
  \end{subfigure}
  \hfill
  \begin{subfigure}{0.48\linewidth}
    \centering
    \includegraphics[width=\linewidth]{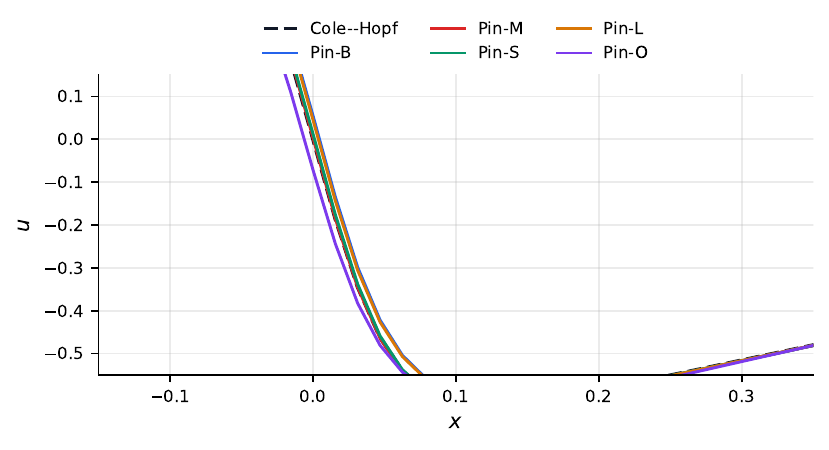}
    \caption{Zoom near the steep front.}
    \label{fig:ch5-pinn-b}
  \end{subfigure}

  \vspace{0.25em}
  \begin{subfigure}{0.325\linewidth}
    \centering
    \includegraphics[width=\linewidth]{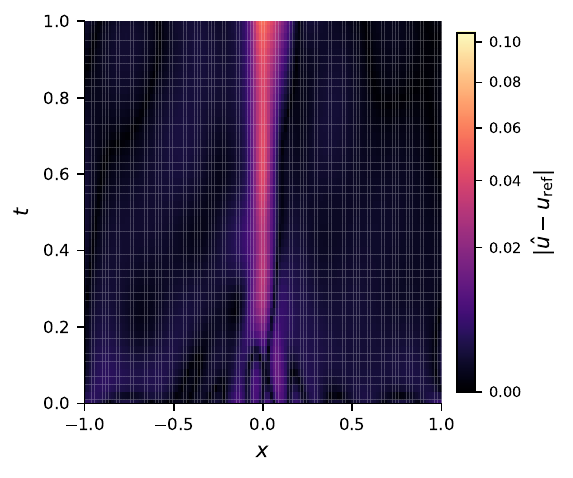}
    \caption{\texttt{Pin-B} absolute error.}
    \label{fig:ch5-pinn-c}
  \end{subfigure}
  \hfill
  \begin{subfigure}{0.325\linewidth}
    \centering
    \includegraphics[width=\linewidth]{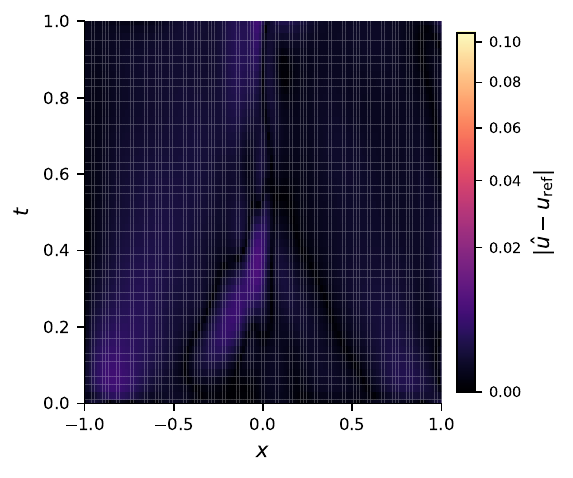}
    \caption{\texttt{Pin-M} absolute error.}
    \label{fig:ch5-pinn-d}
  \end{subfigure}
  \hfill
  \begin{subfigure}{0.325\linewidth}
    \centering
    \includegraphics[width=\linewidth]{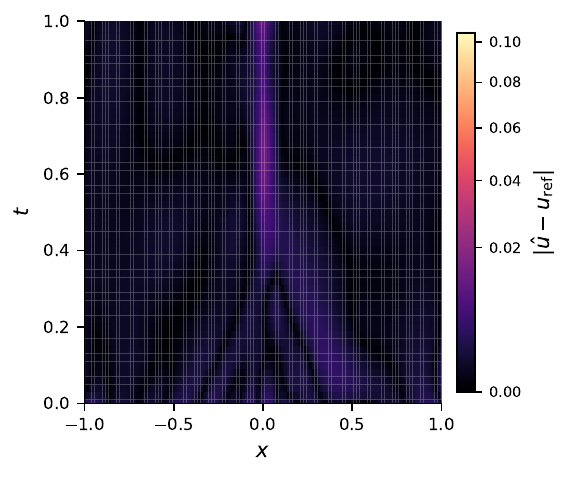}
    \caption{\texttt{Pin-S} absolute error.}
    \label{fig:ch5-pinn-e}
  \end{subfigure}

  \vspace{0.25em}
  \begin{center}
  \begin{subfigure}{0.325\linewidth}
    \centering
    \includegraphics[width=\linewidth]{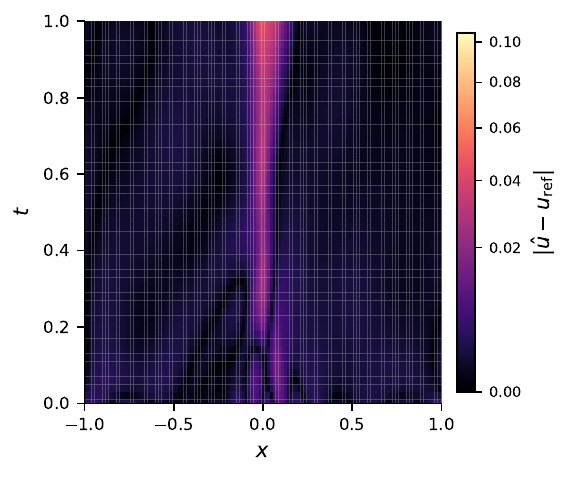}
    \caption{\texttt{Pin-L} absolute error.}
    \label{fig:ch5-pinn-f}
  \end{subfigure}
  \hspace{0.04\linewidth}
  \begin{subfigure}{0.325\linewidth}
    \centering
    \includegraphics[width=\linewidth]{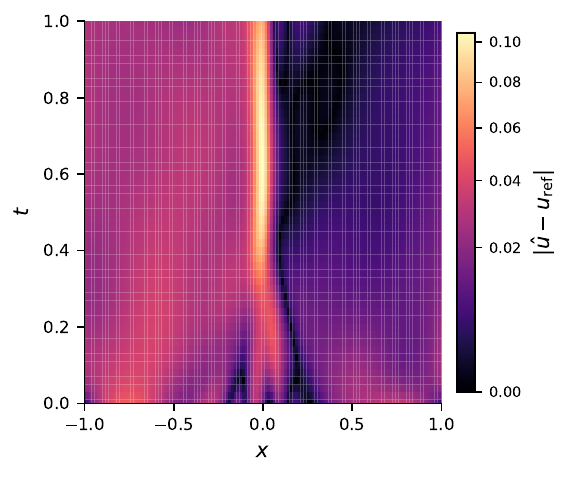}
    \caption{\texttt{Pin-O} absolute error.}
    \label{fig:ch5-pinn-g}
  \end{subfigure}
  \end{center}
  \end{minipage}}

  \caption{Burgers demonstration with $\nu=0.02$.
  Panels~(c)--(g) show space--time absolute-error fields $|\hat{u}-u_{\mathrm{ref}}|$ on a shared color scale.}
  \label{fig:ch5-pinn-burgers}
\end{figure}

The baseline \texttt{Pin-B} uses a multilayer perceptron of width $[64,64,64,64]$ with $\tanh$ activations, identity input features, uniform collocation with $256$ interior points, $64$ boundary points, and $64$ initial points, fixed loss weights, Adam for $3000$ steps then L-BFGS for $40$ outer steps, and coordinate automatic differentiation.
Accuracy is measured by relative $L^2$ error against a Cole--Hopf reference on a held-out space--time grid.
\texttt{Pin-M} replaces identity features by a random Fourier-feature map with $16$ frequencies and scale $0.25$.
\texttt{Pin-S} replaces the frozen uniform interior batch by residual-adaptive sampling under the same point budget.
\texttt{Pin-L} replaces fixed loss weights by registered self-adaptive weighting with weight learning rate $0.01$.
\texttt{Pin-O} removes the L-BFGS stage and keeps the same $3000$ Adam steps only.
All relative errors lie below $0.1$, with the largest gain from Fourier features and a clear rise when L-BFGS is removed.
Figure~\ref{fig:ch5-pinn-burgers} shows the solution at $t=1$, a zoom near the steep front, and space--time absolute-error fields for all five recipes.

\subsection{Neural-operator demonstration}
\label{sec:ex-operator}
The operator demonstration uses the same controlled comparison style on a parametric family rather than a single field.
The family is the periodic Poisson problem $-\Delta u=f$ on $[0,2\pi)^2$.
Each instance draws a Fourier-series solution with harmonics from order one to four, and the forcing is the exact negative Laplacian of that series.
Periodic fluxes cancel and the series has no zero mode, so the spatial mean of $f$ is zero and the manufactured reference $u$ is zero-mean.
Because $u$ and $u+C$ satisfy the same equation, a physics residual alone leaves the additive constant free.
The hybrid recipe pins that constant through nodal observations.
An FNO backbone learns the map $f\mapsto u$ from $32$ training forcings and is evaluated on $16$ held-out forcings.
Three recipes share one physics-only finite-difference baseline \texttt{Op-B}.
Table~\ref{tab:ch5-op-pairs} reports two held-out errors, each averaged over the sixteen forcings.
The raw relative $L^2$ compares prediction and reference fields directly.
The mean-free relative $L^2$ subtracts the spatial mean of each field on every instance before the same relative norm, so it scores the shape of $u$ up to the additive constant.

\begin{figure}[H]
  \centering
  \captionsetup[subfigure]{skip=1.5pt,font=scriptsize}
  \resizebox{0.78\linewidth}{!}{\begin{minipage}{\linewidth}
  \begin{subfigure}{\linewidth}
    \centering
    \includegraphics[width=\linewidth]{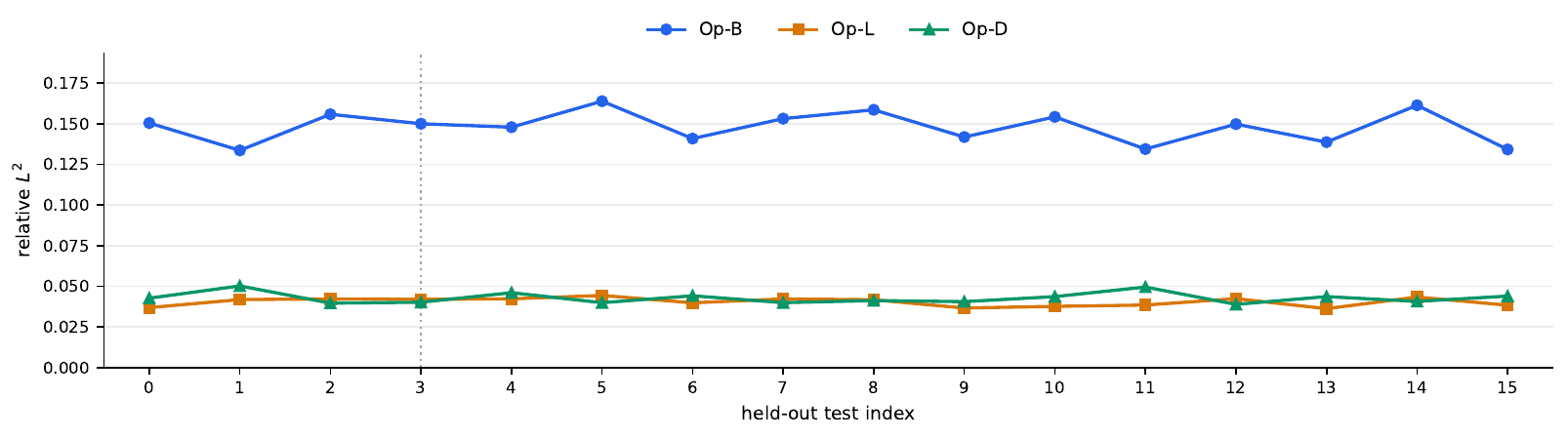}
    \caption{Per-instance raw relative $L^2$ for recipes \texttt{Op-B}, \texttt{Op-L}, and \texttt{Op-D} over all sixteen tests.}
    \label{fig:ch5-op-a}
  \end{subfigure}

  \vspace{0.25em}
  \begin{center}
  \begin{subfigure}{0.325\linewidth}
    \centering
    \includegraphics[width=\linewidth]{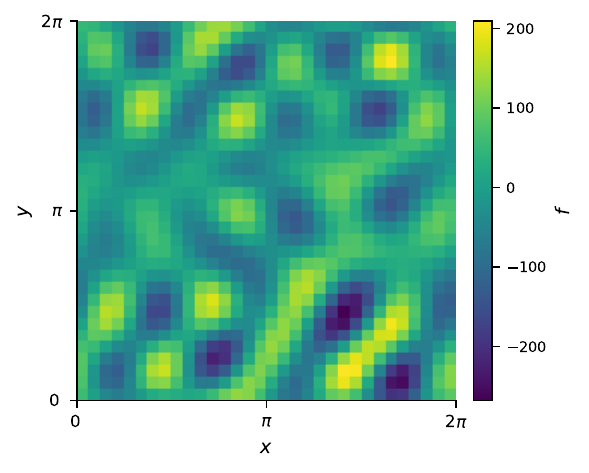}
    \caption{Forcing $f$ on the median-\texttt{Op-B} instance.}
    \label{fig:ch5-op-b}
  \end{subfigure}
  \hspace{0.04\linewidth}
  \begin{subfigure}{0.325\linewidth}
    \centering
    \includegraphics[width=\linewidth]{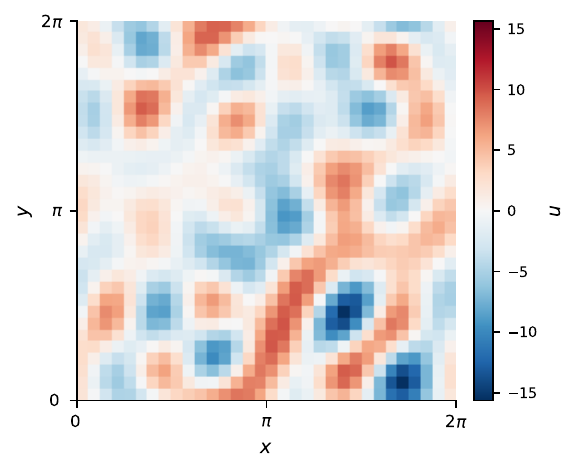}
    \caption{Reference $u$ on the same instance.}
    \label{fig:ch5-op-c}
  \end{subfigure}
  \end{center}

  \vspace{0.25em}
  \begin{subfigure}{0.325\linewidth}
    \centering
    \includegraphics[width=\linewidth]{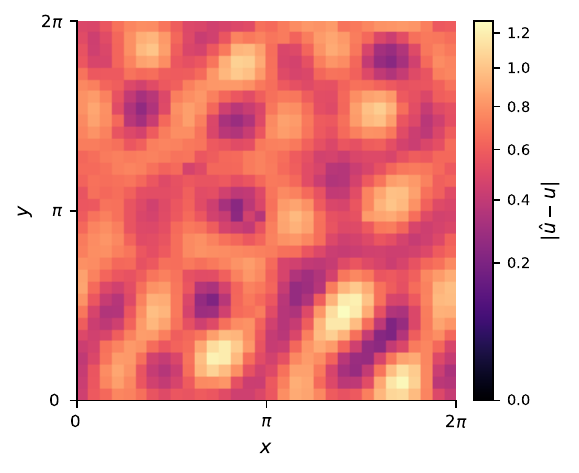}
    \caption{\texttt{Op-B} absolute error.}
    \label{fig:ch5-op-d}
  \end{subfigure}
  \hfill
  \begin{subfigure}{0.325\linewidth}
    \centering
    \includegraphics[width=\linewidth]{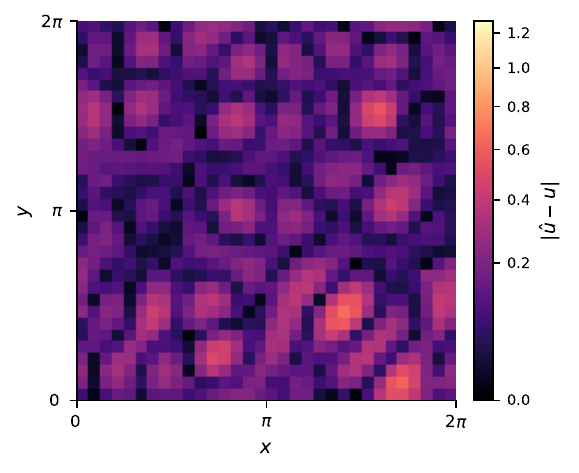}
    \caption{\texttt{Op-L} absolute error.}
    \label{fig:ch5-op-e}
  \end{subfigure}
  \hfill
  \begin{subfigure}{0.325\linewidth}
    \centering
    \includegraphics[width=\linewidth]{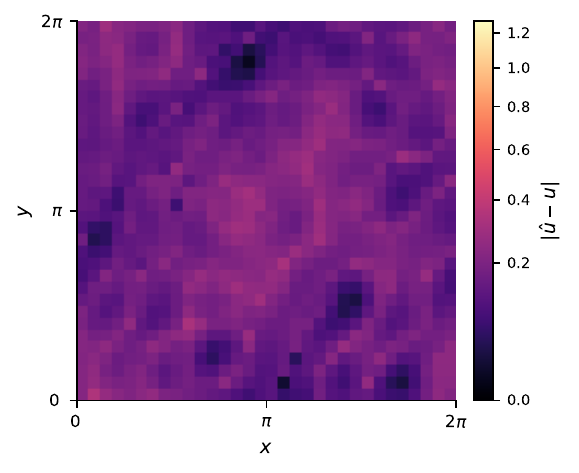}
    \caption{\texttt{Op-D} absolute error.}
    \label{fig:ch5-op-f}
  \end{subfigure}
  \end{minipage}}

  \caption{Operator demonstration on the held-out periodic Poisson set.
  Panels~(b)--(f) use the median-\texttt{Op-B} held-out instance; panels~(d)--(f) share one absolute-error color scale.}
  \label{fig:ch5-op-fields}
\end{figure}

\begin{center}
\begin{minipage}{\linewidth}
  \footnotesize
  \captionof{table}{Operator recipes on the periodic Poisson family with an FNO backbone.
  Errors are means over $16$ held-out forcings.}
  \label{tab:ch5-op-pairs}
  \setlength{\tabcolsep}{2.5pt}
  \begin{tabularx}{\linewidth}{@{}l l >{\raggedright\arraybackslash}X c c@{}}
    \toprule
    Recipe & Changed component & Registered change vs \texttt{Op-B} & Rel.\ $L^2$ & Mean-free Rel.\ $L^2$ \\
    \midrule
    Op-B & baseline & physics-only loss, finite-difference derivatives & $1.48\times10^{-1}$ & $3.89\times10^{-2}$ \\
    Op-L & loss & hybrid physics-plus-data, 16 observation nodes & $4.04\times10^{-2}$ & $3.99\times10^{-2}$ \\
    Op-D & differentiation & spectral derivatives & $4.28\times10^{-2}$ & $1.23\times10^{-2}$ \\
    \bottomrule
  \end{tabularx}
\end{minipage}
\end{center}

The shared recipe uses a $32\times32$ grid, width $24$, three layers, ten modes, and Adam for $1500$ steps at learning rate $3\times10^{-3}$.
\texttt{Op-L} changes only the loss assembly to a hybrid physics-plus-data loss with sixteen observation nodes per instance.
\texttt{Op-D} changes only the grid differentiation path to spectral differentiation.
For \texttt{Op-L}, the raw relative $L^2$ drops sharply while the mean-free score stays near the \texttt{Op-B} level.
The nodal observations therefore fix the free additive constant against the zero-mean manufactured solution, and the finite-difference physics residual already accounts for most of the resolved field shape.
\texttt{Op-D} improves both scores, with the larger gain on the mean-free error, so spectral differentiation tightens the field shape beyond constant alignment.
Figure~\ref{fig:ch5-op-fields} shows the per-instance raw relative $L^2$ over all sixteen tests in Fig.~\ref{fig:ch5-op-a}, the median-\texttt{Op-B} forcing and reference in Figs.~\ref{fig:ch5-op-b}--\ref{fig:ch5-op-c}, and the corresponding absolute-error fields in Figs.~\ref{fig:ch5-op-d}--\ref{fig:ch5-op-f}.

\subsection{Agent-assisted workflow demonstration}
\label{sec:ex-agent}
The third case records an agent-assisted configurator session on a two-dimensional plate-with-hole problem in plane-stress linear elasticity.
A square plate on $[0,2]^2\,\mathrm{m}$ contains a circular hole of radius $0.25\,\mathrm{m}$ centered at $(1,1)\,\mathrm{m}$.
The solid is isotropic with $E=210\,\mathrm{GPa}$ and $\nu=0.3$.
The left edge is fixed, the right edge carries a uniform uniaxial traction of $1\,\mathrm{MPa}$, and the remaining outer edges and the hole boundary are traction-free.
Unknowns are the displacement components $u_x$ and $u_y$.
Unless noted otherwise, the field panels that follow show seed~$0$ predictions against an FEM reference on a probe mesh with $h=0.025$, and that FEM solution is used only for post-hoc evaluation.

The session starts from the documentation-embedded configurator with the prefilled plate statement and a short clarifying prompt.
Figure~\ref{fig:ch5-plate-understand} shows the Problem Understanding and Solver Planning panels returned by the live interfaces.

\begin{figure}[H]
  \centering
  \includegraphics[width=0.65\linewidth]{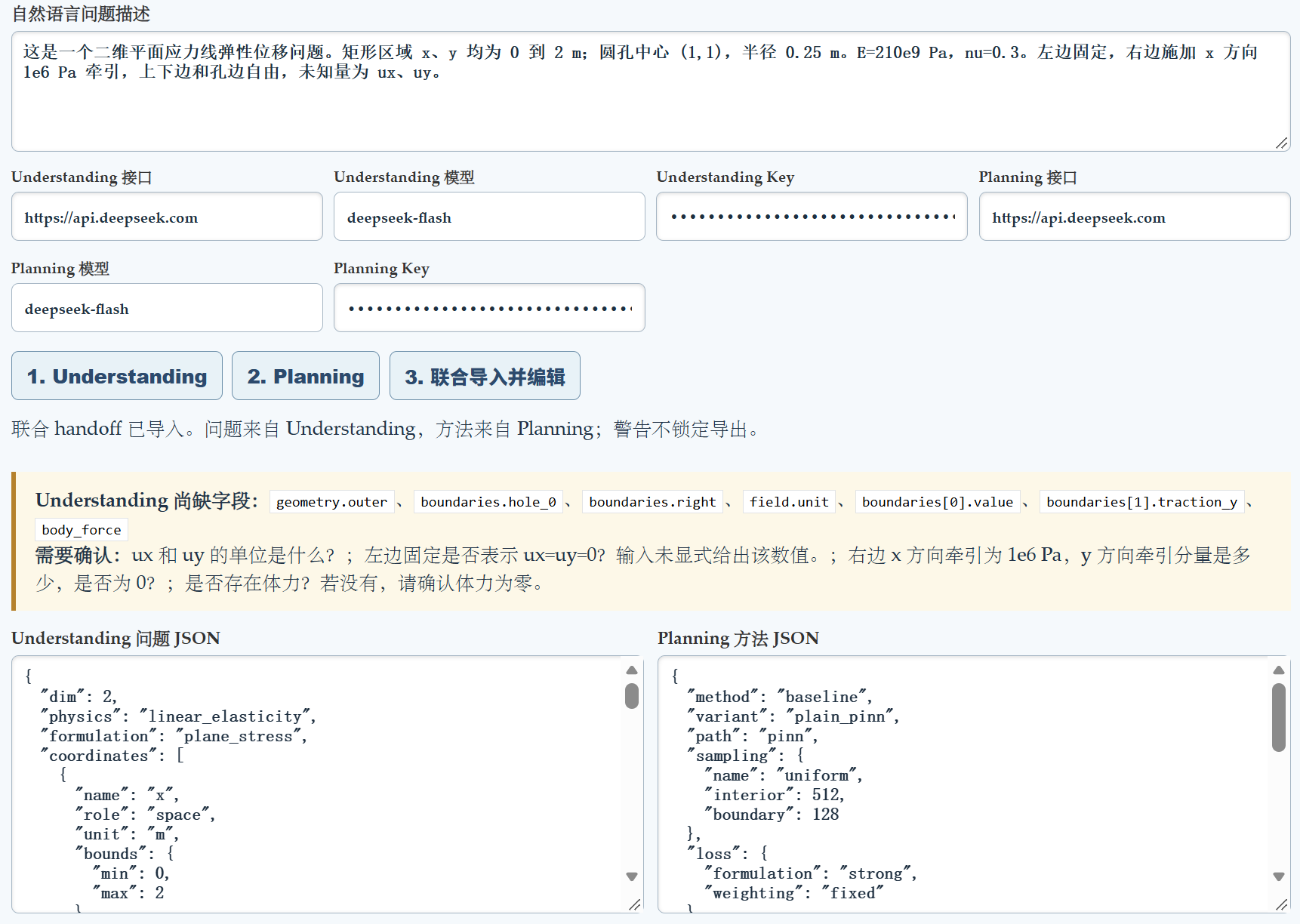}
  \caption{Problem Understanding Agent and Solver Planning Agent stage for the plate-with-hole case.}
  \label{fig:ch5-plate-understand}
\end{figure}

\begin{figure}[H]
  \centering
  \includegraphics[width=0.65\linewidth]{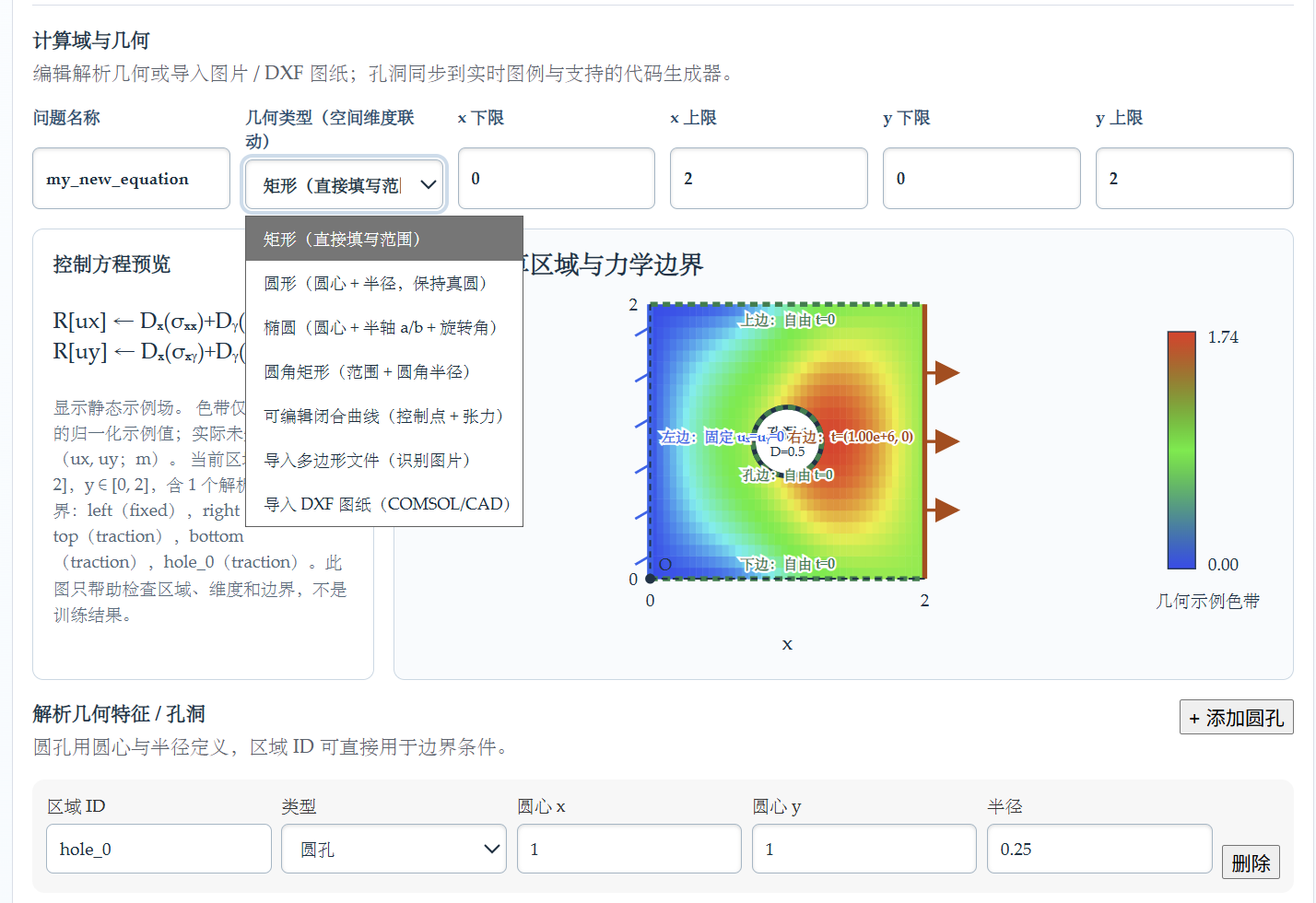}
  \caption{Configurator PDE sketch and domain preview after joint import and manual edits.}
  \label{fig:ch5-plate-sketch}
\end{figure}

The Problem Understanding Agent turns the natural-language statement into a structured problem description aligned with \texttt{ProblemSpec} fields such as geometry, physics, and boundary conditions.
It flags missing mapped outer geometry and the hole-boundary row, and prompts the user to supply the incomplete fields.
The Solver Planning Agent then proposes a baseline plain-PINN recipe as a structured \texttt{MethodSpec}.
It selects a multilayer-perceptron model with $\tanh$ activations, uniform collocation, fixed loss weights, Adam, and coordinate automatic differentiation for the strong-form equilibrium residual, together with a width-$64$ depth-$3$ network and $5000$ Adam steps.
Both drafts are stored as machine-readable JSON for subsequent import into the configurator.
After joint import, the researcher completes the hole center and the traction-free top, bottom, and hole edges in the configurator, and raises the display budget from the planned width-$64$ depth-$3$ network with $5000$ Adam steps to a width-$64$ depth-$4$ network with $10000$ Adam steps.
Figure~\ref{fig:ch5-plate-sketch} shows the configurator equation preview and geometry sketch after these edits.

\begin{figure}[H]
  \centering
  \includegraphics[width=0.65\linewidth]{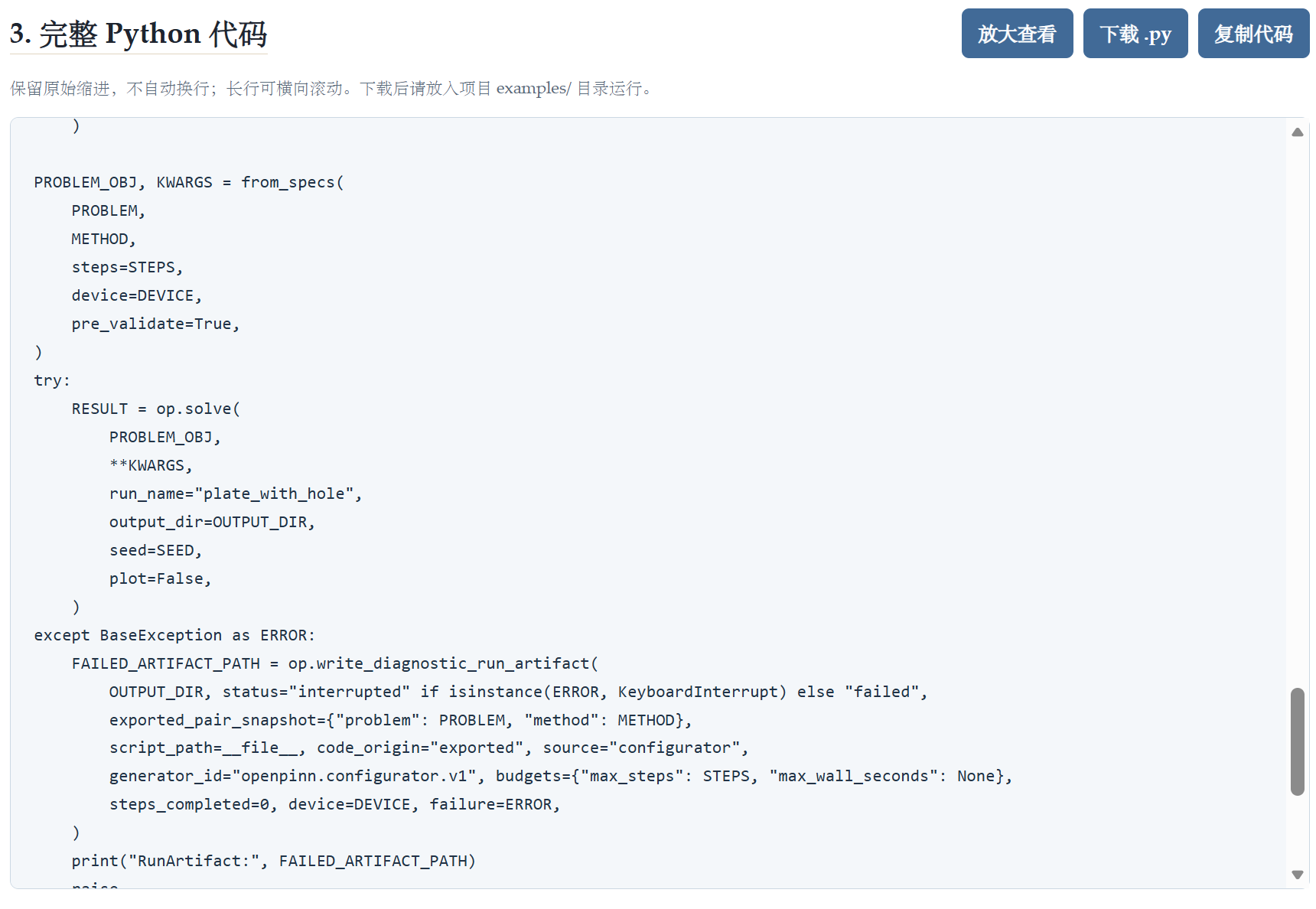}
  \caption{Code Building Agent export for the validated plate-with-hole specifications.}
  \label{fig:ch5-plate-codegen}
\end{figure}

The Code Building Agent assembles a training script from the validated specifications by deterministic template export.
The researcher downloads that script and runs training locally.
Figure~\ref{fig:ch5-plate-codegen} shows the exported Python project panel.

The researcher runs the downloaded script for $10000$ Adam steps on CPU in SI units under seed~$0$.
Training writes a run directory with loss history, metrics, status records, and a diagnostic run artifact.
The researcher submits that run artifact to the Diagnostic Evaluation Agent.
The Diagnostic Evaluation Agent returns assessment \texttt{concerns}.
Its deterministic track marks the run valid but unhealthy with outcome \texttt{budget\_exhausted}.
It reports a final total loss of $2.19\times10^{22}$ that falls only about $1.6\%$ from the initial value over the full budget, PDE residual RMSE near $1.71\times10^{10}$, and traction-boundary RMSEs of order $10^{10}$.
The semantic report offers non-binding repair hints on residual scaling, Neumann traction enforcement, and adding a trusted reference.
Figure~\ref{fig:ch5-plate-diag1} shows this first-round diagnostic report.

\begin{figure}[H]
  \centering
  \includegraphics[width=0.75\linewidth]{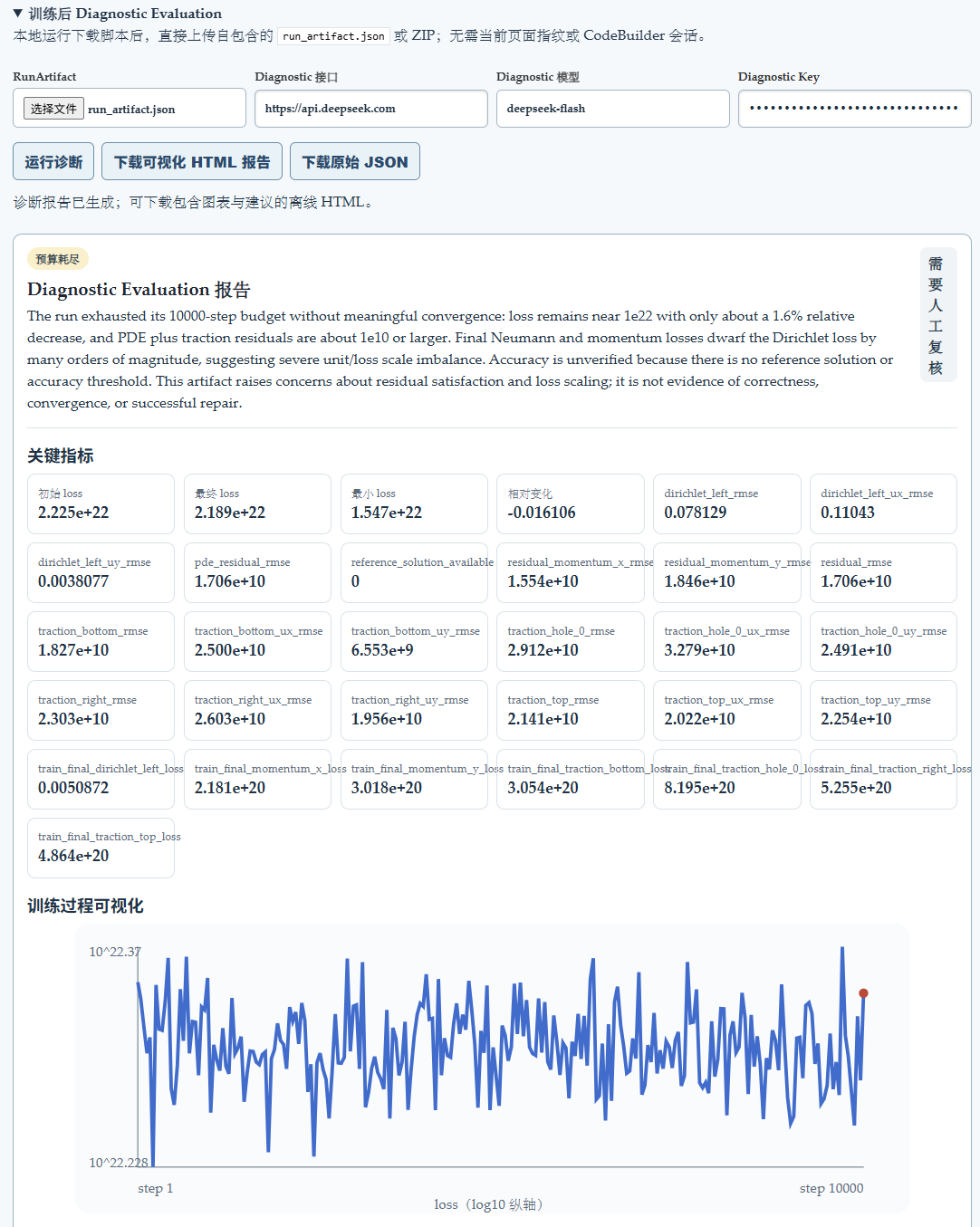}
  \caption{First-round Diagnostic Evaluation Agent report for the SI Adam plate-with-hole run.}
  \label{fig:ch5-plate-diag1}
\end{figure}

Figures~\ref{fig:ch5-plate-si-disp} and~\ref{fig:ch5-plate-si-stress} compare the SI Adam displacement and stress fields with the FEM reference.
The predicted displacements are of order $10^{-1}\,\mathrm{m}$, while the FEM displacements are of order $10^{-5}\,\mathrm{m}$.
The displacement relative $L^2$ on the FEM probe mesh is $3.65\times10^{4}$.

\begin{figure}[H]
  \centering
  \includegraphics[width=0.7\linewidth]{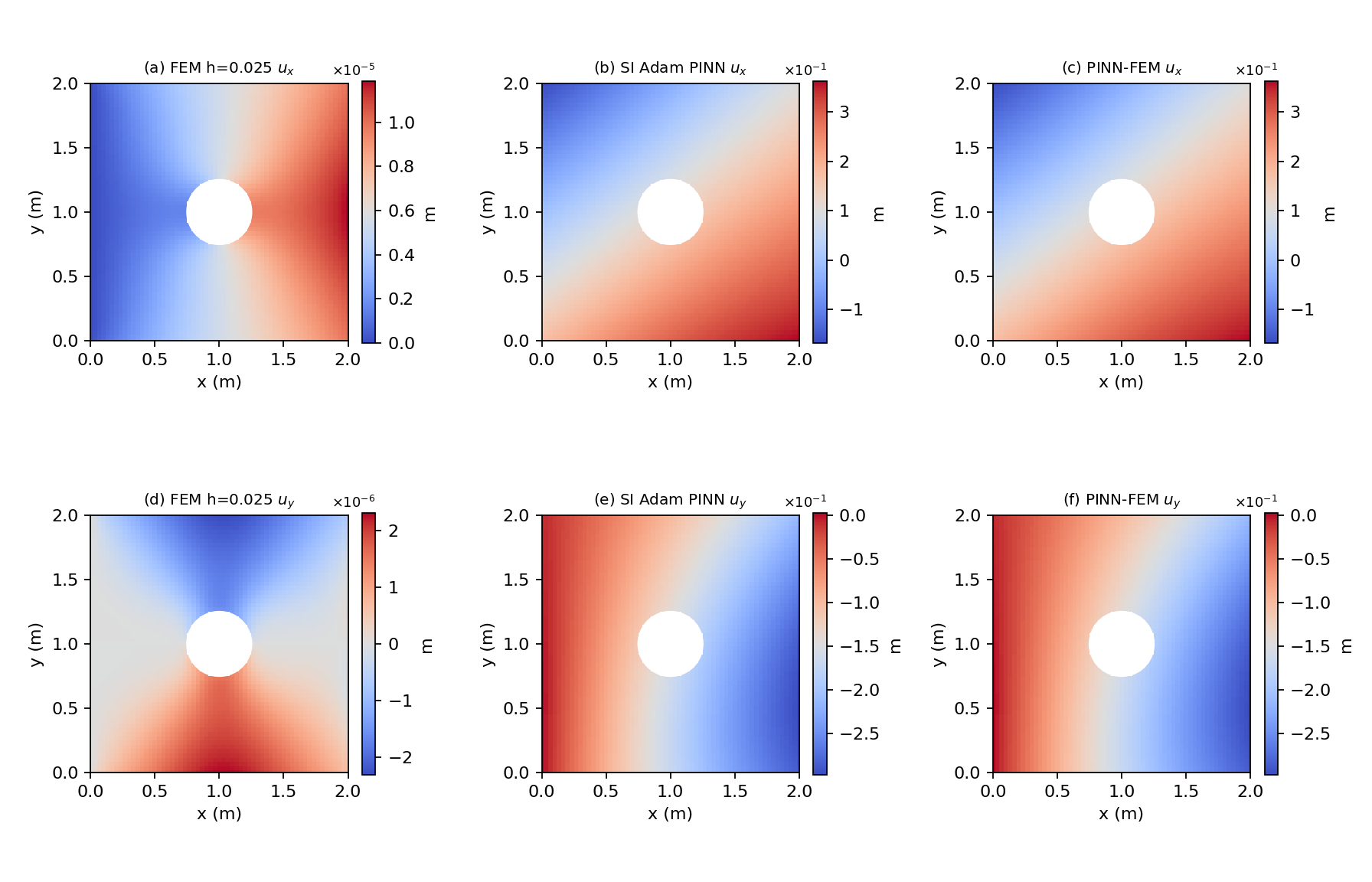}
  \caption{First-round diagnosis.
  SI Adam displacement after the failed SI training run.}
  \label{fig:ch5-plate-si-disp}
\end{figure}

\begin{figure}[H]
  \centering
  \includegraphics[width=0.7\linewidth]{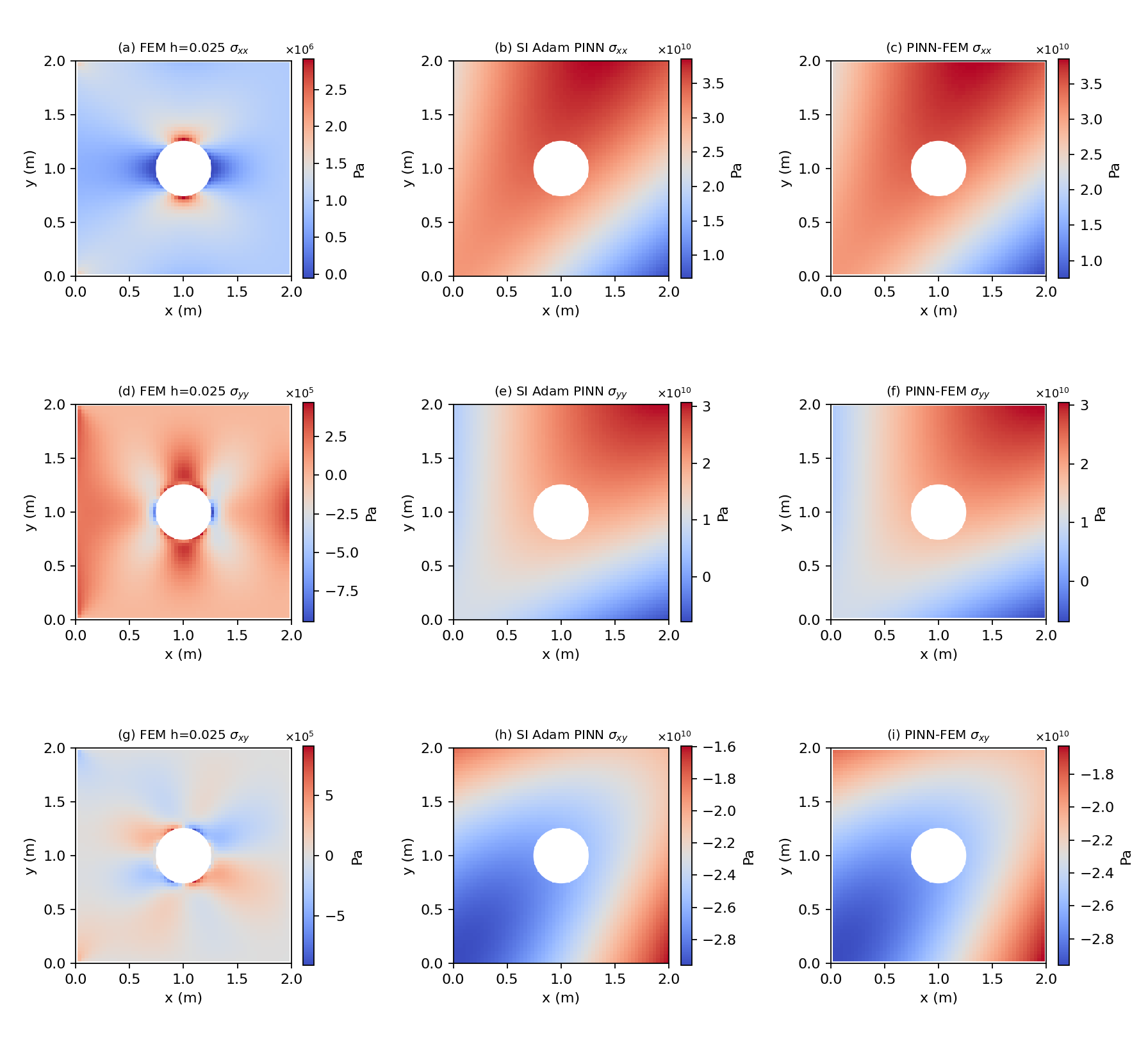}
  \caption{First-round diagnosis.
  SI Adam stress for $\sigma_{xx}$, $\sigma_{yy}$, and $\sigma_{xy}$ after the failed SI training run.}
  \label{fig:ch5-plate-si-stress}
\end{figure}

Following the Diagnostic Evaluation Agent hints on residual scaling, a researcher-led investigation attributes the SI failure to float32 Adam second-moment overflow under the large residual scale.
A fair nondimensional repair keeps the same network and Adam settings and changes only the scales $L=1\,\mathrm{m}$, $S=10^6\,\mathrm{Pa}$, and $U=SL/E$.
Three seeds with $10000$ Adam steps reach displacement relative $L^2$ errors of $0.758$, $0.765$, and $0.775$ against a held-out FEM probe mesh with $h=0.025$.
Figures~\ref{fig:ch5-plate-adam-disp} and~\ref{fig:ch5-plate-adam-stress} compare the nondimensional Adam seed-$0$ displacement and stress fields with the FEM reference.

\begin{figure}[H]
  \centering
  \includegraphics[width=0.7\linewidth]{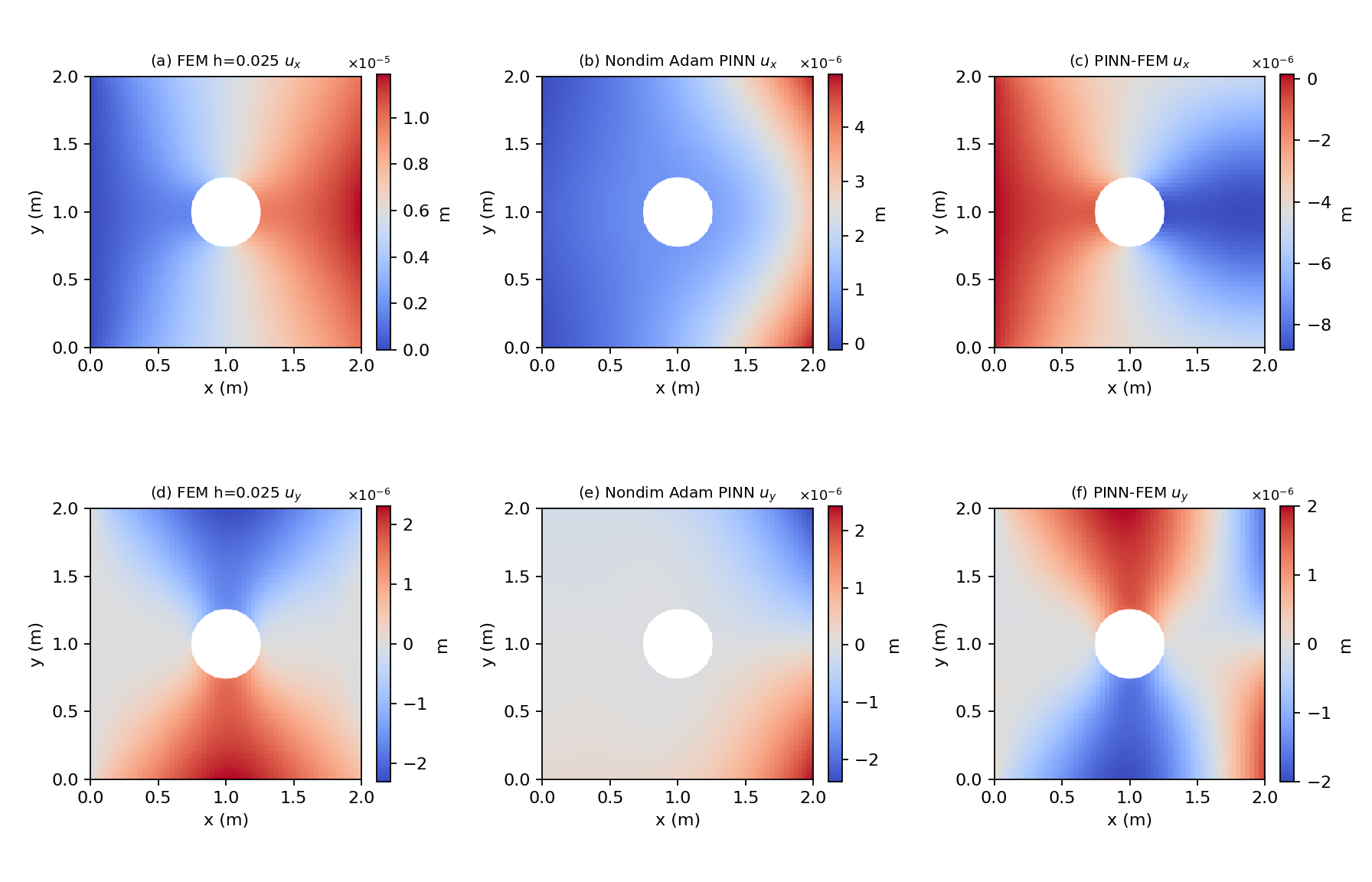}
  \caption{Second-round diagnosis.
  Nondimensional Adam displacement after the residual-scale repair.}
  \label{fig:ch5-plate-adam-disp}
\end{figure}

\begin{figure}[H]
  \centering
  \includegraphics[width=0.7\linewidth]{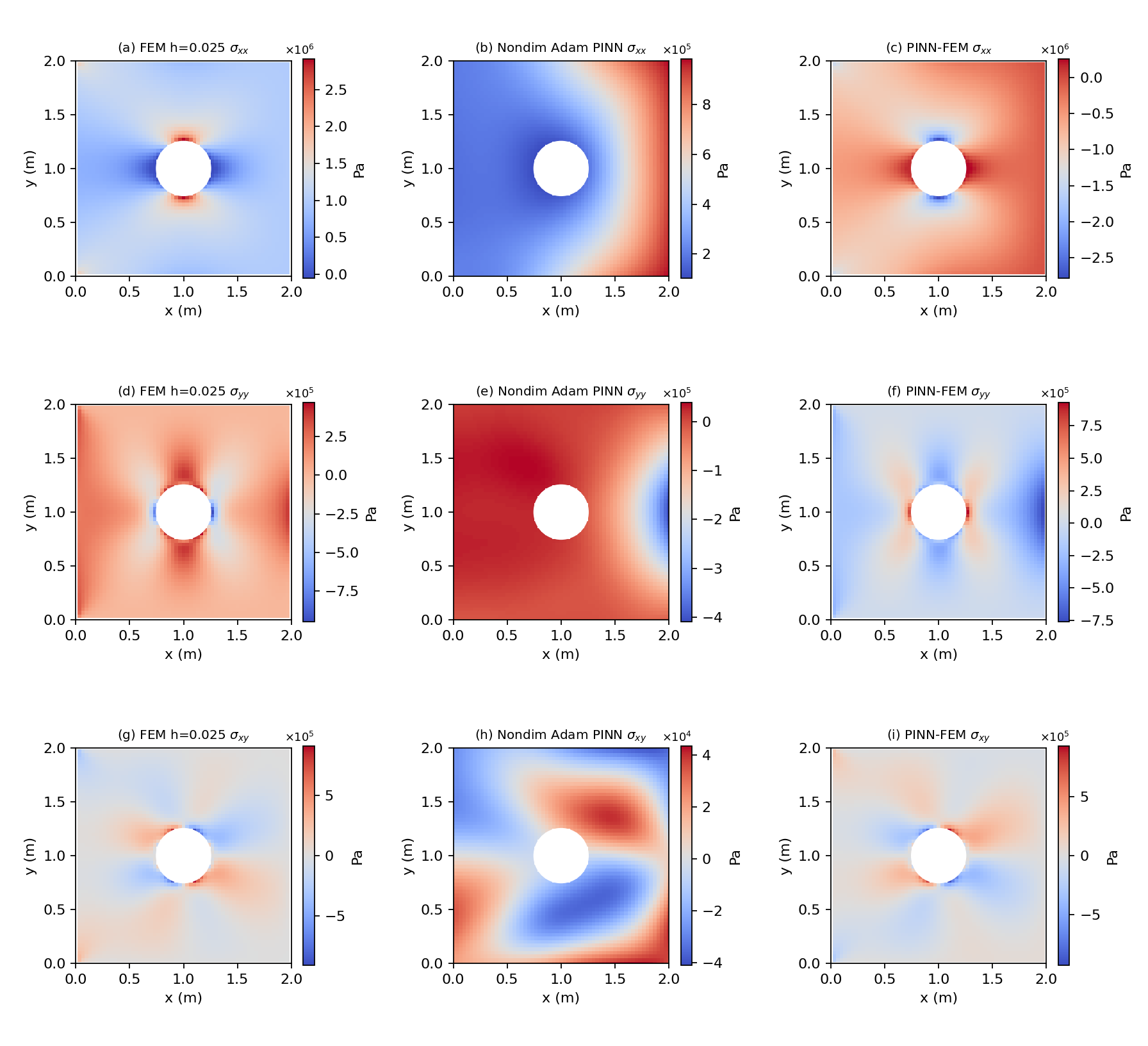}
  \caption{Second-round diagnosis.
  Nondimensional Adam stress for $\sigma_{xx}$, $\sigma_{yy}$, and $\sigma_{xy}$ after the residual-scale repair.}
  \label{fig:ch5-plate-adam-stress}
\end{figure}

A second call to the Diagnostic Evaluation Agent on the nondimensional Adam seed-$0$ artifact again returns \texttt{concerns}.
It reports elevated $x$-momentum and hole-traction residuals, and offers non-binding repair hints to review the dominant $x$-momentum residual, inspect hole-traction $u_x$ error together with that residual.
Guided by those hints, the researcher designs a stronger optimization continuation.
The network, loss assembly, and nondimensional scales stay unchanged.
Each seed warm-starts from its nondimensional Adam checkpoint, replaces the Adam stage by the registered L-BFGS optimizer, and freezes one larger uniform collocation batch of $2048$ interior and $512$ boundary points with no resampling during the L-BFGS stage.
Relative to the preceding Adam recipe of $512$ interior and $128$ boundary points redrawn each step, this fixed larger batch gives L-BFGS a stable residual landscape for local refinement of the already nondimensionalized problem.
Each seed then runs $260$ L-BFGS outer steps.
The final seed-$0$ displacement relative $L^2$ falls from $0.758$ under nondimensional Adam to $3.51\times10^{-2}$ on the FEM probe mesh, with $3.52\times10^{-2}$ on a shifted held-out probe set.
The corresponding stress relative $L^2$ falls from $0.713$ to $3.66\times10^{-2}$, and the nondimensional $x$-momentum residual RMSE falls from $0.494$ to $0.113$.
Seeds~$1$ and~$2$ reach relative $L^2$ values of $4.13\times10^{-2}$ and $6.31\times10^{-2}$ on the same FEM probe mesh.
Figure~\ref{fig:ch5-plate-disp} compares the final seed-$0$ L-BFGS displacement fields with the FEM reference. 
Figure~\ref{fig:ch5-plate-stress} compares the corresponding stress components $\sigma_{xx}$, $\sigma_{yy}$, and $\sigma_{xy}$.

The diagnostic path therefore ran in two stages.
The first diagnosis flagged anomalous residual scales after SI Adam training failed, and the researcher nondimensionalized the problem under the same Adam recipe.
The second diagnosis reported elevated $x$-momentum and hole-edge traction residuals, and the researcher continued from the nondimensional Adam checkpoints with fixed large-batch L-BFGS until all three seeds met the displacement target.
Across the full agent-assisted workflow, the Problem Understanding and Solver Planning Agents drafted the specifications, the Code Building Agent exported the local training script, and the Diagnostic Evaluation Agent twice located the residual-scale and residual-imbalance cues that directed those researcher-led repairs.

\begin{figure}[H]
  \centering
  \includegraphics[width=0.7\linewidth]{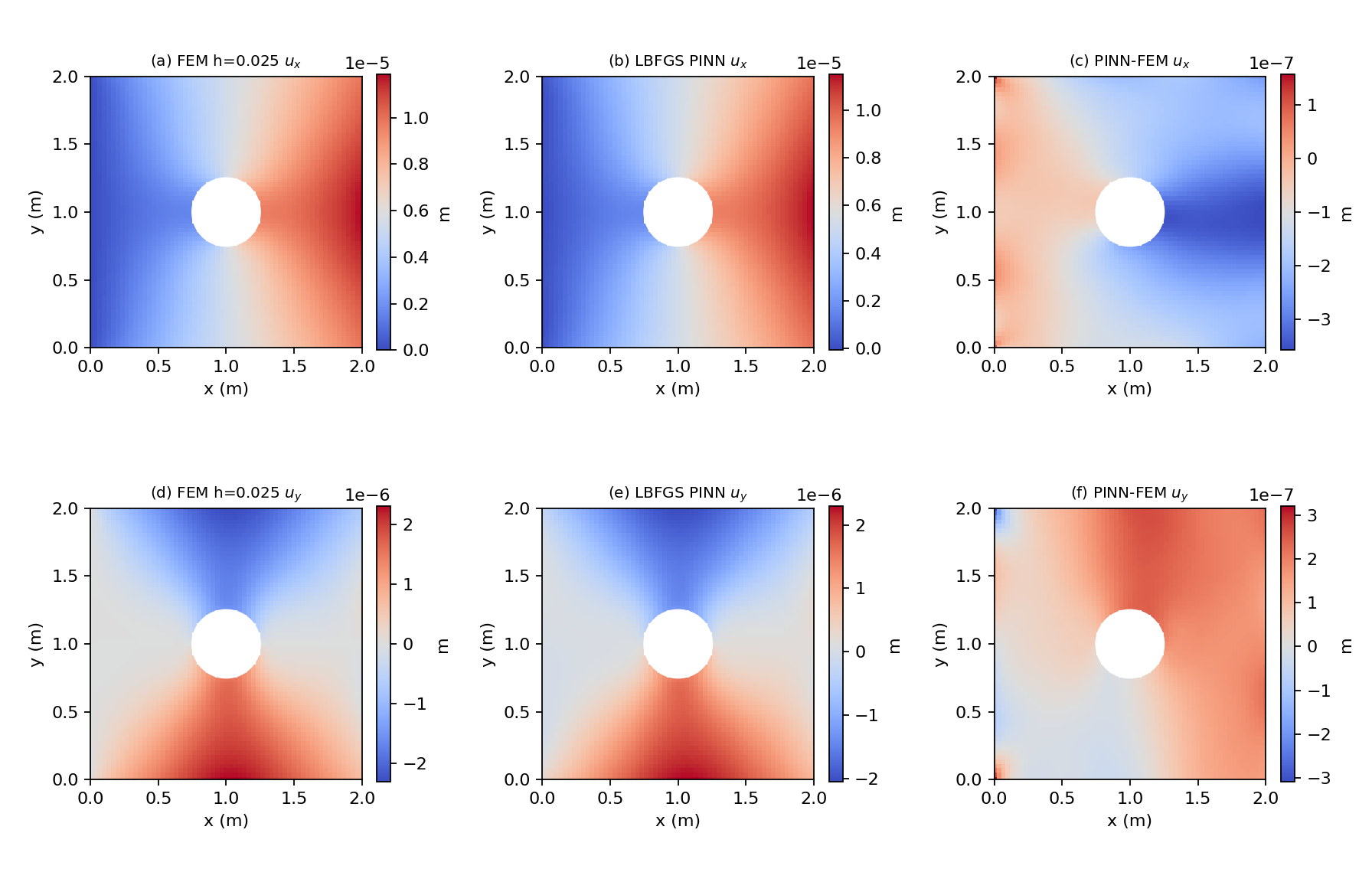}
  \caption{Final result after the second-round diagnosis and fixed-batch L-BFGS continuation.
  Displacement fields.}
  \label{fig:ch5-plate-disp}
\end{figure}

\begin{figure}[H]
  \centering
  \includegraphics[width=0.7\linewidth]{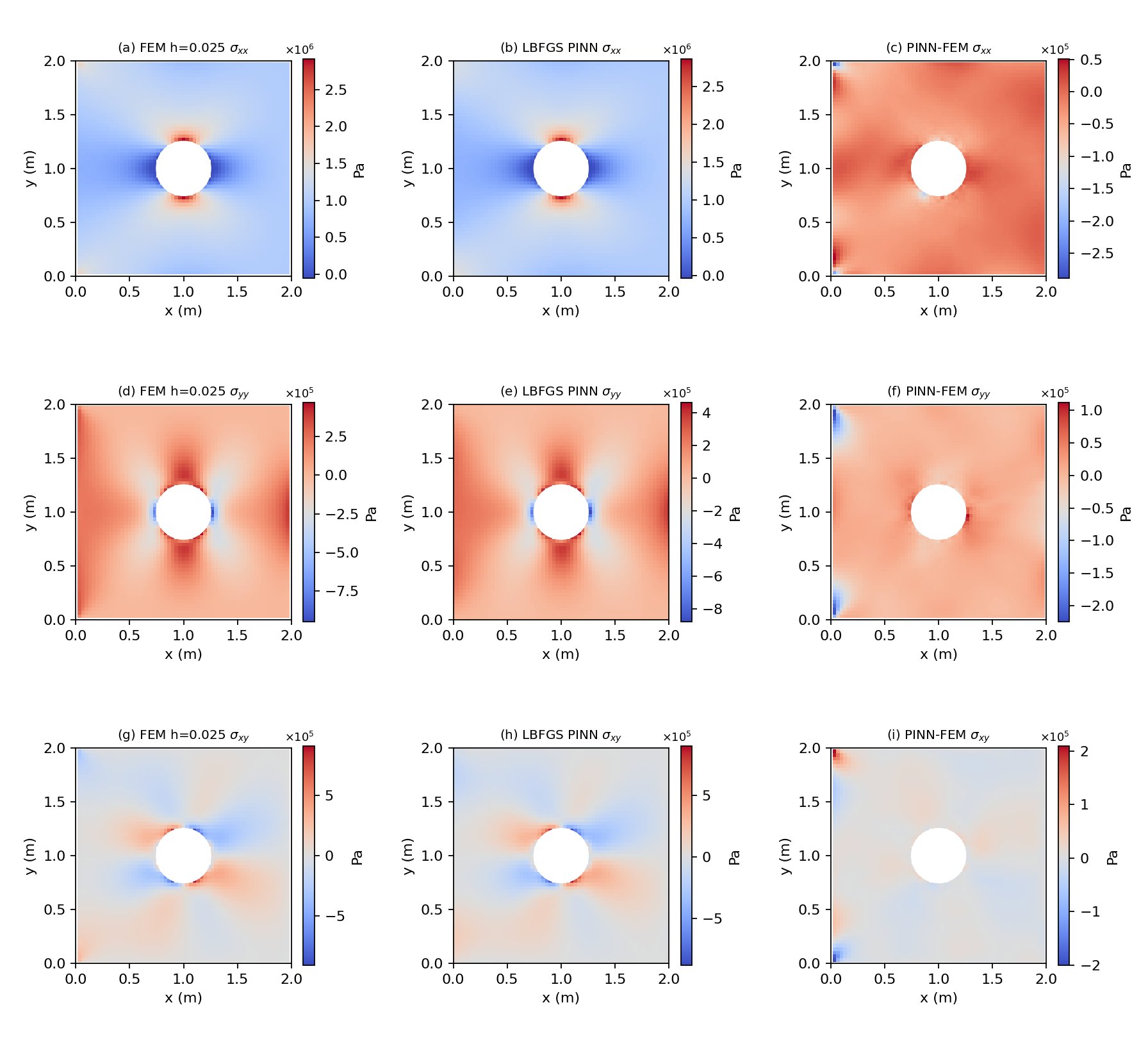}
  \caption{Final result after the second-round diagnosis and fixed-batch L-BFGS continuation.
  Stress fields for $\sigma_{xx}$, $\sigma_{yy}$, and $\sigma_{xy}$.}
  \label{fig:ch5-plate-stress}
\end{figure}

Figure~\ref{fig:ch5-plate-loss} shows the nondimensional loss histories for Adam and the subsequent L-BFGS stage.

\begin{figure}[H]
  \centering
  \includegraphics[width=0.78\linewidth]{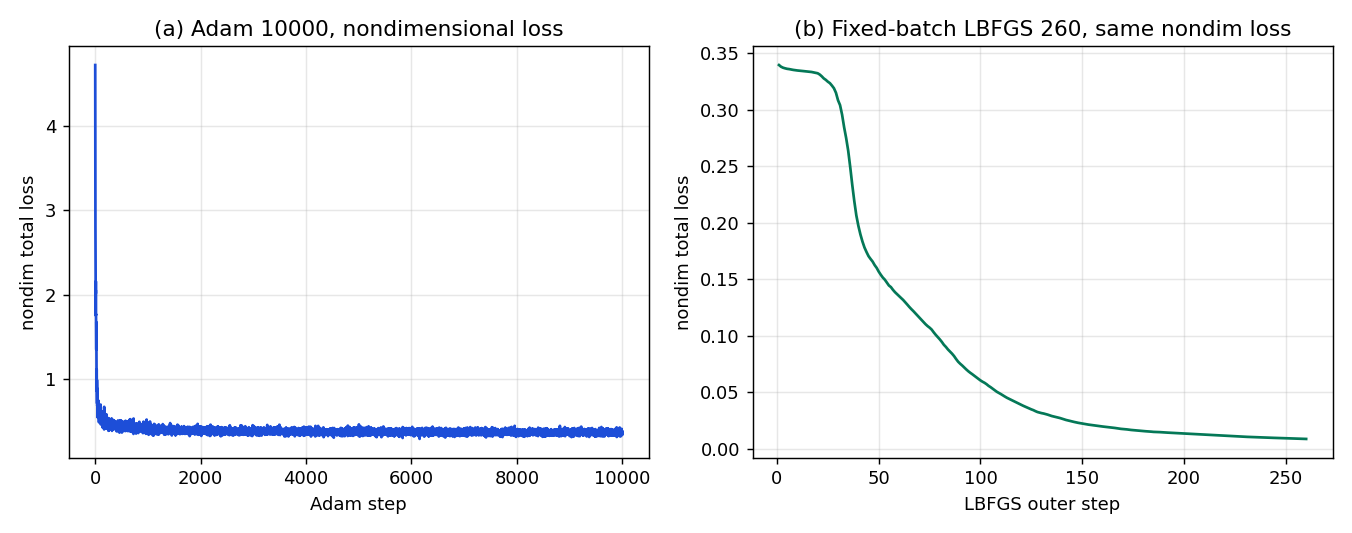}
  \caption{Nondimensional total loss for Adam $10000$ and the warm-started fixed-batch L-BFGS continuation.}
  \label{fig:ch5-plate-loss}
\end{figure}

\section{Limitations and Outlook}
\label{sec:limitations}

OpenPINN/PINO's first demonstration scope focuses on smooth tutorial PDEs and registered operator families for single-axis method comparison, together with the shipped validation-to-codegen path.
Section~\ref{sec:ex-agent} exercises early Problem Understanding, Solver Planning, and Diagnostic Evaluation interfaces on a plate-with-hole elasticity case.
Those calls still required human completion of missing geometry and boundary fields.
Complex contact, strong geometric nonlinearity, three-dimensional unstructured operators, and arbitrary model--derivative cross products are not claimed.
The shipped Specification Validation Agent and deterministic Code Building Agent cover the middle of the agent-assisted workflow for structured configurator and YAML inputs.
Closed-loop specification adjustment remains incomplete relative to the target workflow.

Future work includes hardening understanding, planning, and diagnostic evaluation toward ready handoffs and closed-loop specification adjustment, expanding shared benchmarks and reference comparisons across application domains, strengthening provenance in the specifications, and carefully extending operator physics on structured grids without informal cross-adapters between coordinate and grid derivative paths.

\section{Conclusion}
\label{sec:conclusion}

We presented OpenPINN/PINO as an agent-assisted scientific computing platform built around an overall preprocessing–computation–postprocessing architecture and a five-module agent-assisted workflow.
Shared \texttt{ProblemSpec} and \texttt{MethodSpec} contracts encode the physical problem and solution method, drive PINN, neural-operator, and custom backends that train locally, and support postprocessing into metrics, visualizations, and experiment records.
The agent workflow helps researchers formulate the problem, plan a solver, obtain runnable local code, and diagnose failed training.
Upstream agents draft \texttt{ProblemSpec} and \texttt{MethodSpec}, flag missing fields, and propose baseline recipes.
Validated specifications are assembled into deterministic training scripts for local execution.
Diagnostic evaluation then reviews run artifacts and surfaces residual-scale and residual-imbalance cues that guide researcher-led repairs.
We hope OpenPINN/PINO helps researchers and educators build, share, and verify physics-informed learning workflows with less friction and more accountability.

\section*{Data and Code Availability}
The OpenPINN/PINO source code is available at \url{https://github.com/junelu0214/OpenPINN-PINO} under the Apache-2.0 license~\cite{openpinn_upstream}.

\clearpage
\renewcommand{\refname}{References}
\bibliographystyle{unsrt}
\bibliography{refs}

\clearpage
\pdfbookmark[0]{中文全文}{chinese-fulltext}

\makeatletter
\let\bilingual@oldlabel\label
\let\bilingual@oldref\ref
\let\bilingual@oldpageref\pageref
\renewcommand{\label}[1]{\bilingual@oldlabel{zh:#1}}
\renewcommand{\ref}[1]{\bilingual@oldref{zh:#1}}
\renewcommand{\pageref}[1]{\bilingual@oldpageref{zh:#1}}
\let\ltx@label\label
\@ifundefined{subcaption@label}{}{\let\subcaption@label\label}
\renewcommand{\theHsection}{zh.\thesection}
\renewcommand{\theHsubsection}{zh.\thesection.\thesubsection}
\renewcommand{\theHsubsubsection}{zh.\thesection.\thesubsection.\thesubsubsection}
\renewcommand{\theHfigure}{zh.\thefigure}
\renewcommand{\theHtable}{zh.\thetable}
\renewcommand{\theHequation}{zh.\theequation}
\renewcommand{\theHfootnote}{zh.\thefootnote}
\gdef\@extra@b@citeb{.zh}
\makeatother

\setcounter{section}{0}
\setcounter{subsection}{0}
\setcounter{subsubsection}{0}
\setcounter{figure}{0}
\setcounter{table}{0}
\setcounter{equation}{0}
\setcounter{footnote}{0}
\renewcommand{\thefootnote}{\arabic{footnote}}

\makeatletter
\let\maketitle\bilingual@maketitle
\let\@maketitle\bilingual@internalmaketitle
\let\title\bilingual@title
\let\author\bilingual@author
\let\affil\bilingual@affil
\let\date\bilingual@date
\let\thanks\bilingual@thanks
\let\and\bilingual@and
\setcounter{authors}{0}
\setcounter{affil}{0}
\newaffiltrue
\gdef\AB@authors{}
\gdef\AB@authlist{}
\gdef\AB@affillist{}
\gdef\AB@author{\noexpand\AB@blk@and}
\gdef\@thanks{}
\gdef\@date{}
\gdef\@title{}
\renewcommand\@author{\ifx\AB@affillist\AB@empty\AB@author\else
  \ifnum\value{affil}>\value{Maxaffil}\def\rlap##1{##1}%
  \AB@authlist\\[\affilsep]\AB@affillist
  \else\AB@authors\fi\fi}
\renewcommand\AB@affilsep{\protect\Affilfont}
\makeatother

\renewcommand\Authands{, }
\title{OpenPINN/PINO：面向智能体辅助物理信息神经网络与神经算子的可扩展开源框架}
\author[1,2]{\mbox{王熙}}
\author[1]{\mbox{陆骏}}
\author[3]{\mbox{张城}}
\author[1,4]{\mbox{庄晓莹}\thanks{通讯作者。邮箱：\texttt{xiaoying.zhuang@gmail.com}}}
\author[4,5]{Timon Rabczuk}
\author[1]{\mbox{朱合华}}
\author[1]{\mbox{武威}}
\author[3]{\mbox{李培楠}}
\author[6]{\mbox{赵吉东}}
\author[2]{\mbox{尹振宇}\thanks{通讯作者。邮箱：\texttt{zhenyu.yin@polyu.edu.hk}}}
\author[7]{\mbox{孙浩}\thanks{通讯作者。邮箱：\texttt{haosun@ruc.edu.cn}}}
\affil[1]{同济大学土木工程学院，上海 200092}
\affil[2]{香港理工大学土木及环境工程学系，香港 999077}
\affil[3]{东华大学环境科学与工程学院，上海 201620}
\affil[4]{复旦大学计算力学与人工智能交叉研究所，上海 200433}
\affil[5]{魏玛包豪斯大学结构力学研究所，德国魏玛 99423}
\affil[6]{香港科技大学土木与环境工程学系，香港 999077}
\affil[7]{中国人民大学高瓴人工智能学院，北京 100872}

\date{}

\maketitle

\renewcommand{\abstractname}{摘要}
\renewcommand{\figurename}{图}
\renewcommand{\tablename}{表}
\begin{abstract}
物理信息学习中的诸多挑战催生了大量改进方法，但碎片化实现妨碍了系统组织、集成与比较。
我们提出OpenPINN/PINO，一个面向物理信息神经网络与神经算子的可扩展开源PyTorch框架。
两套共享规格将问题设定、本地计算与评估贯通起来。
智能体支持问题表述、求解器规划与诊断。
问题规格定义几何、控制方程、本构关系、参数、初边值条件，以及可选数据。
方法规格配置模型、采样、损失、优化与微分。
研究者可注册新的问题模板与方法组件，并在固定问题上组合或比较Fourier特征、残差自适应采样与自适应性损失加权等策略。
智能体将自然语言描述转写为规格，标出缺失信息，并提出求解计划。
框架校验这些规格，并确定性生成本地训练与测试代码。
诊断智能体审阅训练日志与评估结果，识别问题并建议修订。
三个算例展示方法选择与智能体诊断如何影响精度。
对粘性Burgers方程，Fourier特征与残差自适应采样相对共享基线各使相对$L^2$误差约降低$75\%$。
对周期Poisson问题，谱微分使16个留出荷载上的平均去均值相对$L^2$误差约降低$68\%$。
对带孔平板，相继的智能体诊断引导了无量纲化与优化。
初始训练失败后，三次运行均达到$10\%$目标，位移相对$L^2$误差为$3.51\%$--$6.31\%$。

\end{abstract}

\medskip
\noindent\textbf{关键词：}物理信息神经网络，神经算子，科学机器学习，智能体辅助计算，计算力学

\section{引言}
\label{sec:introduction}

物理信息神经网络（physics-informed neural networks, PINNs）通过自动微分将控制方程嵌入神经网络训练，已成为正反问题偏微分方程（PDEs）求解的标准无网格方法~\cite{raissi2019pinn,lu2021deepxde}。
神经算子将同一科学机器学习议程从单个PDE实例扩展到实例族上的映射~\cite{lu2021deeponet,li2021fno}。
尽管进展迅速，完整工作流的组装对许多团队仍很困难。
研究者须定义物理问题与数值方法，并在不断增长的模型、采样器、损失与优化器集合中做出选择。

开源库如 SciANN 表明，紧凑且贴近数学表述的 API 使 PINN 可教可扩展~\cite{haghighat2021sciann}。
算子生态方面，DeepXDE 与 NeuralOperator 亦并行成熟~\cite{lu2021deepxde,kossaifi2026neuraloperator}。
更新的工作线使用语言模型与多智能体系统，从自然语言提示起草求解器代码。
这些线索在实践中仍仅松散耦合。
问题定义、方法变体、PINN 与算子后端，以及智能体生成的脚本，往往仍分散在不同库与工作流中。

我们提出OpenPINN/PINO，一个面向智能体辅助物理信息学习的开源PyTorch框架。
共享的问题规格与方法规格契约刻画物理问题与求解方法，驱动在用户机器上训练的PINN、神经算子与自定义后端，并支持后处理为指标、图形与实验记录。
方法生态沿五条轴组织已注册组件：模型、采样、损失、优化与微分。
研究者可跨后端互换这些轴，并通过同一接口注册新组件。
智能体辅助工作流覆盖问题理解、求解器规划、规格校验、确定性代码构建与诊断评估。
当上游智能体可用时，用户可从自然语言进入；也可通过文档内嵌配置器与YAML项目，从校验与代码构建起步。

\section{PINN与神经算子}
\label{sec:algorithms}


\subsection{深度神经网络}
前馈网络通过对坐标或传感器输入施加堆叠仿射变换与逐元非线性，实现到场量的复合映射。
权重与偏置是可训练参数。
物理信息求解器与算子学习器复用这一网络代数作为构建块，但目标映射不同。
PINN逼近单实例场\(u(\mathbf{x})\)。
神经算子逼近实例族上的映射\(a\mapsto u\)。

\subsection{物理信息神经网络}
考虑定义在区域$\Omega$上的一般稳态或时变PDE，
\begin{equation}
  \mathcal{N}[u](\mathbf{x}) = f(\mathbf{x}),
  \qquad \mathbf{x} \in \Omega,
\end{equation}
以及边界或初始条件$\mathcal{B}[u]=g$，定义在$\partial\Omega$或$t=0$。
PINN用神经网络$\hat{u}(\cdot;\boldsymbol{\theta})$替代未知场$u$。
自动微分通过对计算图施加链式法则，精确求取网络输出对输入的导数。
方法再在配点集$\mathcal{T}_f\subset\Omega$与$\mathcal{T}_b\subset\partial\Omega$上惩罚方程与条件残差：
\begin{equation}
  \mathcal{L}(\boldsymbol{\theta})
  =
  w_f\,\mathcal{L}_f(\boldsymbol{\theta};\mathcal{T}_f)
  +
  w_b\,\mathcal{L}_b(\boldsymbol{\theta};\mathcal{T}_b)
  +
  w_d\,\mathcal{L}_d(\boldsymbol{\theta};\mathcal{T}_d),
\end{equation}
其中$\mathcal{L}_f$度量PDE残差，$\mathcal{L}_b$度量边界或初始失配，$\mathcal{L}_d$是观测集$\mathcal{T}_d$上的可选数据项。
训练以一阶或拟牛顿优化器最小化$\mathcal{L}$。
正演求解使用已知系数与条件。
反演或混合设定保留部分参数自由，并在存在数据项时与$\boldsymbol{\theta}$联合辨识~\cite{raissi2019pinn}。

相对基于网格的离散化，PINN模板在残差层面无网格，对多种PDE类型共享同一代码路径，并通过扩展损失吸收反问题。
精度仍依赖采样、损失权重、网络容量与优化，因此这些轴上的方法选择在实践中至关重要。

\subsection{神经算子}
神经算子逼近函数空间之间的映射，
\begin{equation}
  \mathcal{G}:\; a \mapsto u,
\end{equation}
其中$a$可编码系数、荷载或边界数据，$u$为相应解场。
在实例族上训练完成后，对新的$a$求值即可得到替代求解，无需从零重启完整PINN优化。

DeepONet通过分支网络编码$a$、主干网络编码查询坐标来分解算子，再组合其潜特征以预测$u$~\cite{lu2021deeponet}。
Fourier神经算子在Fourier空间参数化全局积分核，并天然与结构化网格配对~\cite{li2021fno}。
在此类网格上，物理信息训练的导数可用结构化数组上的谱微分或有限差分算子形成，作为逐点自动微分的替代。
物理信息算子学习在配点网格或点集上增加残差惩罚，因此训练可使用数据项、物理项，或二者兼用。

\section{相关工作}
\label{sec:related}


\subsection{物理信息与算子学习库}
开源库使神经PDE求解器便于编写、教学与扩展。
表~\ref{tab:library-survey}按时间顺序汇总代表性库，并给出发表年份、框架与文档侧重点。

\begin{table}[htbp]
  \centering
  \small
  \caption{代表性物理信息与算子学习库。
  年份取本综述所用首次公开时间（期刊在线、会议在线或首次arXiv发布）。
  NVIDIA Modulus承接同一产品线中的SimNet。}
  \label{tab:library-survey}
  \setlength{\tabcolsep}{3pt}
  \begin{tabularx}{\linewidth}{@{}l c l >{\raggedright\arraybackslash}X@{}}
    \toprule
    库 & 年份 & 框架 & 侧重点 \\
    \midrule
    NeuroDiffEq~\cite{chen2020neurodiffeq} & 2020 & Python / PyTorch & 神经微分方程求解器 \\
    SciANN~\cite{haghighat2021sciann} & 2020 & Keras / TensorFlow & 声明式科学与PINN API \\
    DeepXDE~\cite{lu2021deepxde,lu2021deeponet} & 2021 & TensorFlow / PyTorch / JAX & 强形式PINN、DeepONet、几何API \\
    TensorDiffEq~\cite{mcclenny2021tensordiffeq} & 2021 & TensorFlow & 多GPU正反问题PINN \\
    SimNet / Modulus~\cite{hennigh2021simnet} & 2021 & PyTorch & 生产后端、多物理模板 \\
    IDRLnet~\cite{peng2021idrlnet} & 2021 & PyTorch & 模块化PINN问题与模型组件 \\
    NeuralPDE~\cite{zubov2021neuralpde} & 2021 & Julia / SciML & 自动化PINN构建 \\
    JAXPI~\cite{wang2023jaxpi} & 2023 & JAX & 稳定PINN的训练实践 \\
    NeuralOperator~\cite{kossaifi2026neuraloperator,li2021fno} & 2024 & PyTorch & 函数空间算子、FNO族 \\
    \bottomrule
  \end{tabularx}
\end{table}

NeuroDiffEq在Python中封装神经微分方程求解器~\cite{chen2020neurodiffeq}。
SciANN以声明式风格为科学与物理信息深度学习提供Keras与TensorFlow封装~\cite{haghighat2021sciann}。
DeepXDE将几何、残差、初边值条件与训练循环置于贴近问题数学表述的紧凑API之后，并覆盖强形式PINN以及DeepONet等算子学习路径~\cite{lu2021deepxde,lu2021deeponet}。
TensorDiffEq面向可扩展的多GPU正反问题PINN训练~\cite{mcclenny2021tensordiffeq}。
NVIDIA SimNet其后演进为Modulus，侧重物理--机器学习工作流的生产后端与多物理模板~\cite{hennigh2021simnet}。
IDRLnet提供面向PyTorch的PINN库，具有模块化问题与模型组件~\cite{peng2021idrlnet}。
在Julia SciML生态中，NeuralPDE自动化PINN构建，并附带误差逼近工具~\cite{zubov2021neuralpde}。
JAXPI等面向训练的工具包伴随稳定PINN优化的文档化最佳实践~\cite{wang2023jaxpi}。
NeuralOperator提供学习函数空间映射的PyTorch库，包括Fourier神经算子及相关参数化PDE族架构~\cite{kossaifi2026neuraloperator,li2021fno}。

总体而言，这些库降低了单后端PINN与算子实验的门槛。

\subsection{智能体与语言模型辅助求解器}
更新的系统线使用语言模型与多智能体工作流，从高层问题描述起草、形式化或调谐物理信息及更广的SciML求解器。
表~\ref{tab:agent-survey}按时间顺序汇总代表性工作。

\begin{table}[htbp]
  \centering
  \small
  \caption{面向PDE与物理信息学习的代表性智能体与语言模型辅助系统。
  年份取本综述所用首次公开时间。}
  \label{tab:agent-survey}
  \setlength{\tabcolsep}{3pt}
  \begin{tabularx}{\linewidth}{@{}l c >{\raggedright\arraybackslash}X@{}}
    \toprule
    系统 & 年份 & 侧重点 \\
    \midrule
    PINNsAgent~\cite{wuwu2025pinnsagent} & 2025 & 以PINN为数值基底的LLM驱动自动化PDE代理 \\
    PDE-Controller~\cite{soroco2025pdecontroller} & 2025 & PDE的LLM自动形式化与推理 \\
    Lang-PINN~\cite{he2025langpinn} & 2025 & 从语言到PINN工作流的多智能体路径 \\
    ATHENA~\cite{toscano2025athena} & 2025 & 面向分层数值算法的智能体团队 \\
    PDE-Agent~\cite{liu2025pdeagent} & 2025 & 工具链增强的多智能体PDE求解 \\
    OpInf-LLM~\cite{wang2026opinfllm} & 2026 & 经算子推断的LLM辅助参数化PDE求解 \\
    AgenticSciML~\cite{jiang2026agenticsciml} & 2026 & 面向SciML架构发现的协作智能体 \\
    EvoPINN~\cite{yin2026evopinn} & 2026 & 可执行PINN算法的智能体发现 \\
    \bottomrule
  \end{tabularx}
\end{table}

PINNsAgent围绕PINN训练这一数值基底，用LLM自动化PDE代理建模~\cite{wuwu2025pinnsagent}。
PDE-Controller使用LLM对PDE支配系统进行自动形式化与推理~\cite{soroco2025pdecontroller}。
Lang-PINN协调智能体，将自然语言问题陈述解释为PDE表述、架构、损失与训练工作流~\cite{he2025langpinn}。
ATHENA组织智能体团队，进行数值算法的分层演化设计~\cite{toscano2025athena}。
PDE-Agent将PDE求解表述为经多智能体循环调用工具链~\cite{liu2025pdeagent}。
OpInf-LLM将语言模型与算子推断耦合，用于参数化PDE族~\cite{wang2026opinfllm}。
AgenticSciML研究协作多智能体对SciML架构、损失与训练策略的发现~\cite{jiang2026agenticsciml}。
EvoPINN面向可执行PINN算法设计的智能体搜索~\cite{yin2026evopinn}。

\section{OpenPINN/PINO框架}
\label{sec:openpinn}


\subsection{概述}
相对第~\ref{sec:related}节综述的库与智能体系统，OpenPINN/PINO强调跨PINN与算子后端的共享问题规格与方法规格契约、用于受控比较的已注册方法组件，以及校验规格并构建可审计本地训练项目的智能体。
设计由前处理、计算与后处理构成的整体架构，并在同一契约上配套五模块智能体辅助工作流。
方法生态与定性权衡见第~\ref{sec:methods}节。

\subsection{整体架构}
OpenPINN/PINO按商业数值工作流常用的三大阶段组织：前处理、计算与后处理。
前处理再次承载主要科学决策。
计算是在所选后端上的本地训练与测试。
后处理将运行产物转化为指标、图形与可复用记录。
同一套机器可读契约服务全部三阶段，使智能体输出、文档内嵌配置器与YAML项目可互换。
图~\ref{fig:arch}概括三阶段、共享问题规格与方法规格契约、三种计算后端，以及后处理产物。

\begin{figure}[htbp]
  \centering
  \includegraphics[width=\linewidth]{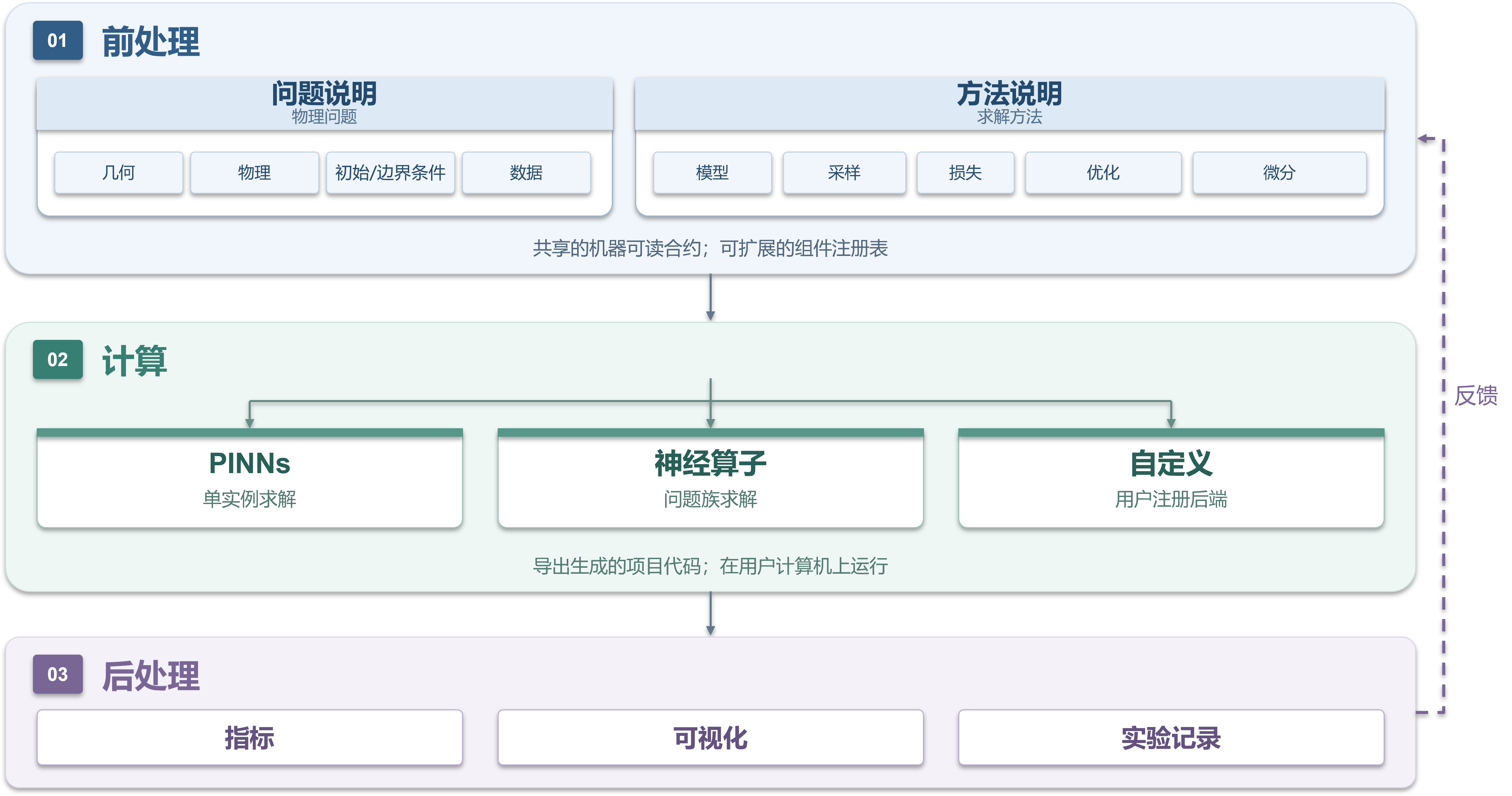}
  \caption{OpenPINN/PINO整体架构。
  前处理将物理问题与求解方法记录为共享规格。
  计算在PINN、神经算子或自定义后端上本地运行。
  后处理生成指标、可视化与实验记录，并可选择反馈至规格。}
  \label{fig:arch}
\end{figure}

\paragraph{物理问题设定。}
物理问题在问题规格中记录为相互独立的字段组。
\begin{itemize}
  \item \textbf{几何}定义计算域、坐标约定，以及网格或点集提示。
  \item \textbf{物理}定义控制方程、本构模型与材料或方程参数，包括留待辨识的自由参数。
  \item \textbf{初边值}定义初始条件、边界条件、荷载及其他闭合问题的约束。
  \item \textbf{数据}定义可选的实测或仿真观测，用于反演、数据驱动或混合设定，并作为测试与比较的留出参考。
\end{itemize}
这些字段组跨PINN、算子与自定义后端保持不变。
新物理通过扩展相应字段组或注册可复用模板进入，无需改写架构其余部分。

\paragraph{求解方法设定。}
数值方法在方法规格中记录为第二组独立字段。
\begin{itemize}
  \item \textbf{模型}选择网络族、宽度与深度、激活函数，以及硬约束封装。
  \item \textbf{采样}选择配点、边界与数据采样器，包括固定或自适应策略。
  \item \textbf{损失}选择残差、边界、初始与数据项及其权重或平衡规则。
  \item \textbf{优化}选择优化器、学习率调度、迭代预算与多阶段训练协议。
  \item \textbf{微分}选择导数形成方式，包括面向逐点模型的坐标自动微分，以及面向网格原生模型的结构化网格导数算子。
\end{itemize}
每组字段由组件注册表支撑。
用户可通过注册表互换采样器、损失或导数路径，从而保持方法面的通用性与可扩展性。

\paragraph{计算后端。}
训练与测试通过三种消费同一套已校验规格的后端在本地运行。
\begin{itemize}
  \item \textbf{PINN}在单个物理问题实例上，相对方程与条件残差训练场网络，并在同一实例上评估该网络。
  \item \textbf{神经算子}在实例族上训练映射，并在新输入上评估已学算子，无需重启完整单实例优化。
  \item \textbf{自定义}接受用户注册的求解器、模型或训练循环；它们实现所需接口，并在可能时仍经规格校验与确定性导出。
\end{itemize}
训练、测试、检查点与产物写出均在用户本地机器上运行。
导出项目是带钉死规格与随机种子的普通Python包，因此计算可在机器间与批量研究中保持可移植。

\paragraph{后处理。}
后处理读取本地训练与测试产生的产物，并转化为可审计的科学结果。
设计面含三组：指标与参考比较，图形或表格的可视化与导出，以及存储配置钉、关键指标与决策笔记的实验记录。
当相应后处理器可用时，评估应用相关的导出量。

\subsection{智能体辅助工作流}
智能体辅助工作流组织为五个模块，将问题从接入推进到可本地运行的训练代码，并可选地进入运行后评估阶段。
图~\ref{fig:workflow}概括五模块、侧向输入、本地运行时，以及可解释调整的反馈路径。

\begin{figure}[htbp]
  \centering
  \includegraphics[width=\linewidth]{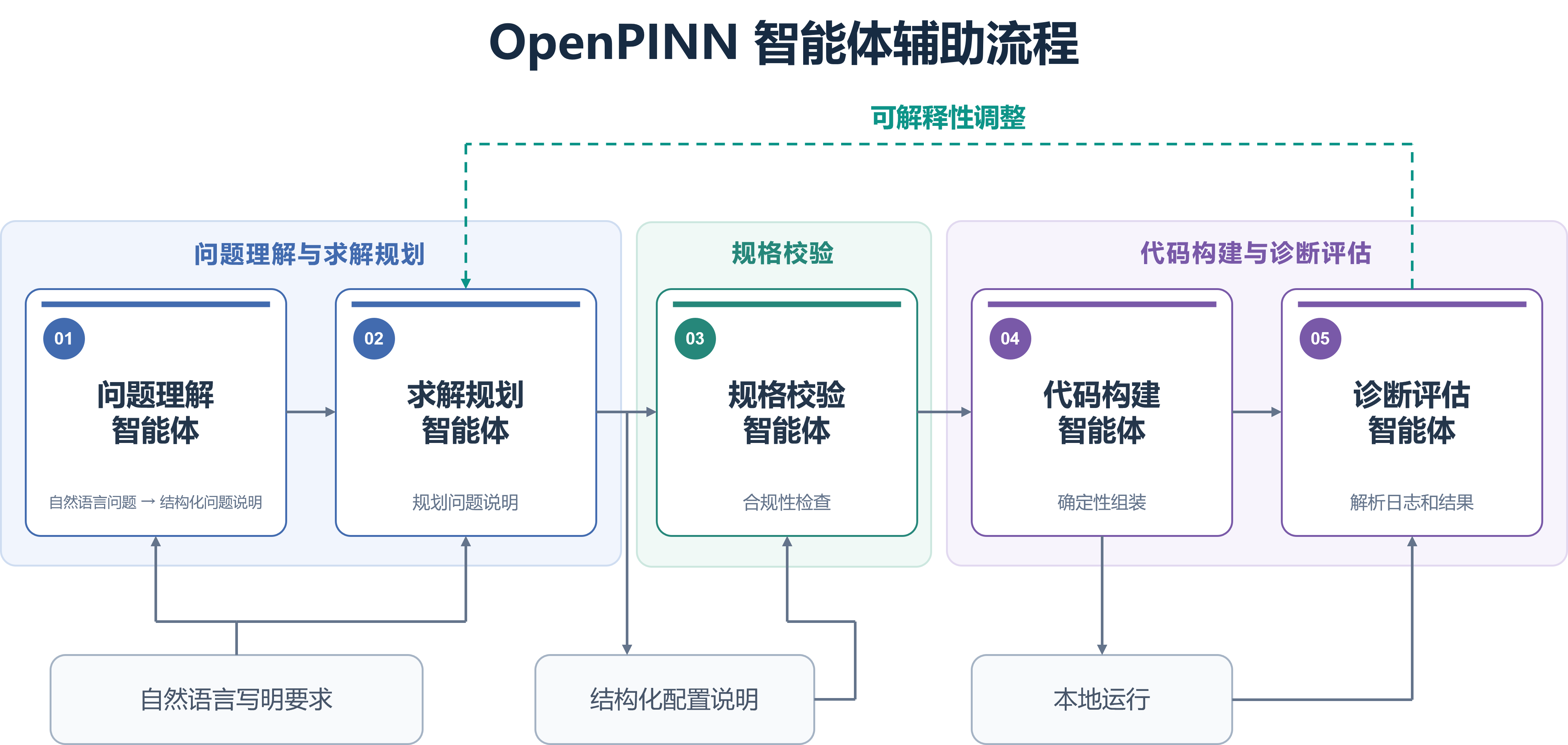}
  \caption{OpenPINN/PINO智能体辅助工作流。
  自然语言与方法需求进入理解与规划。
  配置器或YAML规格可在规格校验处汇入。
  代码构建导出本地训练与测试代码。
  诊断评估解析运行产物，返回调整建议，或写出实验报告。}
  \label{fig:workflow}
\end{figure}

\begin{enumerate}
  \item \textbf{问题理解智能体。}
  该模块接受待解问题的自然语言描述，并生成与问题规格字段对齐的结构化问题定义，涵盖几何、物理与初边值等。

  \item \textbf{求解器规划智能体。}
  给定该结构化理解，该模块提出求解计划，并推荐合适的模型、采样、损失与优化器。
  它也可摄入用户对求解方案的显式要求，并编码为与方法规格对齐的结构化方法配置。

  \item \textbf{规格校验智能体。}
  该模块检查问题理解与求解器规划的输出是否足够完整、一致，以便后续训练。
  这些检查在启动任何数值任务之前，充当结构化规格的提交前合规门。
  已交付校验器覆盖配置器与YAML输入这一角色，这些输入已提供结构化规格。

  \item \textbf{代码构建智能体。}
  校验成功后，该模块将已注册模板与组件组装为用户本地运行的训练与测试代码。
  一旦已校验规格固定，组装是确定性的，从而使等价输入产生可审计、可重复的导出。

  \item \textbf{诊断评估智能体。}
  本地训练与测试之后，该模块解析训练过程日志与测试结果文件。
  当解析结果未达到既定目标时，它提出对问题或方法规格的可解释调整，以便再次进入规划或校验。
  当目标达到时，它汇编实验报告，记录该次运行的配置、关键指标与可复现钉。
\end{enumerate}

用户可在理解与规划可用时经自然语言进入工作流，也可经配置器或YAML直接提供结构化规格，并从规格校验与代码构建起步。

\label{sec:methods}


\subsection{方法生态}
OpenPINN/PINO将数值方法在方法规格中记录为五个独立组：模型、采样、损失、优化与微分。
每组由组件注册表支撑，使各轴可独立互换。
新模型、采样器、损失、优化器或微分路径可通过同一接口注册，从而随时间扩展方法生态。
本小节描述这些方法组及指导方法选择的实践权衡。
定量比较见第~\ref{sec:examples}节的演示算例。

\subsubsection{模型}
模型组接纳单实例PINN的逐点场网络，以及问题族映射的算子骨干。
表~\ref{tab:model-families}汇总注册表拟承载的经典族，以及各族最常使用的设定。

\begin{table}[htbp]
  \centering
  \small
  \caption{OpenPINN/PINO模型轴的经典模型族。
  条目为可扩展注册表的文献原型，而非发布清单。}
  \label{tab:model-families}
  \setlength{\tabcolsep}{3pt}
  \renewcommand{\tabularxcolumn}[1]{m{#1}}
  \begin{tabularx}{\linewidth}{@{}l l >{\raggedright\arraybackslash}X@{}}
    \toprule
    族 & 典型形式 & 常见设定 \\
    \midrule
    Pointwise MLP & MLP & 单实例强形式PINN~\cite{raissi2019pinn} \\
    Spectral embedding & Fourier features & 多尺度或高频场~\cite{tancik2020fourierfeatures,wang2021fourierbias} \\
    Hard constraints & Trial-solution / distance wrappers & 精确满足选定边界条件~\cite{lagaris1998annpde,lu2021hardconstraints} \\
    Domain decomposition & cPINN, XPINN-style multi-network & 大域或多区域~\cite{jagtap2020cpinn,jagtap2020xpinn} \\
    Branch--trunk operators & DeepONet & 不规则传感器、查询坐标~\cite{lu2021deeponet} \\
    Graph operators & GNO, multipole GNO & 不规则网格、多尺度图~\cite{li2020gno,li2020mgno} \\
    Spectral / kernel operators & FNO, low-rank kernels & 结构化网格、参数族~\cite{li2021fno,kovachki2023neuraloperator} \\
    Emerging bases & Physics-informed KANs & PINN中的替代一元基~\cite{wang2025pikan} \\
    \bottomrule
  \end{tabularx}
\end{table}

对PINN，普通多层感知机仍是从坐标到场量的基线映射~\cite{raissi2019pinn}。
Fourier特征嵌入改变同一网络族的输入表示，并扩大多尺度场上可表示的频率内容~\cite{tancik2020fourierfeatures,wang2021fourierbias}。
硬约束封装改变试探表示，使选定边界条件内置于网络输出，遵循经典ANN-PDE构造及后续PINN硬约束设计，代价是几何相关的构造~\cite{lagaris1998annpde,lu2021hardconstraints}。
守恒型与扩展型PINN等区域分解布局，用空间或时空子域上的耦合子网络替代单一全局网络~\cite{jagtap2020cpinn,jagtap2020xpinn}。
物理信息Kolmogorov--Arnold网络说明更新的一元基如何进入同一逐点槽位~\cite{wang2025pikan}。

对算子学习，DeepONet通过输入函数上的分支网络与查询坐标上的主干网络分解映射，适于不规则传感器布局~\cite{lu2021deeponet}。
图神经算子在图离散上学习积分核，当空间网格非结构化时自然适用~\cite{li2020gno}。
多极图神经算子加入多尺度图粗化，以加速大图上的长程相互作用~\cite{li2020mgno}。
Fourier神经算子在Fourier空间参数化全局核，并与结构化网格及参数化PDE族配对~\cite{li2021fno}。
当偏好积分核的紧凑分解时，低秩核参数化同属神经算子族~\cite{kovachki2023neuraloperator}。
当控制问题具有能量结构时，深度能量法直接最小化总势能，对变分固体力学设定往往更合适~\cite{samaniego2020dem}。
同一变分思想延伸到算子学习中的变分物理信息神经算子~\cite{eshaghi2025vino}，以及带可选有限元热启动的物理信息Transolver预训练~\cite{wang2026pfem}。
GeoPT、AB-UPT与Transolver++等基于Transformer的算子面向工程中常见的复杂几何，并在流体与飞行器基准上报告了较强结果~\cite{wu2026geopt,alkin2025abupt,luo2025transolverpp}。

\subsubsection{采样}
采样选择评估残差与约束项的配点、边界与数据位置。
表~\ref{tab:sampling-families}将固定非自适应设计与按空间指示组织的自适应族分开，并包含控制哪些时间切片处于活动状态的时间课程。

\begin{table}[htbp]
  \centering
  \small
  \caption{OpenPINN/PINO采样轴的代表性采样族。
  空间自适应行按用于放置或重权点的指示组织。
  时间课程限制或扩展配点集的活动时间支撑。}
  \label{tab:sampling-families}
  \setlength{\tabcolsep}{3pt}
  \renewcommand{\tabularxcolumn}[1]{m{#1}}
  \begin{tabularx}{\linewidth}{@{}l l >{\raggedright\arraybackslash}X@{}}
    \toprule
    族 & 典型形式 & 常见设定 \\
    \midrule
    Non-adaptive & Uniform, grid, LHS / QMC & 固定配点基线~\cite{raissi2019pinn} \\
    Residual-based adaptive & RAR, RAD, importance sampling & 用PDE残差集中配点~\cite{lu2021deepxde,wu2023adaptivesampling,nabian2021importance} \\
    Gradient-based adaptive & Residual/gradient ASMs & 锐锋与大解梯度~\cite{mao2023gradientadaptive} \\
    Uncertainty-based adaptive & Ensemble-disagreement expansion & 在模型分歧处扩展配点集~\cite{haitsiukevich2023ensemble} \\
    Learned adaptive & DAS-style density networks & 高维残差景观~\cite{tang2023daspinn} \\
    Temporal curriculum & Time marching, growing time windows & 长时程演化PDE~\cite{wight2021adaptive,krishnapriyan2021failuremodes} \\
    \bottomrule
  \end{tabularx}
\end{table}

非自适应采样预先固定配点集。
均匀随机点、等距网格，以及拉丁超立方或拟蒙特卡洛等低差异序列是常用基线~\cite{raissi2019pinn}。
基于残差的自适应采样用当前PDE残差更新位置。
残差自适应加密（RAR）保留已有点，并在残差仍大处添加新点~\cite{lu2021deepxde}。
残差自适应分布（RAD）保持规定的点预算，并从残差知情密度重抽或重权集合~\cite{wu2023adaptivesampling}。
重要性采样同样用残差知情提议偏置小批量~\cite{nabian2021importance}。
基于梯度的自适应采样使用解梯度信息，常与残差一起，将点置于尖锐特征附近~\cite{mao2023gradientadaptive}。
基于不确定性的自适应采样在PINN集成于锚定数据或初始条件之外发生分歧处扩展配点集~\cite{haitsiukevich2023ensemble}。
学习型自适应采样拟合参数密度，如深度自适应采样（DAS），并在高维设定中从该密度抽取新点~\cite{tang2023daspinn}。
失败知情加密是相关的残差指示变体~\cite{gao2023failure}。
时间课程采样将训练限制在活动时间窗，再将该窗推进或扩展至终时，从而在后期点进入损失之前先解析早期配点~\cite{wight2021adaptive,krishnapriyan2021failuremodes}。
该选择改变活动集中出现的\((x,t)\)样本。
保持全部时间可见但下调后期残差的时间因果残差重权记录在损失轴上~\cite{wang2024causal}。

\subsubsection{损失}
损失组将残差、边界、初始与可选数据项组装为标量目标，并记录训练中这些项如何加权。
强形式残差遵循第~\ref{sec:algorithms}节的经典PINN模板。
表~\ref{tab:loss-families}汇总可扩展注册表拟承载的经典加权族。

\begin{table}[htbp]
  \centering
  \small
  \caption{OpenPINN/PINO损失轴的经典损失加权族。
  条目为可扩展注册表的文献原型，而非发布清单。}
  \label{tab:loss-families}
  \setlength{\tabcolsep}{3pt}
  \renewcommand{\tabularxcolumn}[1]{m{#1}}
  \begin{tabularx}{\linewidth}{@{}l l >{\raggedright\arraybackslash}X@{}}
    \toprule
    族 & 典型形式 & 常见设定 \\
    \midrule
    Fixed-weight composite & Hand-tuned scalars & 经典PINN基线~\cite{raissi2019pinn} \\
    Gradient-balanced & Learning-rate annealing & 相互竞争的反向传播梯度~\cite{wang2021gradpathology} \\
    NTK-balanced & Eigenvalue weighting & 损失分量间的谱失衡~\cite{wang2022ntkpinn} \\
    Uncertainty-balanced & Learnable precision, inverse Dirichlet & 自动项缩放~\cite{xiang2022lossbalanced,maddu2022inversedichlet} \\
    Pointwise self-adaptive & SA-PINN masks, residual attention & 空间变化的软约束~\cite{mcclenny2023sapinn,anagnostopoulos2024rba} \\
    Multi-objective schedules & ReLoBRaLo and related rules & 动态相对平衡~\cite{bischof2025multiloss} \\
    Causal / curriculum & Time-causal residual weights & 演化与长时程PDE~\cite{wang2024causal} \\
    \bottomrule
  \end{tabularx}
\end{table}

固定权重为各项赋予常系数，在尺度良好的问题上仍是默认基线~\cite{raissi2019pinn}。
梯度平衡方案根据反向传播梯度幅值更新系数，使残差与边界项更均衡地竞争~\cite{wang2021gradpathology}。
基于NTK的方案由支配各损失分量收敛速率的神经切核特征值设定权重~\cite{wang2022ntkpinn}。
不确定性风格的平衡器引入可学习精度参数或逆Dirichlet权重，随训练进程重缩放各项~\cite{xiang2022lossbalanced,maddu2022inversedichlet}。
逐点自适应方法为各个配点附加可训练或残差驱动掩码，并在局部残差仍大处加强软约束~\cite{mcclenny2023sapinn,anagnostopoulos2024rba}。
相对损失平衡与随机回看等对多目标调度根据近期损失历史更新权重，无需梯度或核谱~\cite{bischof2025multiloss}。
时变问题的因果加权在早期时刻充分解析之前下调后期残差~\cite{wang2024causal}。

\subsubsection{优化}
优化组记录用于最小化已组装损失的优化器、学习率调度、迭代预算与多阶段协议。
表~\ref{tab:optimization-families}主要按每次更新所用导数信息组织经典族，并将混合调度列为串联这些族的协议。

\begin{table}[htbp]
  \centering
  \small
  \caption{OpenPINN/PINO优化轴的经典优化族。
  前三行按导数阶次排列。
  末行记录组合它们的多阶段协议。
  条目为可扩展注册表的文献原型，而非发布清单。}
  \label{tab:optimization-families}
  \setlength{\tabcolsep}{3pt}
  \renewcommand{\tabularxcolumn}[1]{m{#1}}
  \begin{tabularx}{\linewidth}{@{}l l >{\raggedright\arraybackslash}X@{}}
    \toprule
    族 & 典型形式 & 常见设定 \\
    \midrule
    First-order & Adam, AdamW & 随机探索与算子训练~\cite{kingma2015adam,loshchilov2019adamw} \\
    Quasi-Newton & L-BFGS & 光滑PINN目标的全批量精化~\cite{liu1989lbfgs,raissi2019pinn} \\
    Second-order / curvature & Gauss--Newton, natural gradient & 利用Jacobian或Fisher结构的高精度精化~\cite{jnini2026gngd} \\
    Hybrid multi-stage & Adam then L-BFGS & 一阶探索后接拟牛顿抛光~\cite{lu2021deepxde,urban2025pinnopt} \\
    \bottomrule
  \end{tabularx}
\end{table}

Adam与AdamW等一阶方法用带矩估计的小批量梯度更新参数，是早期训练及大数据集上算子学习的常用主力~\cite{kingma2015adam,loshchilov2019adamw}。
有限内存BFGS等拟牛顿方法从梯度历史逼近逆Hessian信息而不形成完整二阶导数张量，在损失景观局部光滑后广泛用于全批量PINN求解~\cite{liu1989lbfgs,raissi2019pinn}。
L-BFGS偏好全批量或大批量评估，并依赖线搜索与内存设置。
二阶与曲率方法显式使用残差Jacobian或类Fisher度量，如Gauss--Newton自然梯度下降，以加速物理信息目标上的高精度精化~\cite{jnini2026gngd}。
精确Newton及更高阶张量方法在PINN文献中出现少得多，主因是在网络尺度上形成并求逆更密的导数张量代价高。
混合调度先运行固定步数的Adam，再切换至L-BFGS~\cite{lu2021deepxde}。
对PINN优化过程的研究表明，仔细调谐的多阶段拟牛顿精化可在经典基准上达到有竞争力的精度~\cite{urban2025pinnopt}。
学习率调度与迭代预算与所选优化器一并存储，使完整训练配方可复现。

\subsubsection{微分}
微分选择如何由模型输出形成PDE残差中的空间或时空导数。
表~\ref{tab:diff-families}汇总可扩展注册表拟承载的经典族，涵盖逐点自动微分、离散与谱网格算子、弱形式残差构造，以及混合数值--自动方案。

\begin{table}[htbp]
  \centering
  \small
  \caption{OpenPINN/PINO微分轴的经典微分族。
  条目为可扩展注册表的文献原型，而非发布清单。}
  \label{tab:diff-families}
  \setlength{\tabcolsep}{3pt}
  \renewcommand{\tabularxcolumn}[1]{m{#1}}
  \begin{tabularx}{\linewidth}{@{}l l >{\raggedright\arraybackslash}X@{}}
    \toprule
    族 & 典型形式 & 常见设定 \\
    \midrule
    Coordinate AD & Forward / reverse-mode AD & 逐点强形式PINN与基于查询的算子~\cite{baydin2018autodiff,raissi2019pinn,wang2021pideeponet} \\
    Numerical differentiation & Finite differences on points or grids & AD替代与校核基线~\cite{chiu2022canpinn} \\
    Spectral grid derivatives & Fourier / spectral differentiation & 周期结构化网格与物理信息FNO风格训练~\cite{li2021fno,li2024pino} \\
    Discrete grid operators & FD / FV-style array stencils & 非周期结构化网格~\cite{li2024pino,chiu2022canpinn} \\
    Weak / variational residuals & Integration by parts, VPINN & 降低网络所需经典导数阶~\cite{kharazmi2021hpvpinn} \\
    Hybrid AD--numerical & Coupled automatic--numerical schemes & 混合精确AD与局部模板以提速或增稳~\cite{chiu2022canpinn} \\
    \bottomrule
  \end{tabularx}
\end{table}

坐标自动微分在计算图上施加链式法则，得到网络输出对输入坐标的精确导数，是强形式逐点PINN的标准路径~\cite{baydin2018autodiff,raissi2019pinn}。
同一坐标路径可扩展到在查询位置评估场的算子模型，包括物理信息DeepONet训练~\cite{wang2021pideeponet}。
在一批输入函数与大量配点上，将每个坐标样本视为独立AD叶会形成多根多叶导数布局，对反向模式AD代价高，并倾向于沿函数批次复制计算图~\cite{leng2024zerocoords}。
结构化网格上的算子学习常用Fourier空间谱微分，或作用在离散场上的有限差分及相关数组模板，替代逐点AD；这与FNO风格骨干及物理信息神经算子自然配对~\cite{li2021fno,li2024pino}。
配点云或网格上的数值微分也可作为独立残差路径，以及CAN-PINN等耦合自动--数值方案中的构建块~\cite{chiu2022canpinn}。
弱形式与变分表述经分部积分将导数移至检验函数，使网络在强形式意义上需要更少的经典导数~\cite{kharazmi2021hpvpinn}。


\section{算例演示}
\label{sec:examples}


OpenPINN/PINO将研究问题记录在问题规格中，将数值配方记录在方法规格中，再沿第~\ref{sec:methods}节综述的方法轴互换已注册组件。
研究者可在所研究问题上开展受控比较，并将注意力保持在方法选择与科学结果上，同时由规格校验与确定性代码导出组装可本地运行的训练路径。
本节报告三个互补演示。
其一在一维粘性Burgers方程上求解，比较五个每次仅改动一个已注册方法组件的PINN配方。
其二在带FNO骨干的参数化Poisson算子族上展示同一风格的受控比较。
其三记录从问题理解智能体到诊断评估智能体的智能体辅助带孔平板弹性工作流，以及达到有限元相对$L^2$低于$0.1$的研究者主导修复路径。
下文全部定量结果来自与预印本材料一并存档的钉死OpenPINN-upstream参考运行。

\subsection{PINN演示}
\label{sec:ex-pinn}
PINN演示采用粘性Burgers方程
\begin{equation}
  u_t + u\,u_x - \nu u_{xx} = 0,
  \qquad x\in[-1,1],\quad t\in[0,1],
\end{equation}
粘性系数$\nu=0.02$，初始条件$u(x,0)=-\sin(\pi x)$，以及齐次Dirichlet边界。
在该固定Burgers设定上比较五个PINN配方。
其中四个从共享基线出发，恰好改动一个已注册方法组件。
表~\ref{tab:ch5-pinn-burgers}汇总五个配方。

\begin{center}
\begin{minipage}{\linewidth}
  \footnotesize
  \captionof{table}{Burgers问题（$\nu=0.02$）上的PINN配方。
  相对$L^2$相对留出$129\times51$时空网格上的Cole--Hopf参考解。}
  \label{tab:ch5-pinn-burgers}
  \setlength{\tabcolsep}{2.5pt}
  \begin{tabularx}{\linewidth}{@{}l l >{\raggedright\arraybackslash}X c c@{}}
    \toprule
    配方 & 改动组件 & 相对\texttt{Pin-B}的已注册改动 & 相对$L^2$ & 相对\texttt{Pin-B}比值 \\
    \midrule
    Pin-B & 基线 & MLP，均匀配点，固定权重，Adam后接L-BFGS & $1.16\times10^{-2}$ & $1.00$ \\
    Pin-M & 模型 & 随机Fourier特征，尺度$0.25$ & $2.82\times10^{-3}$ & $0.244$ \\
    Pin-S & 采样 & 残差自适应内部采样 & $2.96\times10^{-3}$ & $0.255$ \\
    Pin-L & 损失 & 自适应性损失权重 & $9.04\times10^{-3}$ & $0.781$ \\
    Pin-O & 优化 & 仅Adam，无L-BFGS & $4.47\times10^{-2}$ & $3.86$ \\
    \bottomrule
  \end{tabularx}
\end{minipage}
\end{center}

\begin{figure}[H]
  \centering
  \captionsetup[subfigure]{skip=1.5pt,font=scriptsize}
  \resizebox{0.78\linewidth}{!}{\begin{minipage}{\linewidth}
  \begin{subfigure}{0.48\linewidth}
    \centering
    \includegraphics[width=\linewidth]{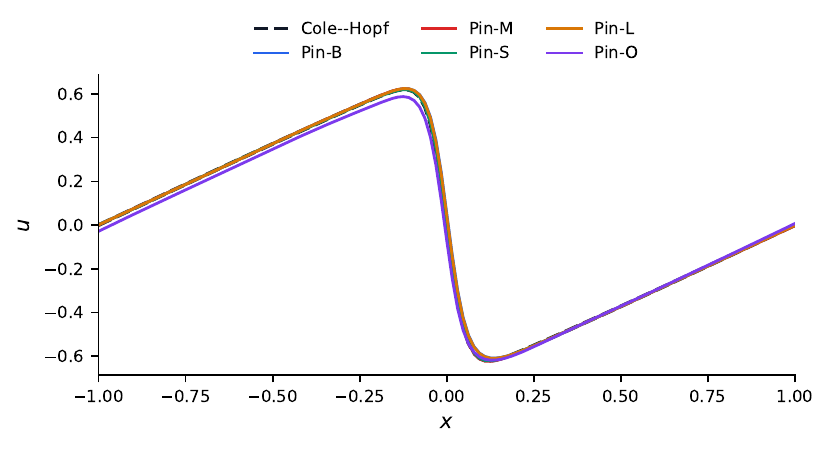}
    \caption{$t=1$处相对Cole--Hopf参考的预测解。}
    \label{fig:ch5-pinn-a}
  \end{subfigure}
  \hfill
  \begin{subfigure}{0.48\linewidth}
    \centering
    \includegraphics[width=\linewidth]{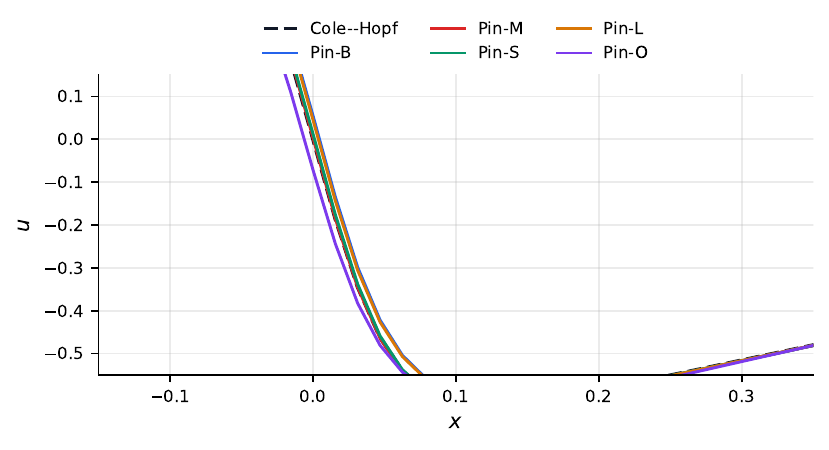}
    \caption{陡锋附近放大。}
    \label{fig:ch5-pinn-b}
  \end{subfigure}

  \vspace{0.25em}
  \begin{subfigure}{0.325\linewidth}
    \centering
    \includegraphics[width=\linewidth]{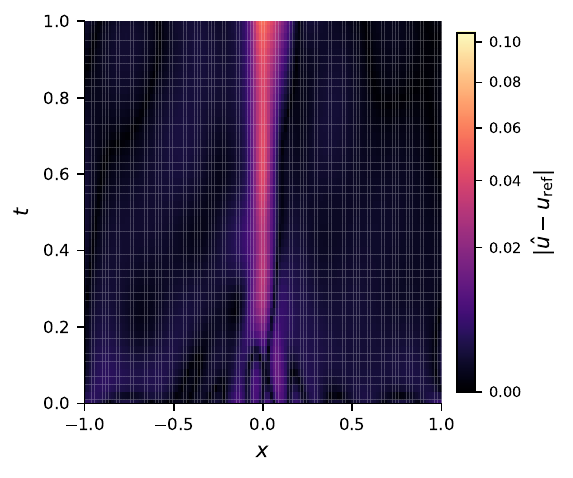}
    \caption{\texttt{Pin-B}绝对误差。}
    \label{fig:ch5-pinn-c}
  \end{subfigure}
  \hfill
  \begin{subfigure}{0.325\linewidth}
    \centering
    \includegraphics[width=\linewidth]{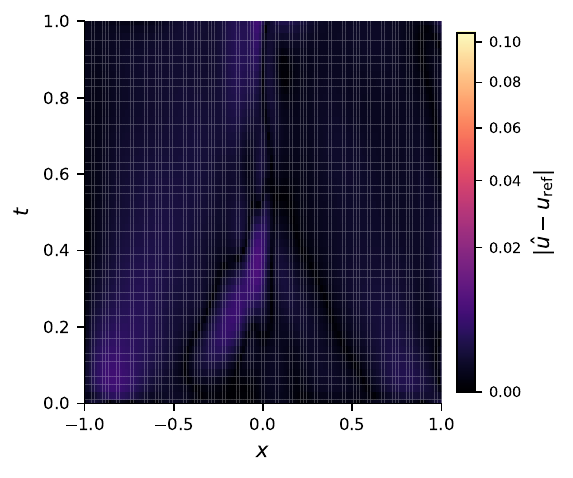}
    \caption{\texttt{Pin-M}绝对误差。}
    \label{fig:ch5-pinn-d}
  \end{subfigure}
  \hfill
  \begin{subfigure}{0.325\linewidth}
    \centering
    \includegraphics[width=\linewidth]{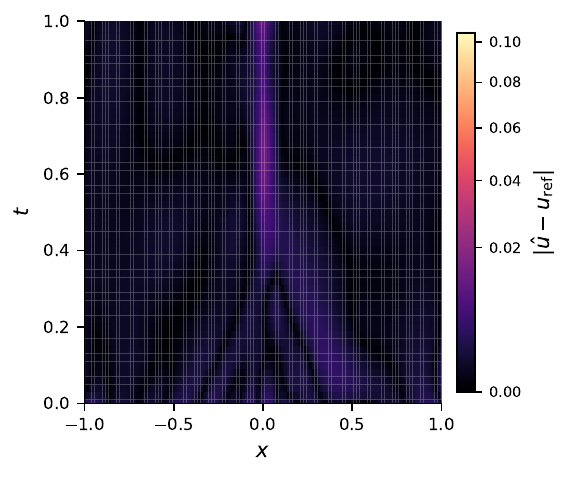}
    \caption{\texttt{Pin-S}绝对误差。}
    \label{fig:ch5-pinn-e}
  \end{subfigure}

  \vspace{0.25em}
  \begin{center}
  \begin{subfigure}{0.325\linewidth}
    \centering
    \includegraphics[width=\linewidth]{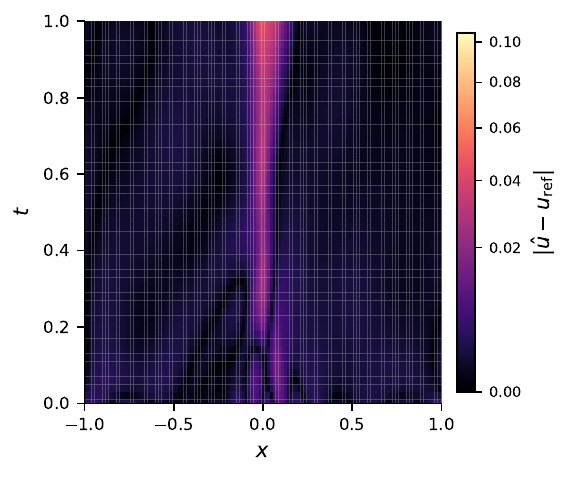}
    \caption{\texttt{Pin-L}绝对误差。}
    \label{fig:ch5-pinn-f}
  \end{subfigure}
  \hspace{0.04\linewidth}
  \begin{subfigure}{0.325\linewidth}
    \centering
    \includegraphics[width=\linewidth]{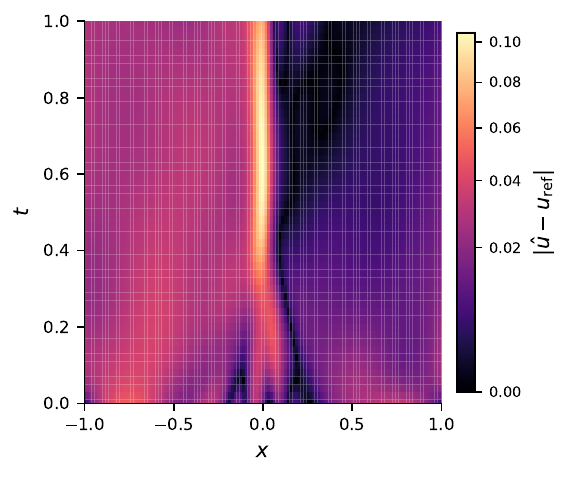}
    \caption{\texttt{Pin-O}绝对误差。}
    \label{fig:ch5-pinn-g}
  \end{subfigure}
  \end{center}
  \end{minipage}}

  \caption{Burgers演示（$\nu=0.02$）。
  子图~(c)--(g)在共享色标上显示时空绝对误差场$|\hat{u}-u_{\mathrm{ref}}|$。}
  \label{fig:ch5-pinn-burgers}
\end{figure}

基线\texttt{Pin-B}使用宽度$[64,64,64,64]$、$\tanh$激活、恒等输入特征的多层感知机，均匀配点含$256$个内部点、$64$个边界点与$64$个初始点，固定损失权重，Adam $3000$步后接L-BFGS $40$外步，以及坐标自动微分。
精度以相对留出时空网格上Cole--Hopf参考解的相对$L^2$误差度量。
\texttt{Pin-M}将恒等特征替换为含$16$个频率、尺度$0.25$的随机Fourier特征映射。
\texttt{Pin-S}在相同点预算下，用残差自适应采样替换冻结的均匀内部批量。
\texttt{Pin-L}用权重学习率$0.01$的已注册自适应性加权替换固定损失权重。
\texttt{Pin-O}去掉L-BFGS阶段，仅保留相同的$3000$步Adam。
全部相对误差均低于$0.1$，最大增益来自Fourier特征，去掉L-BFGS则误差明显上升。
图~\ref{fig:ch5-pinn-burgers}给出$t=1$处解、陡锋附近放大，以及五个配方的时空绝对误差场。

\subsection{神经算子演示}
\label{sec:ex-operator}
算子演示在参数族而非单场上采用同一受控比较风格。
该族为$[0,2\pi)^2$上的周期Poisson问题$-\Delta u=f$。
每个实例抽取一至四阶谐波的Fourier级数解，荷载为该级数的精确负Laplace。
周期通量相互抵消且级数无零模，因此$f$的空间均值为零，制造解$u$亦为零均值。
由于$u$与$u+C$满足同一方程，仅物理残差会使加性常数自由。
混合配方通过节点观测钉死该常数。
FNO骨干从$32$个训练荷载学习映射$f\mapsto u$，并在$16$个留出荷载上评估。
三个配方共享一个仅物理、有限差分基线\texttt{Op-B}。
表~\ref{tab:ch5-op-pairs}报告两项留出误差，均对十六个荷载取平均。
原始相对$L^2$直接比较预测场与参考场。
去均值相对$L^2$在每一实例上先减去各场的空间均值再取同一相对范数，从而在加性常数意义下评分$u$的形状。

\begin{figure}[H]
  \centering
  \captionsetup[subfigure]{skip=1.5pt,font=scriptsize}
  \resizebox{0.78\linewidth}{!}{\begin{minipage}{\linewidth}
  \begin{subfigure}{\linewidth}
    \centering
    \includegraphics[width=\linewidth]{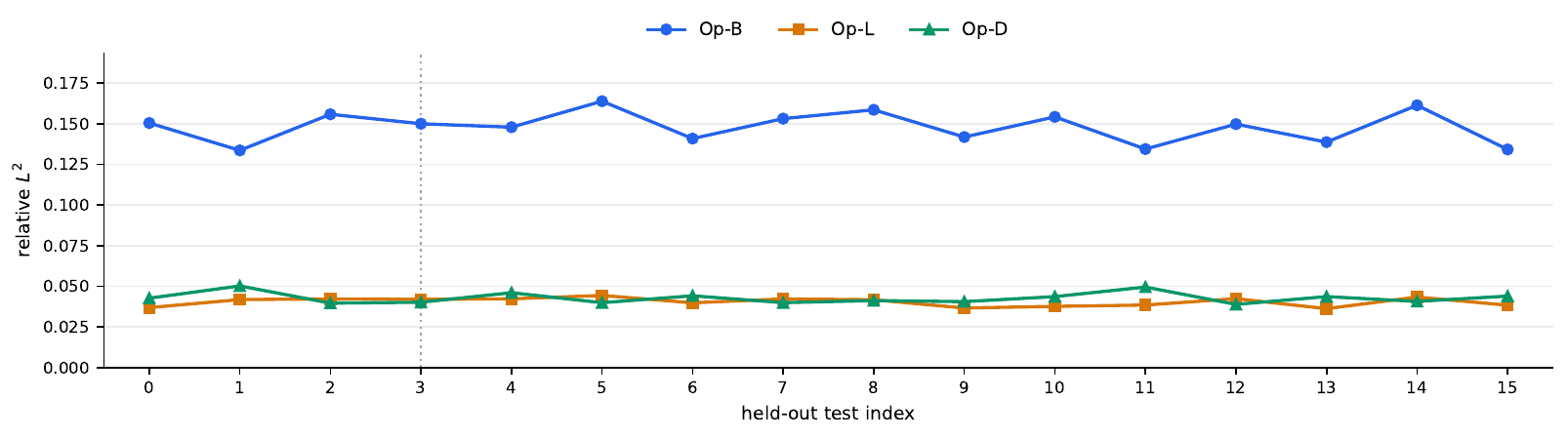}
    \caption{配方\texttt{Op-B}、\texttt{Op-L}与\texttt{Op-D}在全部十六个测试上的逐实例原始相对$L^2$。}
    \label{fig:ch5-op-a}
  \end{subfigure}

  \vspace{0.25em}
  \begin{center}
  \begin{subfigure}{0.325\linewidth}
    \centering
    \includegraphics[width=\linewidth]{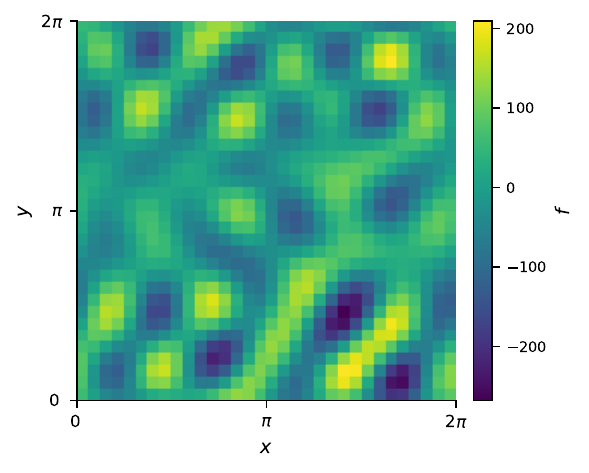}
    \caption{中位\texttt{Op-B}实例上的荷载$f$。}
    \label{fig:ch5-op-b}
  \end{subfigure}
  \hspace{0.04\linewidth}
  \begin{subfigure}{0.325\linewidth}
    \centering
    \includegraphics[width=\linewidth]{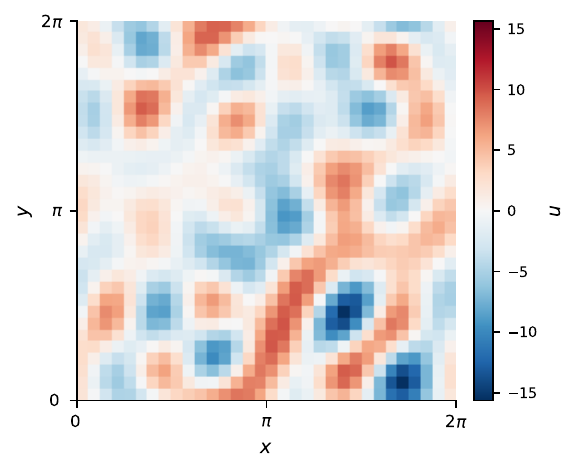}
    \caption{同一实例上的参考$u$。}
    \label{fig:ch5-op-c}
  \end{subfigure}
  \end{center}

  \vspace{0.25em}
  \begin{subfigure}{0.325\linewidth}
    \centering
    \includegraphics[width=\linewidth]{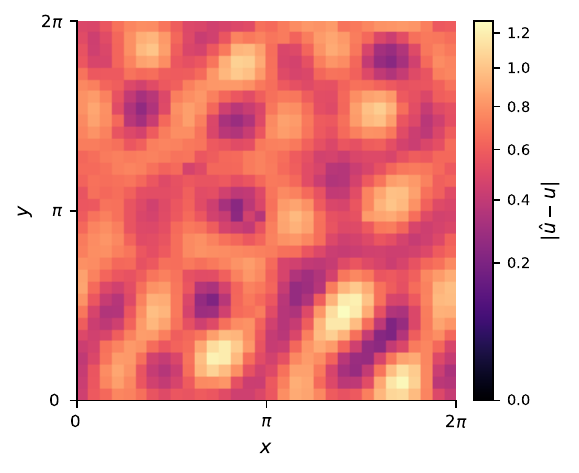}
    \caption{\texttt{Op-B}绝对误差。}
    \label{fig:ch5-op-d}
  \end{subfigure}
  \hfill
  \begin{subfigure}{0.325\linewidth}
    \centering
    \includegraphics[width=\linewidth]{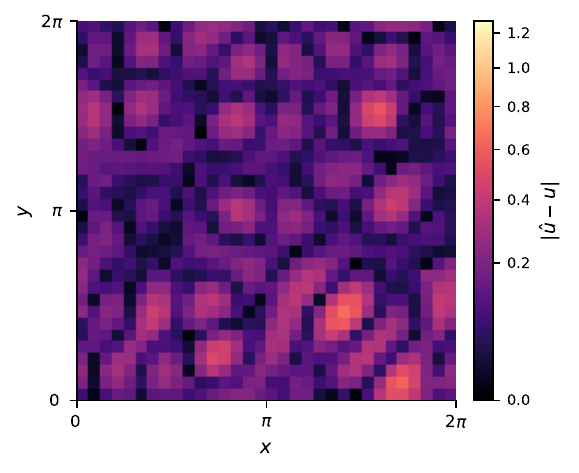}
    \caption{\texttt{Op-L}绝对误差。}
    \label{fig:ch5-op-e}
  \end{subfigure}
  \hfill
  \begin{subfigure}{0.325\linewidth}
    \centering
    \includegraphics[width=\linewidth]{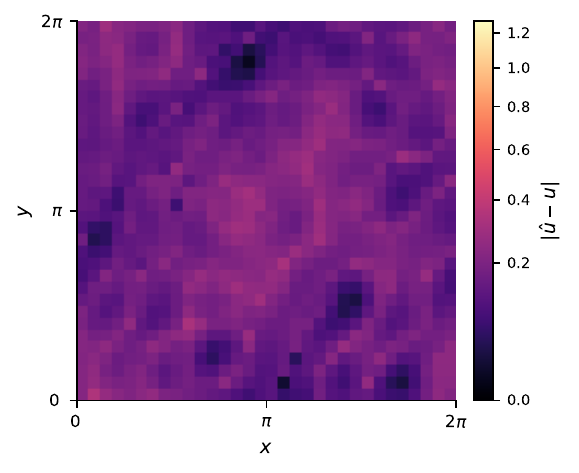}
    \caption{\texttt{Op-D}绝对误差。}
    \label{fig:ch5-op-f}
  \end{subfigure}
  \end{minipage}}

  \caption{留出周期Poisson集上的算子演示。
  子图~(b)--(f)使用中位\texttt{Op-B}留出实例；子图~(d)--(f)共享同一绝对误差色标。}
  \label{fig:ch5-op-fields}
\end{figure}

\begin{center}
\begin{minipage}{\linewidth}
  \footnotesize
  \captionof{table}{带FNO骨干的周期Poisson族上的算子配方。
  误差为对$16$个留出荷载的均值。}
  \label{tab:ch5-op-pairs}
  \setlength{\tabcolsep}{2.5pt}
  \begin{tabularx}{\linewidth}{@{}l l >{\raggedright\arraybackslash}X c c@{}}
    \toprule
    配方 & 改动组件 & 相对\texttt{Op-B}的已注册改动 & 相对$L^2$ & 去均值相对$L^2$ \\
    \midrule
    Op-B & 基线 & 仅物理损失，有限差分导数 & $1.48\times10^{-1}$ & $3.89\times10^{-2}$ \\
    Op-L & 损失 & 混合物理加数据，16个观测节点 & $4.04\times10^{-2}$ & $3.99\times10^{-2}$ \\
    Op-D & 微分 & 谱微分 & $4.28\times10^{-2}$ & $1.23\times10^{-2}$ \\
    \bottomrule
  \end{tabularx}
\end{minipage}
\end{center}

共享配方使用$32\times32$网格、宽度$24$、三层、十个模态，以及学习率$3\times10^{-3}$的Adam共$1500$步。
\texttt{Op-L}仅将损失组装改为每实例十六个观测节点的混合物理加数据损失。
\texttt{Op-D}仅将网格微分路径改为谱微分。
对\texttt{Op-L}，原始相对$L^2$显著下降，而去均值分数仍接近\texttt{Op-B}水平。
因此节点观测相对零均值制造解钉死了自由加性常数，而有限差分物理残差已解释大部分已分辨场形。
\texttt{Op-D}同时改进两项分数，且去均值误差增益更大，因而谱微分在常数对齐之外进一步收紧场形。
图~\ref{fig:ch5-op-fields}给出图~\ref{fig:ch5-op-a}中全部十六个测试的逐实例原始相对$L^2$、图~\ref{fig:ch5-op-b}--\ref{fig:ch5-op-c}中中位\texttt{Op-B}荷载与参考，以及图~\ref{fig:ch5-op-d}--\ref{fig:ch5-op-f}中相应绝对误差场。

\subsection{智能体辅助工作流演示}
\label{sec:ex-agent}
第三例记录二维带孔平板平面应力线弹性问题上的智能体辅助配置器会话。
正方形板定义在$[0,2]^2\,\mathrm{m}$，中心$(1,1)\,\mathrm{m}$处含半径$0.25\,\mathrm{m}$的圆孔。
固体各向同性，$E=210\,\mathrm{GPa}$，$\nu=0.3$。
左缘固定，右缘承受$1\,\mathrm{MPa}$均匀单轴拉力，其余外缘与孔边界无面力。
未知量为位移分量$u_x$与$u_y$。
除非另有说明，后续场图给出种子~$0$预测相对$h=0.025$探测网格上FEM参考的对比，且该FEM解仅用于事后评估。

会话从文档内嵌配置器起步，使用预填平板陈述与简短澄清提示。
图~\ref{fig:ch5-plate-understand}给出实时接口返回的问题理解与求解器规划面板。

\begin{figure}[H]
  \centering
  \includegraphics[width=0.65\linewidth]{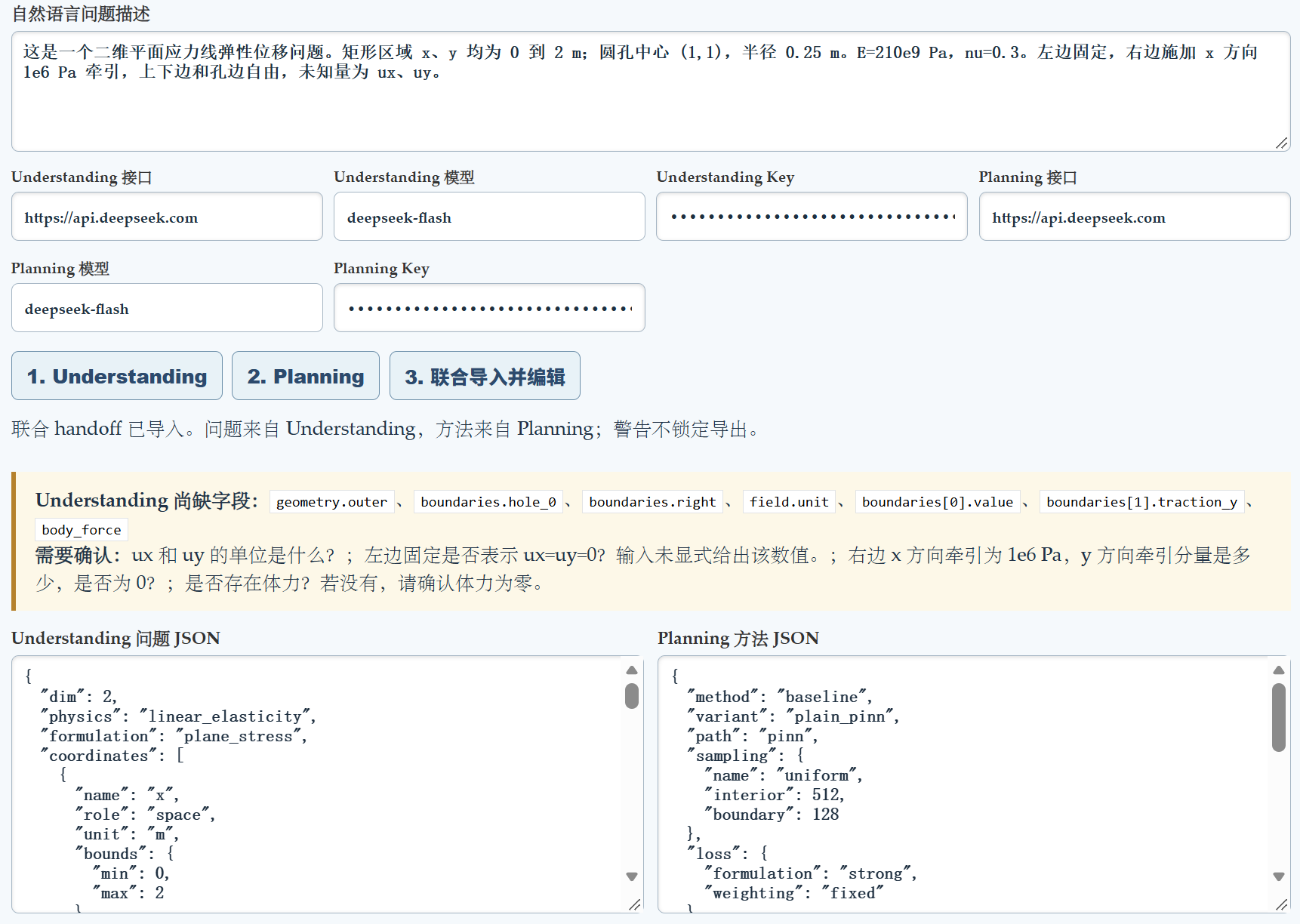}
  \caption{带孔平板算例的问题理解智能体与求解器规划智能体阶段。}
  \label{fig:ch5-plate-understand}
\end{figure}

\begin{figure}[H]
  \centering
  \includegraphics[width=0.65\linewidth]{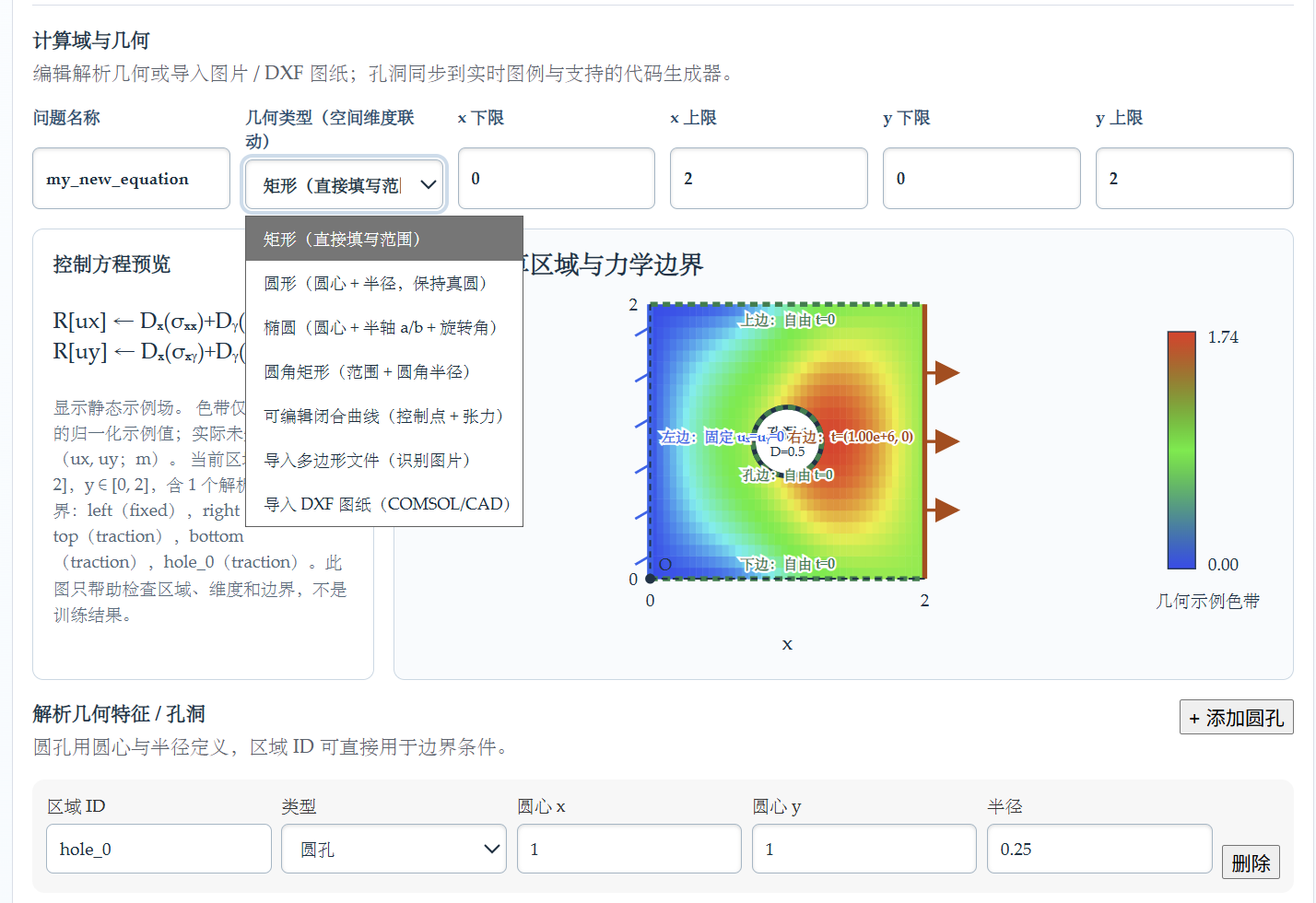}
  \caption{联合导入并经人工编辑后的配置器PDE草图与区域预览。}
  \label{fig:ch5-plate-sketch}
\end{figure}

问题理解智能体将自然语言陈述转为与问题规格字段对齐的结构化问题描述，涵盖几何、物理与边界条件等。
它标出缺失的映射外几何与孔边界行，并提示用户补全不完整字段。
求解器规划智能体随即提出基线普通PINN配方，作为结构化方法规格。
它选择$\tanh$激活的多层感知机模型、均匀配点、固定损失权重、Adam，以及强形式平衡残差的坐标自动微分，网络宽度$64$、深度$3$，Adam共$5000$步。
两份草稿存为机器可读JSON，供随后导入配置器。
联合导入后，研究者在配置器中补全孔心以及无面力的上、下与孔缘，并将展示预算从规划的宽度$64$、深度$3$网络与$5000$步Adam，提高到宽度$64$、深度$4$网络与$10000$步Adam。
图~\ref{fig:ch5-plate-sketch}给出这些编辑后的配置器方程预览与几何草图。

\begin{figure}[H]
  \centering
  \includegraphics[width=0.65\linewidth]{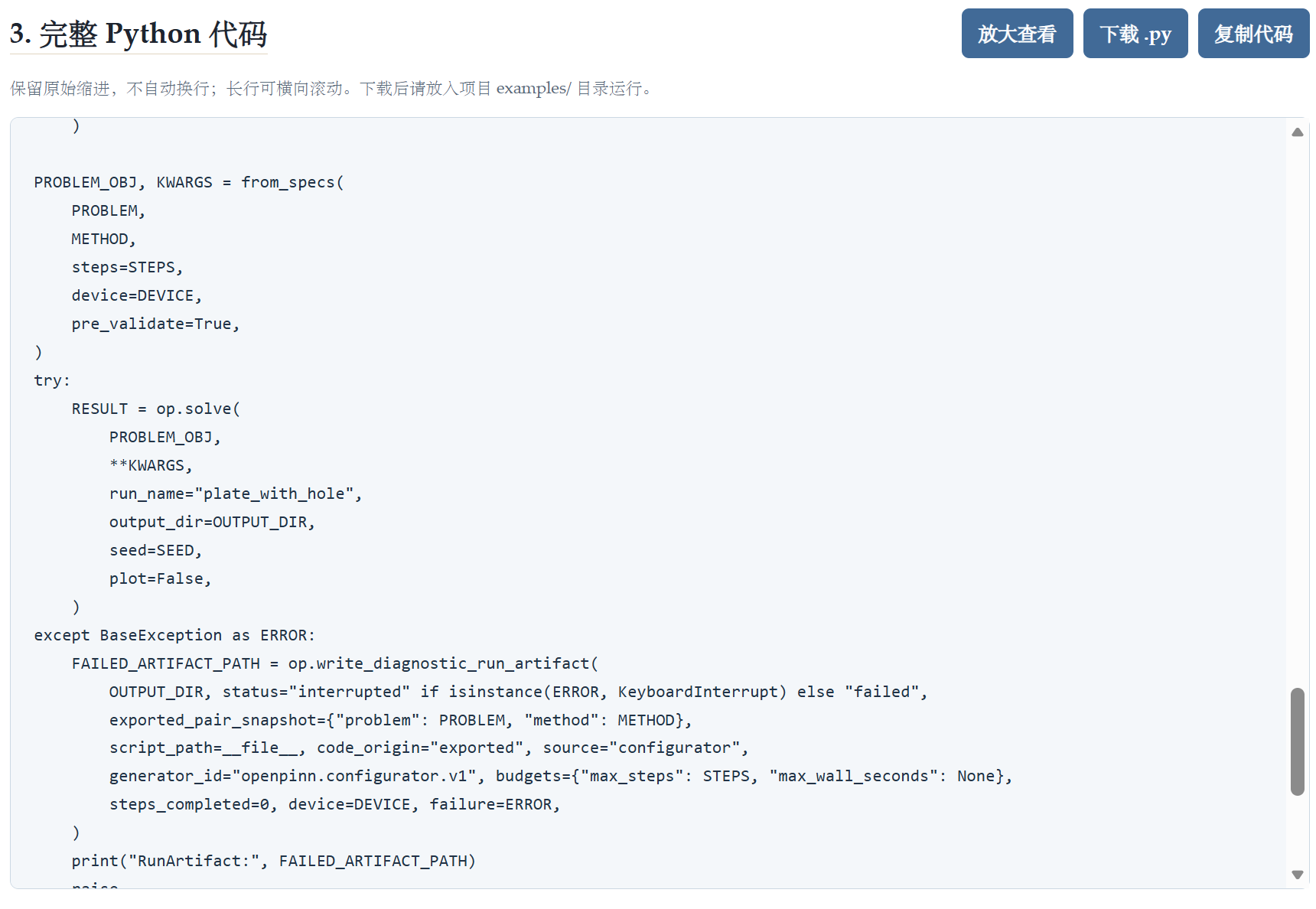}
  \caption{已校验带孔平板规格的代码构建智能体导出。}
  \label{fig:ch5-plate-codegen}
\end{figure}

代码构建智能体经确定性模板导出，由已校验规格组装训练脚本。
研究者下载该脚本并在本地运行训练。
图~\ref{fig:ch5-plate-codegen}给出导出的Python项目面板。

研究者在CPU上以SI单位、种子~$0$运行下载脚本，共$10000$步Adam。
训练写出含损失历史、指标、状态记录与诊断运行产物的运行目录。
研究者将该运行产物提交给诊断评估智能体。
诊断评估智能体返回评估结论为需关注。
其确定性轨道将运行标为有效但不健康，结果为预算耗尽。
它报告最终总损失$2.19\times10^{22}$，相对初值在整段预算内仅下降约$1.6\%$，PDE残差RMSE约$1.71\times10^{10}$，面力边界RMSE量级为$10^{10}$。
语义报告就残差缩放、Neumann面力执行以及加入可信参考给出非约束性修复提示。
图~\ref{fig:ch5-plate-diag1}给出首轮诊断报告。

\begin{figure}[H]
  \centering
  \includegraphics[width=0.75\linewidth]{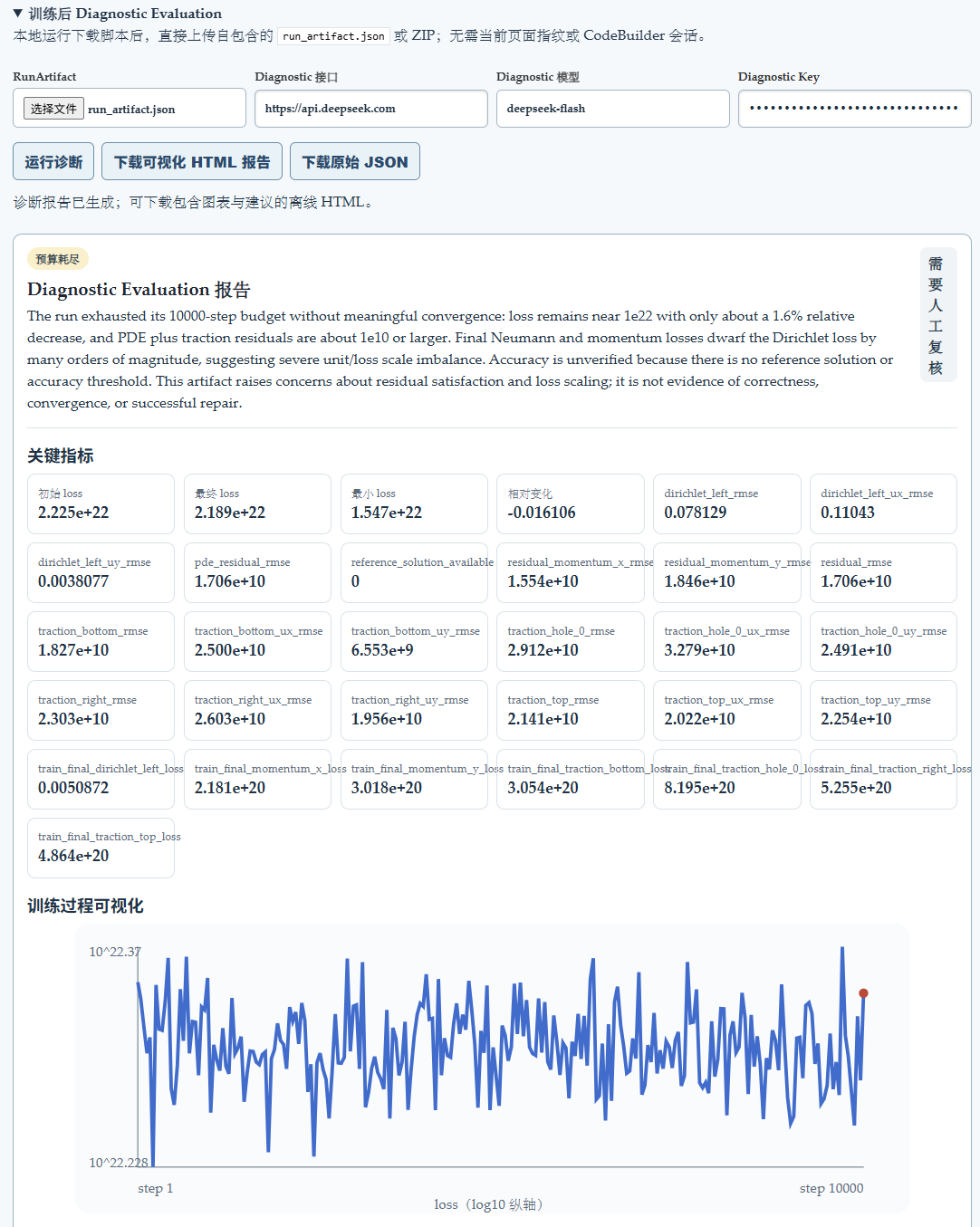}
  \caption{SI Adam带孔平板运行的首轮诊断评估智能体报告。}
  \label{fig:ch5-plate-diag1}
\end{figure}

图~\ref{fig:ch5-plate-si-disp}与图~\ref{fig:ch5-plate-si-stress}将SI Adam位移与应力场与FEM参考比较。
预测位移量级为$10^{-1}\,\mathrm{m}$，而FEM位移量级为$10^{-5}\,\mathrm{m}$。
FEM探测网格上的位移相对$L^2$为$3.65\times10^{4}$。

\begin{figure}[H]
  \centering
  \includegraphics[width=0.7\linewidth]{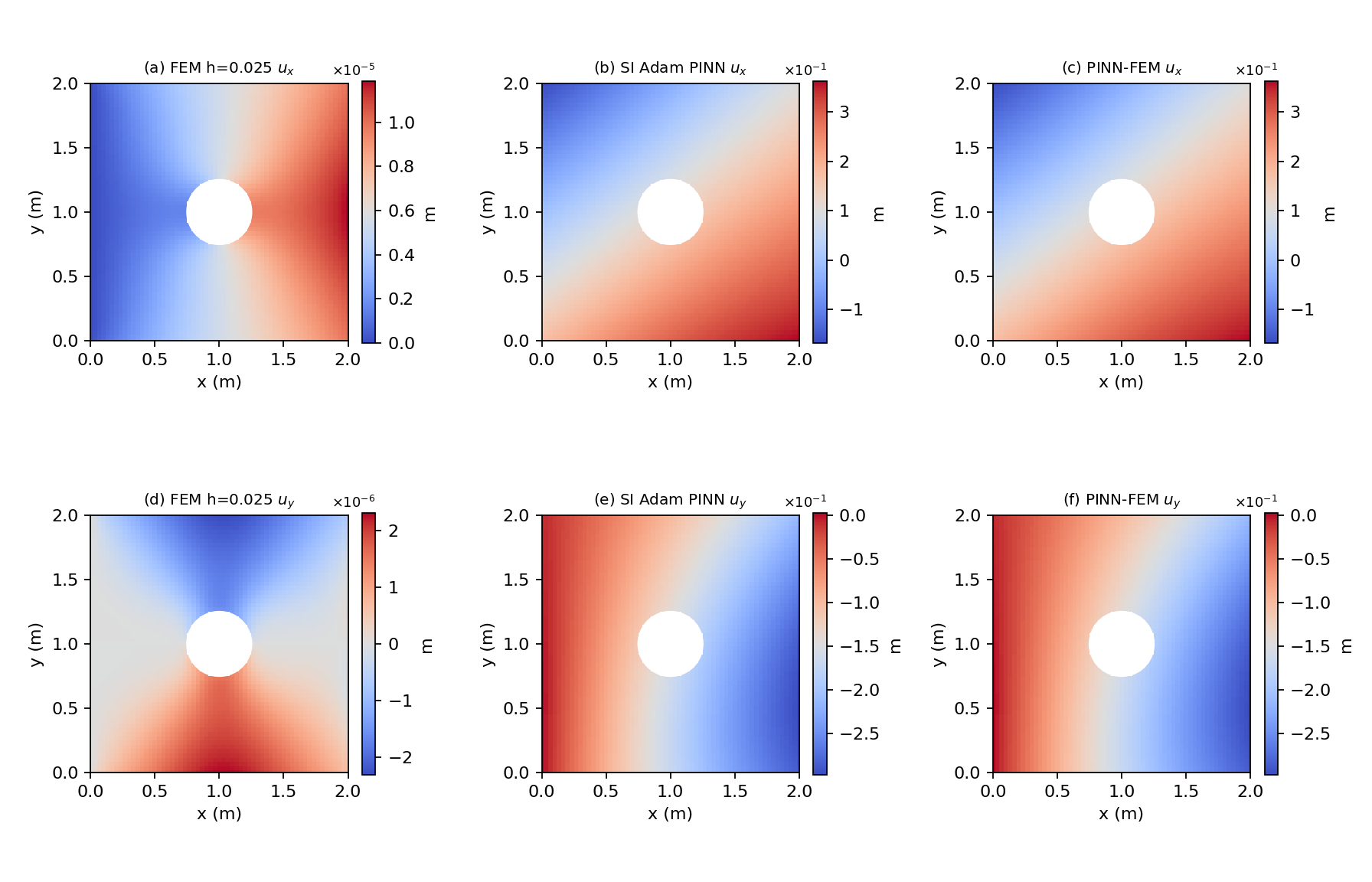}
  \caption{首轮诊断。
  失败的SI训练运行后的SI Adam位移。}
  \label{fig:ch5-plate-si-disp}
\end{figure}

\begin{figure}[H]
  \centering
  \includegraphics[width=0.7\linewidth]{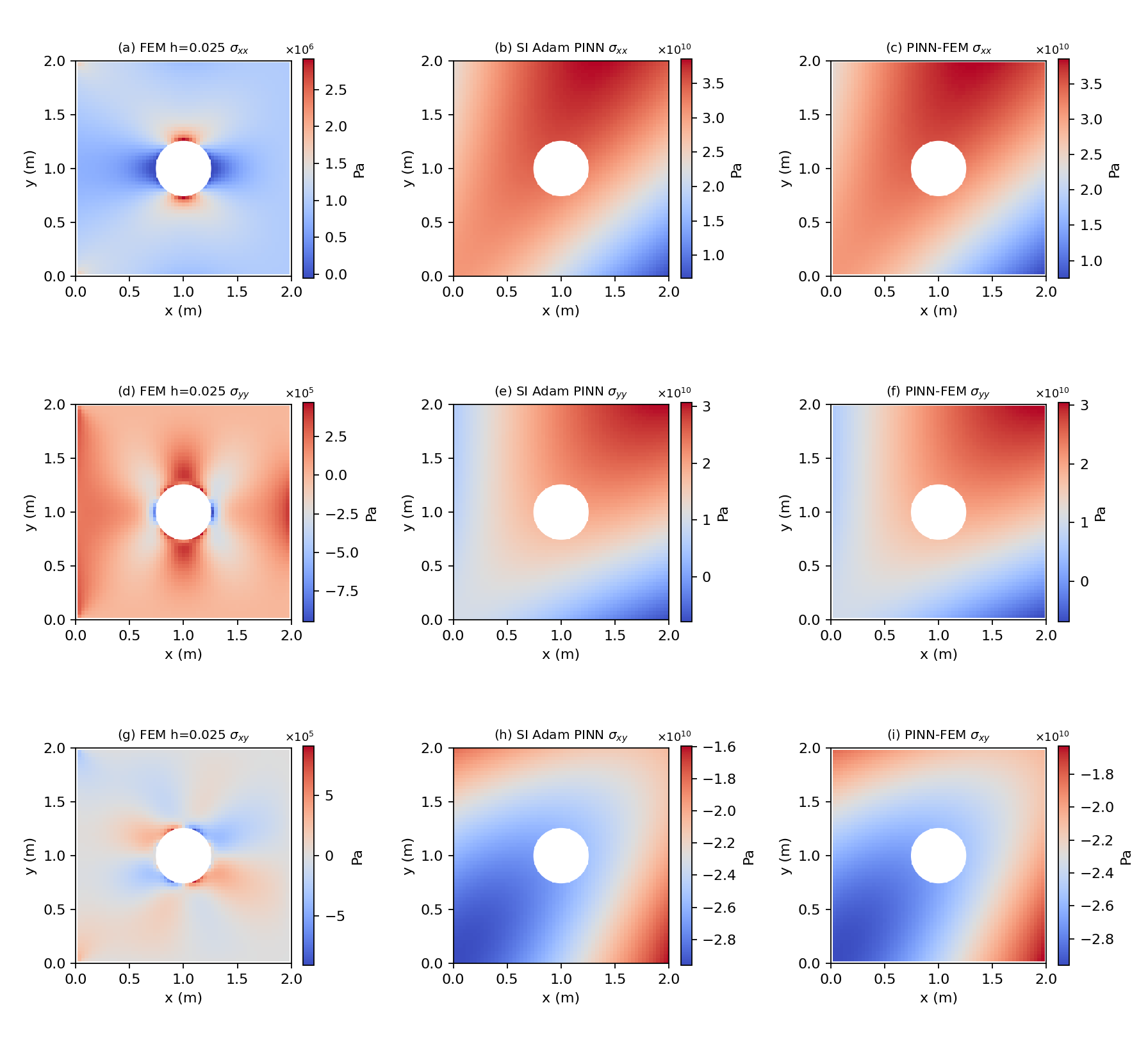}
  \caption{首轮诊断。
  失败的SI训练运行后$\sigma_{xx}$、$\sigma_{yy}$与$\sigma_{xy}$的SI Adam应力。}
  \label{fig:ch5-plate-si-stress}
\end{figure}

遵循诊断评估智能体关于残差缩放的提示，研究者主导的排查将SI失败归因于大残差尺度下float32 Adam二阶矩溢出。
公平的无量纲修复保持同一网络与Adam设置，仅改尺度$L=1\,\mathrm{m}$、$S=10^6\,\mathrm{Pa}$与$U=SL/E$。
三个种子各$10000$步Adam，相对$h=0.025$留出FEM探测网格的位移相对$L^2$误差分别达到$0.758$、$0.765$与$0.775$。
图~\ref{fig:ch5-plate-adam-disp}与图~\ref{fig:ch5-plate-adam-stress}将无量纲Adam种子~$0$的位移与应力场与FEM参考比较。

\begin{figure}[H]
  \centering
  \includegraphics[width=0.7\linewidth]{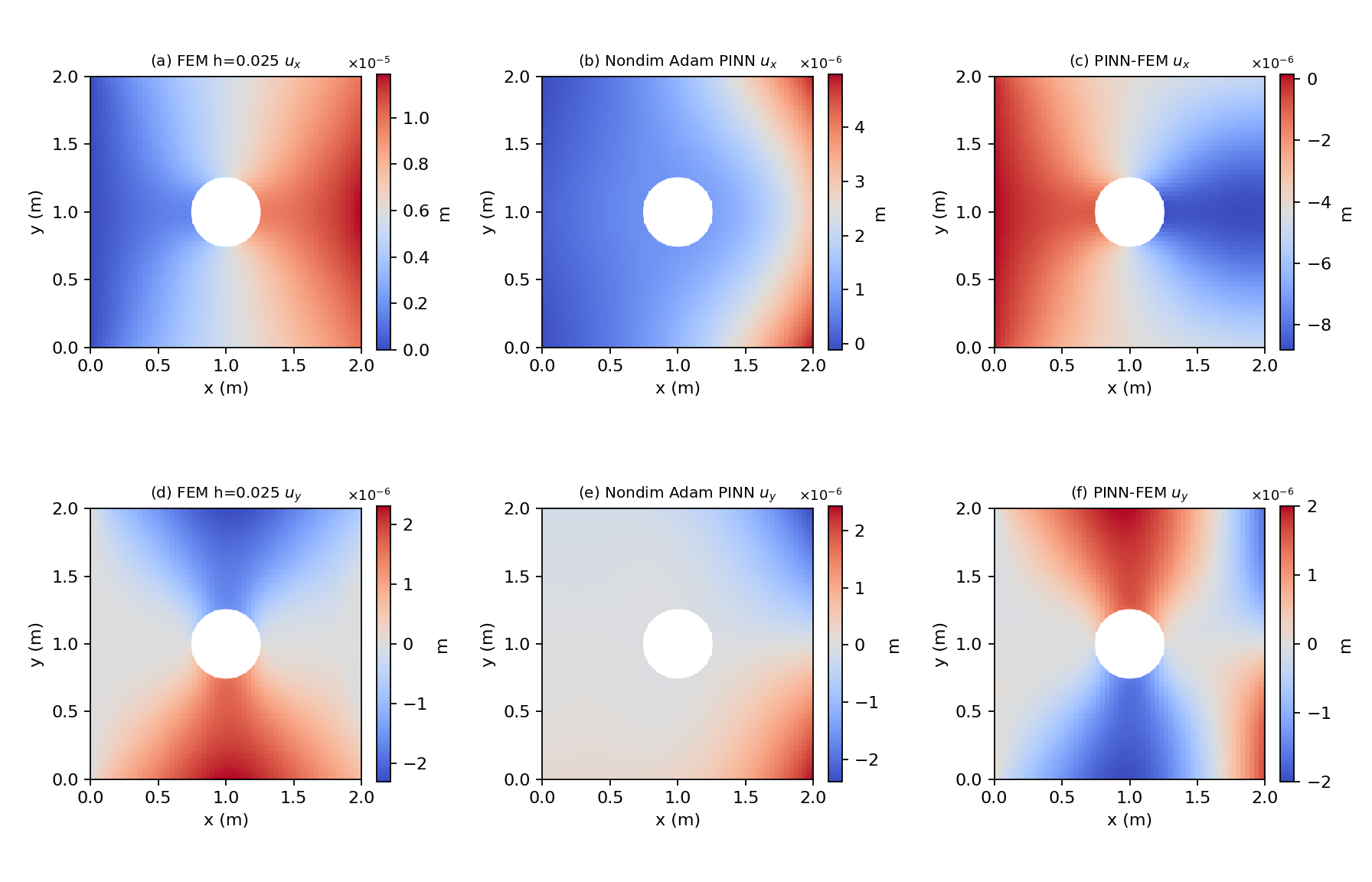}
  \caption{第二轮诊断。
  残差尺度修复后的无量纲Adam位移。}
  \label{fig:ch5-plate-adam-disp}
\end{figure}

\begin{figure}[H]
  \centering
  \includegraphics[width=0.7\linewidth]{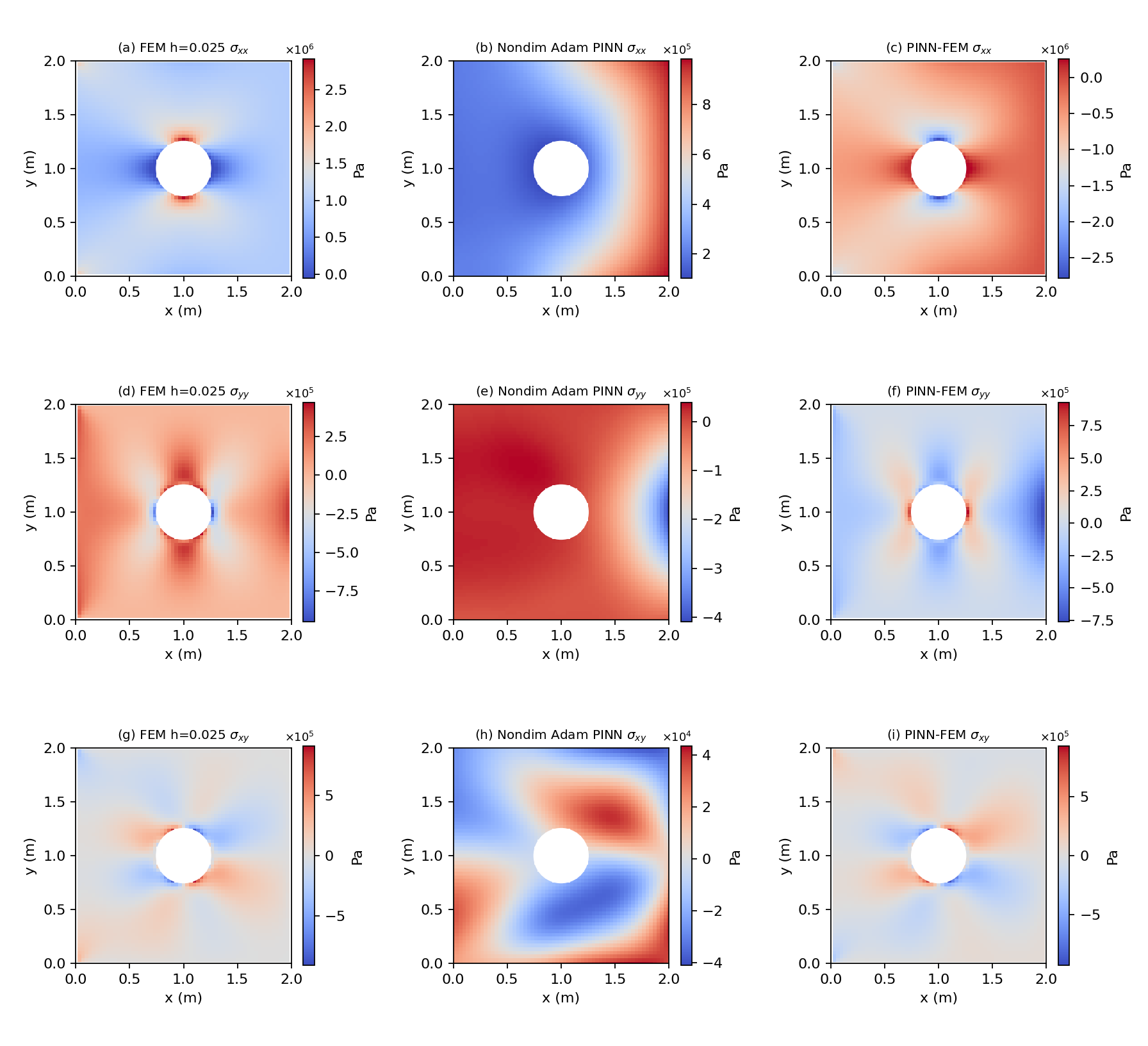}
  \caption{第二轮诊断。
  残差尺度修复后$\sigma_{xx}$、$\sigma_{yy}$与$\sigma_{xy}$的无量纲Adam应力。}
  \label{fig:ch5-plate-adam-stress}
\end{figure}

对无量纲Adam种子~$0$产物再次调用诊断评估智能体，仍返回需关注。
它报告升高的$x$动量与孔面力残差，并给出非约束性修复提示：审查主导$x$动量残差，并结合该残差检查孔面力$u_x$误差。
在这些提示引导下，研究者设计更强的优化延续。
网络、损失组装与无量纲尺度保持不变。
每个种子从其无量纲Adam检查点热启动，用已注册L-BFGS优化器替换Adam阶段，并冻结更大的均匀配点批量，含$2048$个内部点与$512$个边界点，L-BFGS阶段不重采样。
相对此前每步重抽的$512$内部点与$128$边界点Adam配方，该固定更大批量为L-BFGS提供稳定残差景观，以便对已无量纲化问题做局部精化。
每个种子再运行$260$次L-BFGS外步。
最终种子~$0$位移相对$L^2$从无量纲Adam下的$0.758$降至FEM探测网格上的$3.51\times10^{-2}$，在平移留出探测集上为$3.52\times10^{-2}$。
相应应力相对$L^2$从$0.713$降至$3.66\times10^{-2}$，无量纲$x$动量残差RMSE从$0.494$降至$0.113$。
种子~$1$与~$2$在同一FEM探测网格上分别达到相对$L^2$值$4.13\times10^{-2}$与$6.31\times10^{-2}$。
图~\ref{fig:ch5-plate-disp}将最终种子~$0$的L-BFGS位移场与FEM参考比较。
图~\ref{fig:ch5-plate-stress}比较相应应力分量$\sigma_{xx}$、$\sigma_{yy}$与$\sigma_{xy}$。

诊断路径因此分两阶段运行。
首轮诊断在SI Adam训练失败后标出异常残差尺度，研究者在同一Adam配方下对问题无量纲化。
第二轮诊断报告升高的$x$动量与孔缘面力残差，研究者从无量纲Adam检查点以固定大批量L-BFGS延续，直至三个种子均达到位移目标。
贯穿完整智能体工作流，问题理解与求解器规划智能体起草规格，代码构建智能体导出本地训练脚本，诊断评估智能体两次定位残差尺度与残差失衡线索，并引导这些研究者主导的修复。

\begin{figure}[H]
  \centering
  \includegraphics[width=0.7\linewidth]{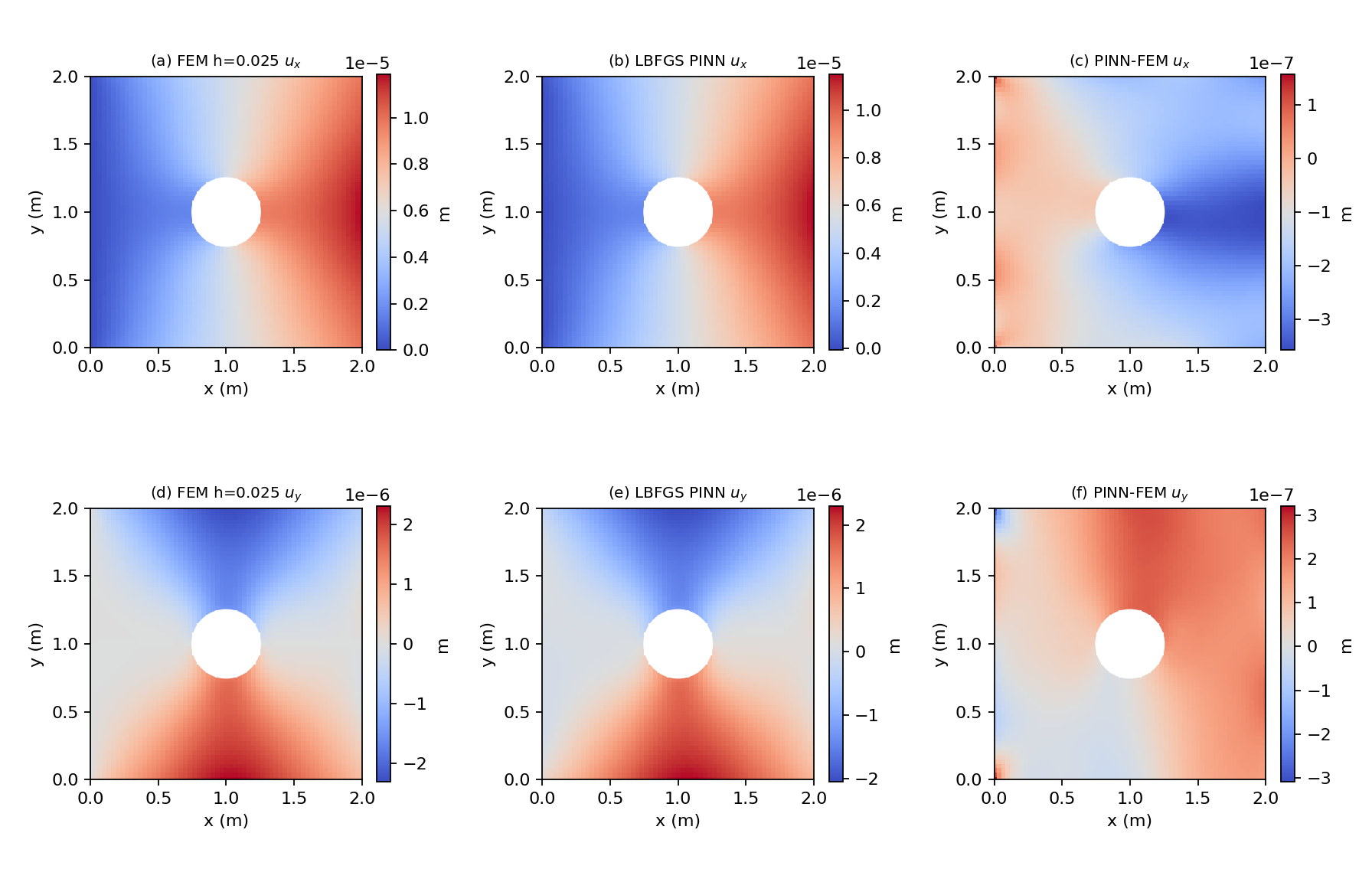}
  \caption{第二轮诊断与固定批量L-BFGS延续后的最终结果。
  位移场。}
  \label{fig:ch5-plate-disp}
\end{figure}

\begin{figure}[H]
  \centering
  \includegraphics[width=0.7\linewidth]{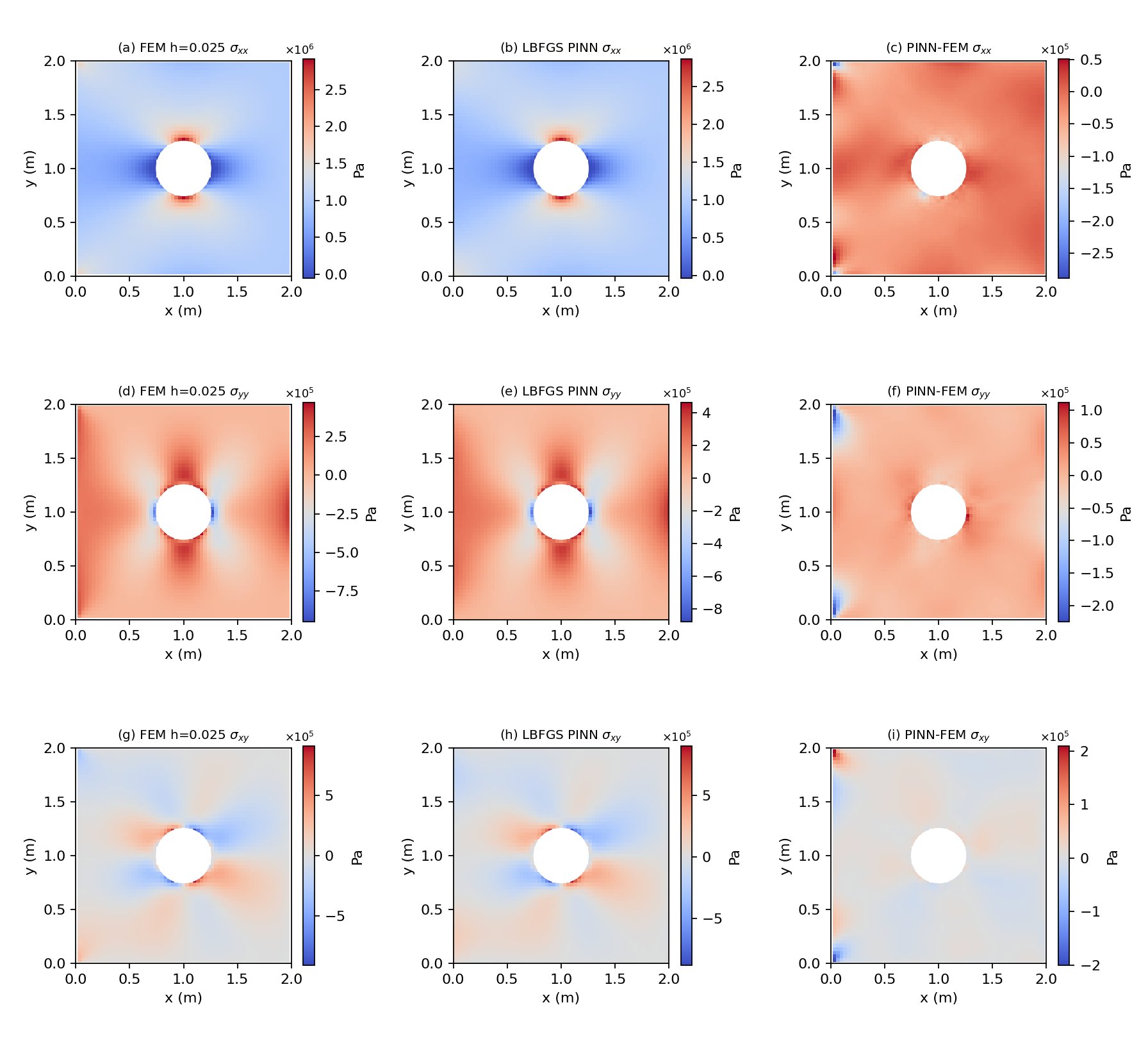}
  \caption{第二轮诊断与固定批量L-BFGS延续后的最终结果。
  $\sigma_{xx}$、$\sigma_{yy}$与$\sigma_{xy}$应力场。}
  \label{fig:ch5-plate-stress}
\end{figure}

图~\ref{fig:ch5-plate-loss}给出Adam及随后L-BFGS阶段的无量纲损失历史。

\begin{figure}[H]
  \centering
  \includegraphics[width=0.78\linewidth]{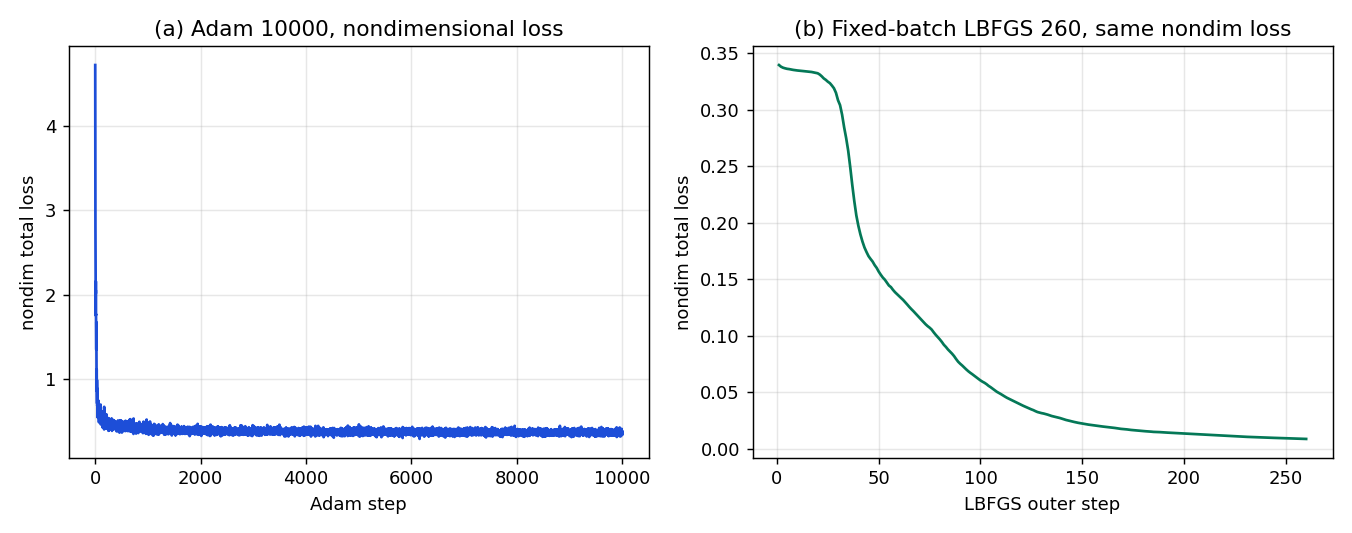}
  \caption{Adam $10000$步与热启动固定批量L-BFGS延续的无量纲总损失。}
  \label{fig:ch5-plate-loss}
\end{figure}

\section{局限与展望}
\label{sec:limitations}

OpenPINN/PINO的首批演示范围聚焦于平滑教程PDE与已注册算子族上的单轴方法比较，以及已交付的校验到代码生成路径。
第~\ref{sec:ex-agent}节在带孔平板弹性算例上演练了早期问题理解、求解器规划与诊断评估接口。
这些调用仍需要人工补全缺失的几何与边界字段。
复杂接触、强几何非线性、三维非结构化算子，以及任意模型--导数交叉乘积不在本文声称范围内。
已交付的规格校验智能体与确定性代码构建智能体覆盖结构化配置器与YAML输入工作流的中段。
相对目标工作流，闭环规格调整仍不完整。

未来工作包括将理解、规划与诊断评估加固到可交接与闭环规格调整，跨应用域扩展共享基准与参考比较，加强规格中的溯源，并在结构化网格上审慎扩展算子物理，避免坐标与网格导数路径之间的非正式交叉适配。

\section{结论}
\label{sec:conclusion}

我们提出OpenPINN/PINO，作为一个围绕前处理–计算–后处理整体架构以及五模块智能体辅助工作流构建的智能体辅助科学计算平台。
共享问题规格与方法规格契约编码物理问题与求解方法，驱动本地训练的PINN、神经算子与自定义后端，并支持后处理为指标、可视化与实验记录。
智能体工作流帮助研究者表述问题、规划求解器、获得可运行本地代码，并诊断失败训练。
上游智能体起草问题规格与方法规格，标出缺失字段，并提出基线配方。
已校验规格被组装为确定性训练脚本供本地执行。
诊断评估随后审阅运行产物，并浮现引导研究者主导修复的残差尺度与残差失衡线索。
我们希望OpenPINN/PINO帮助研究者与教育者以更少摩擦、更多可问责性构建、共享并核验物理信息学习工作流。

\section*{数据与代码可用性}
OpenPINN/PINO源代码以Apache-2.0许可发布于\url{https://github.com/junelu0214/OpenPINN-PINO}~\cite{openpinn_upstream}。

\clearpage
\renewcommand{\refname}{参考文献}
\makeatletter
\def\@bibitem#1{%
  \@skiphyperreftrue\H@item\@skiphyperreffalse
  \Hy@raisedlink{%
    \hyper@anchorstart{cite.#1.zh}\relax\hyper@anchorend
  }%
  \if@filesw
    \immediate\write\@auxout{\string\bibcitezh{#1}{\the\value{\@listctr}}}%
  \fi
  \ignorespaces
}
\def\@lbibitem[#1]#2{%
  \@skiphyperreftrue
  \H@item[%
    \Hy@raisedlink{\hyper@anchorstart{cite.#2.zh}\hyper@anchorend}%
    \@biblabel{#1}\hfill
  ]%
  \@skiphyperreffalse
  \if@filesw
    \immediate\write\@auxout{\string\bibcitezh{#2}{#1}}%
  \fi
  \ignorespaces
}
\makeatother
\input{\jobname.bbl}

\end{document}